\documentclass[longauth]{aa}

\usepackage{euclid}
\usepackage{graphicx}
\usepackage{natbib}
\usepackage{scalerel}

\usepackage{comment}

\usepackage[table]{xcolor}

\usepackage{txfonts}
\usepackage{glossaries}
\usepackage[pdfencoding=auto,psdextra]{hyperref}
\hypersetup{
    colorlinks=true,
    linkcolor=blue,
    filecolor=magenta,      
    urlcolor=blue,
    citecolor=blue
}
\makeatletter
\renewcommand*\aa@pageof{, page \thepage{} of \pageref*{LastPage}}
\makeatother

\usepackage[utf8]{inputenc}

\usepackage[switch, modulo]{lineno}
\renewcommand{\linenumbers}[0]{}
\usepackage{multirow}
\usepackage{bm}         
\usepackage{widetext}
\usepackage{enumitem}

\newacronym{hod}{HOD}{halo occupation distribution}
\newacronym{dic}{DIC}{deviance information criterion}
\newacronym{bao}{BAO}{baryon acoustic oscillation}
\newacronym{cmb}{CMB}{cosmic microwave background}
\newacronym{lss}{LSS}{large-scale structure}
\newacronym{gr}{GR}{general relativity}
\newacronym{flrw}{FLRW}{Friedmann--Lema\^{i}tre--Robertson--Walker}
\newacronym{2pcf}{2PCF}{2-point correlation function}
\newacronym{rsd}{RSD}{redshift-space distortions}
\newacronym{cscs}{CSCS}{Swiss National Supercomputing Center}
\newacronym{fof}{FOF}{friend-of-friend}
\newacronym{pt}{PT}{perturbation theory}
\newacronym{ir}{IR}{infrared}
\newacronym{uv}{UV}{ultraviolet}
\newacronym{fob}{FoB}{figure of bias}
\newacronym{fom}{FoM}{figure of merit}
\newacronym{ap}{AP}{Alcock--Paczynski}
\newacronym{ppd}{PPD}{posterior predictive distribution}
\newacronym{de}{DE}{dark energy}
\newacronym{spt}{SPT}{standard perturbation theory}
\newacronym{ept}{EPT}{Eulerian perturbation theory}
\newacronym{eft}{EFT}{effective field theory of large-scale structure}
\newacronym{vdg}{VDG$_\infty$}{velocity difference-generating function at infinity}
\newacronym{vdgf}{VDG}{velocity difference-generating function}

\newcommand{\rsd}{\gls{rsd}\xspace}
\newcommand{\spt}{\gls{spt}\xspace}
\newcommand{\ept}{\gls{ept}\xspace}
\newcommand{\eft}{\gls{eft}\xspace}
\newcommand{\vdg}{\gls{vdg}\xspace}
\newcommand{\vdgf}{\gls{vdgf}\xspace}
\newcommand{\bao}{\gls{bao}\xspace}
\newcommand{\ir}{\gls{ir}\xspace}
\newcommand{\hod}{\gls{hod}\xspace}
\newcommand{\fob}{\gls{fob}\xspace}
\newcommand{\fom}{\gls{fom}\xspace}
\newcommand{\ppd}{\gls{ppd}\xspace}
\newcommand{\pvalue}{$p$-value\xspace}
\newcommand{\uv}{\gls{uv}\xspace}
\newcommand{\pt}{\gls{pt}\xspace}

\newcommand{\nbody}{\textit{N}-body }

\newcommand{\eul}{{\rm e}}
\newcommand{\imag}{{\rm i}}

\renewcommand{\vec}{\mathbf}

\newcommand{\bdm}{\begin{displaymath}}
\newcommand{\edm}{\end{displaymath}}

\newcommand{\xv}{{\bm x}}
\newcommand{\vv}{{\bm v}}

\newcommand{\kv}{{\bm k}}

\newcommand{\rv}{{\bm r}}
\newcommand{\qv}{{\bm q}}

\newcommand{\kmax}{k_{\rm max}}

\newcommand{\lbox}{L_{\rm box}}

\newcommand{\eg}{{e.g.}~}
\newcommand{\Ms}{\, h^{-1} \, M_\odot}
\newcommand{\Mpc}{\, h^{-1} \, {\rm Mpc}}

\newcommand{\Gpc}{\, h^{-1} \, {\rm Gpc}}

\newcommand{\kMpc}{\, h \, {\rm Mpc}^{-1}}
\newcommand{\kcMpc}{\, h^3 \, {\rm Mpc}^{-3}}
\newcommand{\Hunits}{\, {\rm km \, s^{-1} \, Mpc^{-1}}}

\newcommand{\omegac}{\omega_{\rm c}}
\newcommand{\omegab}{\omega_{\hbox{\hglue 0.5pt}\rm b}}
\newcommand{\deltag}{\delta_{\rm g}}
\newcommand{\As}{A_{\rm s}}
\newcommand{\ns}{n_{\rm s}}
\newcommand{\Pgg}{P_{\rm gg}}
\newcommand{\kF}{k_{\,\rm F}}
\newcommand{\kN}{k_{\,\rm Nyq}}
\newcommand{\dirac}{\delta_{\rm D}}
\newcommand{\deltainit}{\delta_{\rm L}}

\newcommand{\Plin}{P_{\rm L}}
\newcommand{\Pmm}{P_{\rm mm}}

\newcommand{\Pnw}{P_{\rm nw}}
\newcommand{\Pw}{P_{\rm w}}

\newcommand{\bdtwod}{b_{\nabla^{\,2}\delta}}
\newcommand{\bone}{b_1}
\newcommand{\btwo}{b_2}
\newcommand{\gtwo}{\gamma_2}
\newcommand{\gtwoone}{\gamma_{21}}
\newcommand{\czero}{c_0}
\newcommand{\ctwo}{c_2}
\newcommand{\cfour}{c_4}
\newcommand{\cnlo}{c_{\rm nlo}}
\newcommand{\npzero}{N_0^P}
\newcommand{\nptwozero}{N_{20}^P}
\newcommand{\nptwotwo}{N_{22}^P}

\newcommand{\thetav}{\bm{\theta}}

\newcommand{\sv}{ \pmb{s} }

\newcommand{\lamsat}{ \lambda_{\rm sat} }
\newcommand{\fsat}{ f_{\rm sat} }

\newcommand{\avir}{a_{\rm vir}}

\newcommand{\bacco}{{\tt BACCO}\xspace}

\begin{document}
%
%
\title{\Euclid preparation}
\subtitle{Galaxy power spectrum modelling in redshift space}    

\newcommand{\orcid}[1]{} 
\author{Euclid Collaboration: B.~Camacho~Quevedo\orcid{0000-0002-8789-4232}\thanks{\email{bcamacho@sissa.it}}\inst{\ref{aff1},\ref{aff2},\ref{aff3},\ref{aff6},\ref{aff5}}
\and M.~Crocce\orcid{0000-0002-9745-6228}\inst{\ref{aff3},\ref{aff6}}
\and M.~Pellejero~Iba\~nez\orcid{0000-0003-4680-7275}\inst{\ref{aff7}}
\and R.~E.~Angulo\orcid{0000-0003-2953-3970}\inst{\ref{aff8},\ref{aff9}}
\and A.~Pezzotta\orcid{0000-0003-0726-2268}\inst{\ref{aff10}}
\and A.~Eggemeier\orcid{0000-0002-1841-8910}\inst{\ref{aff11}}
\and G.~Gambardella\orcid{0009-0001-1281-1746}\inst{\ref{aff3},\ref{aff6}}
\and C.~Moretti\orcid{0000-0003-3314-8936}\inst{\ref{aff5},\ref{aff1},\ref{aff12}}
\and E.~Sefusatti\orcid{0000-0003-0473-1567}\inst{\ref{aff5},\ref{aff1},\ref{aff12}}
\and A.~Moradinezhad~Dizgah\orcid{0000-0001-8841-9989}\inst{\ref{aff13}}
\and E.~Gaztanaga\orcid{0000-0001-9632-0815}\inst{\ref{aff3},\ref{aff6},\ref{aff14}}
\and M.~Zennaro\orcid{0000-0002-4458-1754}\inst{\ref{aff15}}
\and M.-A.~Breton\inst{\ref{aff16}}
\and A.~Chudaykin\inst{\ref{aff17}}
\and G.~D'Amico\orcid{0000-0002-8183-1214}\inst{\ref{aff18},\ref{aff19}}
\and V.~Desjacques\orcid{0000-0003-2062-8172}\inst{\ref{aff20}}
\and S.~de~la~Torre\inst{\ref{aff21}}
\and P.~Fosalba\orcid{0000-0002-1510-5214}\inst{\ref{aff6},\ref{aff3}}
\and M.~Guidi\orcid{0000-0001-9408-1101}\inst{\ref{aff22},\ref{aff23}}
\and M.~K{\"a}rcher\orcid{0000-0001-5868-647X}\inst{\ref{aff24}}
\and K.~Pardede\orcid{0000-0002-7728-8220}\inst{\ref{aff19}}
\and C.~Porciani\orcid{0000-0002-7797-2508}\inst{\ref{aff11}}
\and A.~Pugno\inst{\ref{aff11}}
\and J.~Salvalaggio\orcid{0000-0002-1431-5607}\inst{\ref{aff5},\ref{aff1},\ref{aff25},\ref{aff12}}
\and E.~Sarpa\orcid{0000-0002-1256-655X}\inst{\ref{aff2},\ref{aff26},\ref{aff5}}
\and A.~Veropalumbo\orcid{0000-0003-2387-1194}\inst{\ref{aff10},\ref{aff27},\ref{aff28}}
\and B.~Altieri\orcid{0000-0003-3936-0284}\inst{\ref{aff29}}
\and S.~Andreon\orcid{0000-0002-2041-8784}\inst{\ref{aff10}}
\and N.~Auricchio\orcid{0000-0003-4444-8651}\inst{\ref{aff23}}
\and C.~Baccigalupi\orcid{0000-0002-8211-1630}\inst{\ref{aff1},\ref{aff5},\ref{aff12},\ref{aff2}}
\and M.~Baldi\orcid{0000-0003-4145-1943}\inst{\ref{aff22},\ref{aff23},\ref{aff30}}
\and S.~Bardelli\orcid{0000-0002-8900-0298}\inst{\ref{aff23}}
\and R.~Bender\orcid{0000-0001-7179-0626}\inst{\ref{aff31},\ref{aff32}}
\and A.~Biviano\orcid{0000-0002-0857-0732}\inst{\ref{aff5},\ref{aff1}}
\and E.~Branchini\orcid{0000-0002-0808-6908}\inst{\ref{aff28},\ref{aff27},\ref{aff10}}
\and M.~Brescia\orcid{0000-0001-9506-5680}\inst{\ref{aff33},\ref{aff34}}
\and S.~Camera\orcid{0000-0003-3399-3574}\inst{\ref{aff35},\ref{aff36},\ref{aff37}}
\and V.~Capobianco\orcid{0000-0002-3309-7692}\inst{\ref{aff37}}
\and C.~Carbone\orcid{0000-0003-0125-3563}\inst{\ref{aff38}}
\and V.~F.~Cardone\inst{\ref{aff39},\ref{aff40}}
\and J.~Carretero\orcid{0000-0002-3130-0204}\inst{\ref{aff41},\ref{aff42}}
\and S.~Casas\orcid{0000-0002-4751-5138}\inst{\ref{aff43},\ref{aff44}}
\and F.~J.~Castander\orcid{0000-0001-7316-4573}\inst{\ref{aff3},\ref{aff6}}
\and M.~Castellano\orcid{0000-0001-9875-8263}\inst{\ref{aff39}}
\and G.~Castignani\orcid{0000-0001-6831-0687}\inst{\ref{aff23}}
\and S.~Cavuoti\orcid{0000-0002-3787-4196}\inst{\ref{aff34},\ref{aff45}}
\and K.~C.~Chambers\orcid{0000-0001-6965-7789}\inst{\ref{aff46}}
\and A.~Cimatti\inst{\ref{aff47}}
\and C.~Colodro-Conde\inst{\ref{aff48}}
\and G.~Congedo\orcid{0000-0003-2508-0046}\inst{\ref{aff7}}
\and L.~Conversi\orcid{0000-0002-6710-8476}\inst{\ref{aff49},\ref{aff29}}
\and Y.~Copin\orcid{0000-0002-5317-7518}\inst{\ref{aff50}}
\and F.~Courbin\orcid{0000-0003-0758-6510}\inst{\ref{aff51},\ref{aff52},\ref{aff4}}
\and H.~M.~Courtois\orcid{0000-0003-0509-1776}\inst{\ref{aff53}}
\and H.~Degaudenzi\orcid{0000-0002-5887-6799}\inst{\ref{aff54}}
\and G.~De~Lucia\orcid{0000-0002-6220-9104}\inst{\ref{aff5}}
\and H.~Dole\orcid{0000-0002-9767-3839}\inst{\ref{aff55}}
\and M.~Douspis\orcid{0000-0003-4203-3954}\inst{\ref{aff55}}
\and F.~Dubath\orcid{0000-0002-6533-2810}\inst{\ref{aff54}}
\and C.~A.~J.~Duncan\orcid{0009-0003-3573-0791}\inst{\ref{aff7}}
\and X.~Dupac\inst{\ref{aff29}}
\and S.~Dusini\orcid{0000-0002-1128-0664}\inst{\ref{aff56}}
\and S.~Escoffier\orcid{0000-0002-2847-7498}\inst{\ref{aff57}}
\and M.~Farina\orcid{0000-0002-3089-7846}\inst{\ref{aff58}}
\and R.~Farinelli\inst{\ref{aff23}}
\and S.~Farrens\orcid{0000-0002-9594-9387}\inst{\ref{aff16}}
\and S.~Ferriol\inst{\ref{aff50}}
\and F.~Finelli\orcid{0000-0002-6694-3269}\inst{\ref{aff23},\ref{aff59}}
\and S.~Fotopoulou\orcid{0000-0002-9686-254X}\inst{\ref{aff60}}
\and N.~Fourmanoit\orcid{0009-0005-6816-6925}\inst{\ref{aff57}}
\and M.~Frailis\orcid{0000-0002-7400-2135}\inst{\ref{aff5}}
\and E.~Franceschi\orcid{0000-0002-0585-6591}\inst{\ref{aff23}}
\and M.~Fumana\orcid{0000-0001-6787-5950}\inst{\ref{aff38}}
\and S.~Galeotta\orcid{0000-0002-3748-5115}\inst{\ref{aff5}}
\and K.~George\orcid{0000-0002-1734-8455}\inst{\ref{aff61}}
\and B.~Gillis\orcid{0000-0002-4478-1270}\inst{\ref{aff7}}
\and C.~Giocoli\orcid{0000-0002-9590-7961}\inst{\ref{aff23},\ref{aff30}}
\and J.~Gracia-Carpio\orcid{0000-0003-4689-3134}\inst{\ref{aff31}}
\and A.~Grazian\orcid{0000-0002-5688-0663}\inst{\ref{aff62}}
\and F.~Grupp\inst{\ref{aff31},\ref{aff32}}
\and L.~Guzzo\orcid{0000-0001-8264-5192}\inst{\ref{aff24},\ref{aff10},\ref{aff63}}
\and S.~V.~H.~Haugan\orcid{0000-0001-9648-7260}\inst{\ref{aff64}}
\and W.~Holmes\inst{\ref{aff65}}
\and F.~Hormuth\inst{\ref{aff66}}
\and A.~Hornstrup\orcid{0000-0002-3363-0936}\inst{\ref{aff67},\ref{aff68}}
\and K.~Jahnke\orcid{0000-0003-3804-2137}\inst{\ref{aff69}}
\and B.~Joachimi\orcid{0000-0001-7494-1303}\inst{\ref{aff70}}
\and S.~Kermiche\orcid{0000-0002-0302-5735}\inst{\ref{aff57}}
\and A.~Kiessling\orcid{0000-0002-2590-1273}\inst{\ref{aff65}}
\and B.~Kubik\orcid{0009-0006-5823-4880}\inst{\ref{aff50}}
\and M.~K\"ummel\orcid{0000-0003-2791-2117}\inst{\ref{aff32}}
\and M.~Kunz\orcid{0000-0002-3052-7394}\inst{\ref{aff17}}
\and H.~Kurki-Suonio\orcid{0000-0002-4618-3063}\inst{\ref{aff71},\ref{aff72}}
\and A.~M.~C.~Le~Brun\orcid{0000-0002-0936-4594}\inst{\ref{aff73}}
\and S.~Ligori\orcid{0000-0003-4172-4606}\inst{\ref{aff37}}
\and P.~B.~Lilje\orcid{0000-0003-4324-7794}\inst{\ref{aff64}}
\and V.~Lindholm\orcid{0000-0003-2317-5471}\inst{\ref{aff71},\ref{aff72}}
\and I.~Lloro\orcid{0000-0001-5966-1434}\inst{\ref{aff74}}
\and G.~Mainetti\orcid{0000-0003-2384-2377}\inst{\ref{aff75}}
\and E.~Maiorano\orcid{0000-0003-2593-4355}\inst{\ref{aff23}}
\and O.~Mansutti\orcid{0000-0001-5758-4658}\inst{\ref{aff5}}
\and S.~Marcin\inst{\ref{aff76}}
\and O.~Marggraf\orcid{0000-0001-7242-3852}\inst{\ref{aff11}}
\and M.~Martinelli\orcid{0000-0002-6943-7732}\inst{\ref{aff39},\ref{aff40}}
\and N.~Martinet\orcid{0000-0003-2786-7790}\inst{\ref{aff21}}
\and F.~Marulli\orcid{0000-0002-8850-0303}\inst{\ref{aff77},\ref{aff23},\ref{aff30}}
\and R.~J.~Massey\orcid{0000-0002-6085-3780}\inst{\ref{aff78}}
\and E.~Medinaceli\orcid{0000-0002-4040-7783}\inst{\ref{aff23}}
\and M.~Melchior\inst{\ref{aff79}}
\and M.~Meneghetti\orcid{0000-0003-1225-7084}\inst{\ref{aff23},\ref{aff30}}
\and E.~Merlin\orcid{0000-0001-6870-8900}\inst{\ref{aff39}}
\and G.~Meylan\inst{\ref{aff80}}
\and A.~Mora\orcid{0000-0002-1922-8529}\inst{\ref{aff81}}
\and M.~Moresco\orcid{0000-0002-7616-7136}\inst{\ref{aff77},\ref{aff23}}
\and L.~Moscardini\orcid{0000-0002-3473-6716}\inst{\ref{aff77},\ref{aff23},\ref{aff30}}
\and C.~Neissner\orcid{0000-0001-8524-4968}\inst{\ref{aff82},\ref{aff42}}
\and S.-M.~Niemi\orcid{0009-0005-0247-0086}\inst{\ref{aff83}}
\and C.~Padilla\orcid{0000-0001-7951-0166}\inst{\ref{aff82}}
\and S.~Paltani\orcid{0000-0002-8108-9179}\inst{\ref{aff54}}
\and F.~Pasian\orcid{0000-0002-4869-3227}\inst{\ref{aff5}}
\and K.~Pedersen\inst{\ref{aff84}}
\and W.~J.~Percival\orcid{0000-0002-0644-5727}\inst{\ref{aff85},\ref{aff86},\ref{aff87}}
\and V.~Pettorino\orcid{0000-0002-4203-9320}\inst{\ref{aff83}}
\and S.~Pires\orcid{0000-0002-0249-2104}\inst{\ref{aff16}}
\and G.~Polenta\orcid{0000-0003-4067-9196}\inst{\ref{aff88}}
\and M.~Poncet\inst{\ref{aff89}}
\and L.~A.~Popa\inst{\ref{aff90}}
\and F.~Raison\orcid{0000-0002-7819-6918}\inst{\ref{aff31}}
\and J.~Rhodes\orcid{0000-0002-4485-8549}\inst{\ref{aff65}}
\and G.~Riccio\inst{\ref{aff34}}
\and F.~Rizzo\orcid{0000-0002-9407-585X}\inst{\ref{aff5}}
\and E.~Romelli\orcid{0000-0003-3069-9222}\inst{\ref{aff5}}
\and M.~Roncarelli\orcid{0000-0001-9587-7822}\inst{\ref{aff23}}
\and R.~Saglia\orcid{0000-0003-0378-7032}\inst{\ref{aff32},\ref{aff31}}
\and Z.~Sakr\orcid{0000-0002-4823-3757}\inst{\ref{aff91},\ref{aff92},\ref{aff93}}
\and A.~G.~S\'anchez\orcid{0000-0003-1198-831X}\inst{\ref{aff31}}
\and D.~Sapone\orcid{0000-0001-7089-4503}\inst{\ref{aff94}}
\and B.~Sartoris\orcid{0000-0003-1337-5269}\inst{\ref{aff32},\ref{aff5}}
\and P.~Schneider\orcid{0000-0001-8561-2679}\inst{\ref{aff11}}
\and A.~Secroun\orcid{0000-0003-0505-3710}\inst{\ref{aff57}}
\and G.~Seidel\orcid{0000-0003-2907-353X}\inst{\ref{aff69}}
\and E.~Sihvola\orcid{0000-0003-1804-7715}\inst{\ref{aff95}}
\and P.~Simon\inst{\ref{aff11}}
\and C.~Sirignano\orcid{0000-0002-0995-7146}\inst{\ref{aff96},\ref{aff56}}
\and G.~Sirri\orcid{0000-0003-2626-2853}\inst{\ref{aff30}}
\and A.~Spurio~Mancini\orcid{0000-0001-5698-0990}\inst{\ref{aff97}}
\and L.~Stanco\orcid{0000-0002-9706-5104}\inst{\ref{aff56}}
\and P.~Tallada-Cresp\'{i}\orcid{0000-0002-1336-8328}\inst{\ref{aff41},\ref{aff42}}
\and D.~Tavagnacco\orcid{0000-0001-7475-9894}\inst{\ref{aff5}}
\and A.~N.~Taylor\inst{\ref{aff7}}
\and I.~Tereno\orcid{0000-0002-4537-6218}\inst{\ref{aff98},\ref{aff99}}
\and N.~Tessore\orcid{0000-0002-9696-7931}\inst{\ref{aff100}}
\and S.~Toft\orcid{0000-0003-3631-7176}\inst{\ref{aff101},\ref{aff102}}
\and R.~Toledo-Moreo\orcid{0000-0002-2997-4859}\inst{\ref{aff103}}
\and F.~Torradeflot\orcid{0000-0003-1160-1517}\inst{\ref{aff42},\ref{aff41}}
\and I.~Tutusaus\orcid{0000-0002-3199-0399}\inst{\ref{aff3},\ref{aff6},\ref{aff92}}
\and J.~Valiviita\orcid{0000-0001-6225-3693}\inst{\ref{aff71},\ref{aff72}}
\and T.~Vassallo\orcid{0000-0001-6512-6358}\inst{\ref{aff5},\ref{aff61}}
\and Y.~Wang\orcid{0000-0002-4749-2984}\inst{\ref{aff104}}
\and J.~Weller\orcid{0000-0002-8282-2010}\inst{\ref{aff32},\ref{aff31}}
\and G.~Zamorani\orcid{0000-0002-2318-301X}\inst{\ref{aff23}}
\and F.~M.~Zerbi\orcid{0000-0002-9996-973X}\inst{\ref{aff10}}
\and E.~Zucca\orcid{0000-0002-5845-8132}\inst{\ref{aff23}}
\and V.~Allevato\orcid{0000-0001-7232-5152}\inst{\ref{aff34}}
\and M.~Ballardini\orcid{0000-0003-4481-3559}\inst{\ref{aff105},\ref{aff106},\ref{aff23}}
\and A.~Boucaud\orcid{0000-0001-7387-2633}\inst{\ref{aff107}}
\and E.~Bozzo\orcid{0000-0002-8201-1525}\inst{\ref{aff54}}
\and C.~Burigana\orcid{0000-0002-3005-5796}\inst{\ref{aff108},\ref{aff59}}
\and R.~Cabanac\orcid{0000-0001-6679-2600}\inst{\ref{aff92}}
\and M.~Calabrese\orcid{0000-0002-2637-2422}\inst{\ref{aff109},\ref{aff38}}
\and A.~Cappi\inst{\ref{aff110},\ref{aff23}}
\and T.~Castro\orcid{0000-0002-6292-3228}\inst{\ref{aff5},\ref{aff12},\ref{aff1},\ref{aff26}}
\and J.~A.~Escartin~Vigo\inst{\ref{aff31}}
\and L.~Gabarra\orcid{0000-0002-8486-8856}\inst{\ref{aff15}}
\and J.~Macias-Perez\orcid{0000-0002-5385-2763}\inst{\ref{aff111}}
\and R.~Maoli\orcid{0000-0002-6065-3025}\inst{\ref{aff112},\ref{aff39}}
\and J.~Mart\'{i}n-Fleitas\orcid{0000-0002-8594-569X}\inst{\ref{aff113}}
\and N.~Mauri\orcid{0000-0001-8196-1548}\inst{\ref{aff47},\ref{aff30}}
\and R.~B.~Metcalf\orcid{0000-0003-3167-2574}\inst{\ref{aff77},\ref{aff23}}
\and P.~Monaco\orcid{0000-0003-2083-7564}\inst{\ref{aff25},\ref{aff5},\ref{aff12},\ref{aff1}}
\and A.~A.~Nucita\inst{\ref{aff114},\ref{aff115},\ref{aff116}}
\and M.~P\"ontinen\orcid{0000-0001-5442-2530}\inst{\ref{aff71}}
\and I.~Risso\orcid{0000-0003-2525-7761}\inst{\ref{aff10},\ref{aff27}}
\and V.~Scottez\orcid{0009-0008-3864-940X}\inst{\ref{aff117},\ref{aff118}}
\and M.~Sereno\orcid{0000-0003-0302-0325}\inst{\ref{aff23},\ref{aff30}}
\and M.~Tenti\orcid{0000-0002-4254-5901}\inst{\ref{aff30}}
\and M.~Tucci\inst{\ref{aff54}}
\and M.~Viel\orcid{0000-0002-2642-5707}\inst{\ref{aff1},\ref{aff5},\ref{aff2},\ref{aff12},\ref{aff26}}
\and M.~Wiesmann\orcid{0009-0000-8199-5860}\inst{\ref{aff64}}
\and Y.~Akrami\orcid{0000-0002-2407-7956}\inst{\ref{aff91},\ref{aff119}}
\and I.~T.~Andika\orcid{0000-0001-6102-9526}\inst{\ref{aff61}}
\and G.~Angora\orcid{0000-0002-0316-6562}\inst{\ref{aff34},\ref{aff105}}
\and M.~Archidiacono\orcid{0000-0003-4952-9012}\inst{\ref{aff24},\ref{aff63}}
\and F.~Atrio-Barandela\orcid{0000-0002-2130-2513}\inst{\ref{aff120}}
\and L.~Bazzanini\orcid{0000-0003-0727-0137}\inst{\ref{aff105},\ref{aff23}}
\and J.~Bel\inst{\ref{aff121}}
\and D.~Bertacca\orcid{0000-0002-2490-7139}\inst{\ref{aff96},\ref{aff62},\ref{aff56}}
\and M.~Bethermin\orcid{0000-0002-3915-2015}\inst{\ref{aff122}}
\and A.~Blanchard\orcid{0000-0001-8555-9003}\inst{\ref{aff92}}
\and L.~Blot\orcid{0000-0002-9622-7167}\inst{\ref{aff123},\ref{aff73}}
\and H.~B\"ohringer\orcid{0000-0001-8241-4204}\inst{\ref{aff31},\ref{aff61},\ref{aff124}}
\and S.~Borgani\orcid{0000-0001-6151-6439}\inst{\ref{aff25},\ref{aff1},\ref{aff5},\ref{aff12},\ref{aff26}}
\and M.~L.~Brown\orcid{0000-0002-0370-8077}\inst{\ref{aff125}}
\and S.~Bruton\orcid{0000-0002-6503-5218}\inst{\ref{aff126}}
\and A.~Calabro\orcid{0000-0003-2536-1614}\inst{\ref{aff39}}
\and F.~Caro\inst{\ref{aff39}}
\and C.~S.~Carvalho\inst{\ref{aff99}}
\and F.~Cogato\orcid{0000-0003-4632-6113}\inst{\ref{aff77},\ref{aff23}}
\and A.~R.~Cooray\orcid{0000-0002-3892-0190}\inst{\ref{aff127}}
\and S.~Davini\orcid{0000-0003-3269-1718}\inst{\ref{aff27}}
\and F.~De~Paolis\orcid{0000-0001-6460-7563}\inst{\ref{aff114},\ref{aff115},\ref{aff116}}
\and G.~Desprez\orcid{0000-0001-8325-1742}\inst{\ref{aff128}}
\and A.~D\'iaz-S\'anchez\orcid{0000-0003-0748-4768}\inst{\ref{aff129}}
\and S.~Di~Domizio\orcid{0000-0003-2863-5895}\inst{\ref{aff28},\ref{aff27}}
\and J.~M.~Diego\orcid{0000-0001-9065-3926}\inst{\ref{aff130}}
\and V.~Duret\orcid{0009-0009-0383-4960}\inst{\ref{aff57}}
\and M.~Y.~Elkhashab\orcid{0000-0001-9306-2603}\inst{\ref{aff5},\ref{aff12},\ref{aff25},\ref{aff1}}
\and A.~Enia\orcid{0000-0002-0200-2857}\inst{\ref{aff23}}
\and Y.~Fang\orcid{0000-0002-0334-6950}\inst{\ref{aff32}}
\and A.~G.~Ferrari\orcid{0009-0005-5266-4110}\inst{\ref{aff30}}
\and A.~Finoguenov\orcid{0000-0002-4606-5403}\inst{\ref{aff71}}
\and A.~Fontana\orcid{0000-0003-3820-2823}\inst{\ref{aff39}}
\and F.~Fontanot\orcid{0000-0003-4744-0188}\inst{\ref{aff5},\ref{aff1}}
\and A.~Franco\orcid{0000-0002-4761-366X}\inst{\ref{aff115},\ref{aff114},\ref{aff116}}
\and K.~Ganga\orcid{0000-0001-8159-8208}\inst{\ref{aff107}}
\and T.~Gasparetto\orcid{0000-0002-7913-4866}\inst{\ref{aff39}}
\and F.~Giacomini\orcid{0000-0002-3129-2814}\inst{\ref{aff30}}
\and F.~Gianotti\orcid{0000-0003-4666-119X}\inst{\ref{aff23}}
\and G.~Gozaliasl\orcid{0000-0002-0236-919X}\inst{\ref{aff131},\ref{aff71}}
\and A.~Gruppuso\orcid{0000-0001-9272-5292}\inst{\ref{aff23},\ref{aff30}}
\and C.~M.~Gutierrez\orcid{0000-0001-7854-783X}\inst{\ref{aff48},\ref{aff132}}
\and A.~Hall\orcid{0000-0002-3139-8651}\inst{\ref{aff7}}
\and C.~Hern\'andez-Monteagudo\orcid{0000-0001-5471-9166}\inst{\ref{aff132},\ref{aff48}}
\and H.~Hildebrandt\orcid{0000-0002-9814-3338}\inst{\ref{aff133}}
\and J.~Hjorth\orcid{0000-0002-4571-2306}\inst{\ref{aff84}}
\and J.~J.~E.~Kajava\orcid{0000-0002-3010-8333}\inst{\ref{aff134},\ref{aff135}}
\and Y.~Kang\orcid{0009-0000-8588-7250}\inst{\ref{aff54}}
\and V.~Kansal\orcid{0000-0002-4008-6078}\inst{\ref{aff136},\ref{aff137}}
\and D.~Karagiannis\orcid{0000-0002-4927-0816}\inst{\ref{aff105},\ref{aff138}}
\and K.~Kiiveri\inst{\ref{aff95}}
\and J.~Kim\orcid{0000-0003-2776-2761}\inst{\ref{aff15}}
\and C.~C.~Kirkpatrick\inst{\ref{aff95}}
\and S.~Kruk\orcid{0000-0001-8010-8879}\inst{\ref{aff29}}
\and L.~Legrand\orcid{0000-0003-0610-5252}\inst{\ref{aff139},\ref{aff140}}
\and M.~Lembo\orcid{0000-0002-5271-5070}\inst{\ref{aff141}}
\and F.~Lepori\orcid{0009-0000-5061-7138}\inst{\ref{aff142}}
\and G.~Leroy\orcid{0009-0004-2523-4425}\inst{\ref{aff143},\ref{aff78}}
\and G.~F.~Lesci\orcid{0000-0002-4607-2830}\inst{\ref{aff77},\ref{aff23}}
\and J.~Lesgourgues\orcid{0000-0001-7627-353X}\inst{\ref{aff43}}
\and T.~I.~Liaudat\orcid{0000-0002-9104-314X}\inst{\ref{aff144}}
\and M.~Magliocchetti\orcid{0000-0001-9158-4838}\inst{\ref{aff58}}
\and F.~Mannucci\orcid{0000-0002-4803-2381}\inst{\ref{aff145}}
\and C.~J.~A.~P.~Martins\orcid{0000-0002-4886-9261}\inst{\ref{aff146},\ref{aff147}}
\and L.~Maurin\orcid{0000-0002-8406-0857}\inst{\ref{aff55}}
\and M.~Miluzio\inst{\ref{aff29},\ref{aff148}}
\and A.~Montoro\orcid{0000-0003-4730-8590}\inst{\ref{aff3},\ref{aff6}}
\and G.~Morgante\inst{\ref{aff23}}
\and S.~Nadathur\orcid{0000-0001-9070-3102}\inst{\ref{aff14}}
\and K.~Naidoo\orcid{0000-0002-9182-1802}\inst{\ref{aff14},\ref{aff69}}
\and A.~Navarro-Alsina\orcid{0000-0002-3173-2592}\inst{\ref{aff11}}
\and S.~Nesseris\orcid{0000-0002-0567-0324}\inst{\ref{aff91}}
\and L.~Pagano\orcid{0000-0003-1820-5998}\inst{\ref{aff105},\ref{aff106}}
\and D.~Paoletti\orcid{0000-0003-4761-6147}\inst{\ref{aff23},\ref{aff59}}
\and F.~Passalacqua\orcid{0000-0002-8606-4093}\inst{\ref{aff96},\ref{aff56}}
\and K.~Paterson\orcid{0000-0001-8340-3486}\inst{\ref{aff69}}
\and L.~Patrizii\inst{\ref{aff30}}
\and A.~Pisani\orcid{0000-0002-6146-4437}\inst{\ref{aff57}}
\and D.~Potter\orcid{0000-0002-0757-5195}\inst{\ref{aff149}}
\and G.~W.~Pratt\inst{\ref{aff16}}
\and S.~Quai\orcid{0000-0002-0449-8163}\inst{\ref{aff77},\ref{aff23}}
\and M.~Radovich\orcid{0000-0002-3585-866X}\inst{\ref{aff62}}
\and K.~Rojas\orcid{0000-0003-1391-6854}\inst{\ref{aff76}}
\and W.~Roster\orcid{0000-0002-9149-6528}\inst{\ref{aff31}}
\and S.~Sacquegna\orcid{0000-0002-8433-6630}\inst{\ref{aff150}}
\and M.~Sahl\'en\orcid{0000-0003-0973-4804}\inst{\ref{aff151}}
\and D.~B.~Sanders\orcid{0000-0002-1233-9998}\inst{\ref{aff46}}
\and A.~Schneider\orcid{0000-0001-7055-8104}\inst{\ref{aff149}}
\and D.~Sciotti\orcid{0009-0008-4519-2620}\inst{\ref{aff39},\ref{aff40}}
\and E.~Sellentin\inst{\ref{aff152},\ref{aff153}}
\and L.~C.~Smith\orcid{0000-0002-3259-2771}\inst{\ref{aff154}}
\and K.~Tanidis\orcid{0000-0001-9843-5130}\inst{\ref{aff155}}
\and C.~Tao\orcid{0000-0001-7961-8177}\inst{\ref{aff57}}
\and F.~Tarsitano\orcid{0000-0002-5919-0238}\inst{\ref{aff156},\ref{aff54}}
\and G.~Testera\inst{\ref{aff27}}
\and R.~Teyssier\orcid{0000-0001-7689-0933}\inst{\ref{aff157}}
\and S.~Tosi\orcid{0000-0002-7275-9193}\inst{\ref{aff28},\ref{aff10},\ref{aff27}}
\and A.~Troja\orcid{0000-0003-0239-4595}\inst{\ref{aff96},\ref{aff56}}
\and D.~Vergani\orcid{0000-0003-0898-2216}\inst{\ref{aff23}}
\and F.~Vernizzi\orcid{0000-0003-3426-2802}\inst{\ref{aff158}}
\and G.~Verza\orcid{0000-0002-1886-8348}\inst{\ref{aff159},\ref{aff160}}
\and P.~Vielzeuf\orcid{0000-0003-2035-9339}\inst{\ref{aff57}}
\and S.~Vinciguerra\orcid{0009-0005-4018-3184}\inst{\ref{aff21}}
\and N.~A.~Walton\orcid{0000-0003-3983-8778}\inst{\ref{aff154}}
\and A.~H.~Wright\orcid{0000-0001-7363-7932}\inst{\ref{aff133}}}
										   
\institute{IFPU, Institute for Fundamental Physics of the Universe, via Beirut 2, 34151 Trieste, Italy\label{aff1}
\and
SISSA, International School for Advanced Studies, Via Bonomea 265, 34136 Trieste TS, Italy\label{aff2}
\and
Institute of Space Sciences (ICE, CSIC), Campus UAB, Carrer de Can Magrans, s/n, 08193 Barcelona, Spain\label{aff3}
\and
Institut d'Estudis Espacials de Catalunya (IEEC),  Edifici RDIT, Campus UPC, 08860 Castelldefels, Barcelona, Spain\label{aff6}
\and
INAF-Osservatorio Astronomico di Trieste, Via G. B. Tiepolo 11, 34143 Trieste, Italy\label{aff5}
\and
Institute for Astronomy, University of Edinburgh, Royal Observatory, Blackford Hill, Edinburgh EH9 3HJ, UK\label{aff7}
\and
Donostia International Physics Center (DIPC), Paseo Manuel de Lardizabal, 4, 20018, Donostia-San Sebasti\'an, Guipuzkoa, Spain\label{aff8}
\and
IKERBASQUE, Basque Foundation for Science, 48013, Bilbao, Spain\label{aff9}
\and
INAF-Osservatorio Astronomico di Brera, Via Brera 28, 20122 Milano, Italy\label{aff10}
\and
Universit\"at Bonn, Argelander-Institut f\"ur Astronomie, Auf dem H\"ugel 71, 53121 Bonn, Germany\label{aff11}
\and
INFN, Sezione di Trieste, Via Valerio 2, 34127 Trieste TS, Italy\label{aff12}
\and
Laboratoire d'Annecy-le-Vieux de Physique Theorique, CNRS \& Universite Savoie Mont Blanc, 9 Chemin de Bellevue, BP 110, Annecy-le-Vieux, 74941 ANNECY Cedex, France\label{aff13}
\and
Institute of Cosmology and Gravitation, University of Portsmouth, Portsmouth PO1 3FX, UK\label{aff14}
\and
Department of Physics, Oxford University, Keble Road, Oxford OX1 3RH, UK\label{aff15}
\and
Universit\'e Paris-Saclay, Universit\'e Paris Cit\'e, CEA, CNRS, AIM, 91191, Gif-sur-Yvette, France\label{aff16}
\and
Universit\'e de Gen\`eve, D\'epartement de Physique Th\'eorique and Centre for Astroparticle Physics, 24 quai Ernest-Ansermet, CH-1211 Gen\`eve 4, Switzerland\label{aff17}
\and
Dipartimento di Scienze Matematiche, Fisiche e Informatiche, Universit\`a di Parma, Viale delle Scienze 7/A 43124 Parma, Italy\label{aff18}
\and
INFN Gruppo Collegato di Parma, Viale delle Scienze 7/A 43124 Parma, Italy\label{aff19}
\and
Technion Israel Institute of Technology, Israel\label{aff20}
\and
Aix-Marseille Universit\'e, CNRS, CNES, LAM, Marseille, France\label{aff21}
\and
Dipartimento di Fisica e Astronomia, Universit\`a di Bologna, Via Gobetti 93/2, 40129 Bologna, Italy\label{aff22}
\and
INAF-Osservatorio di Astrofisica e Scienza dello Spazio di Bologna, Via Piero Gobetti 93/3, 40129 Bologna, Italy\label{aff23}
\and
Dipartimento di Fisica "Aldo Pontremoli", Universit\`a degli Studi di Milano, Via Celoria 16, 20133 Milano, Italy\label{aff24}
\and
Dipartimento di Fisica - Sezione di Astronomia, Universit\`a di Trieste, Via Tiepolo 11, 34131 Trieste, Italy\label{aff25}
\and
ICSC - Centro Nazionale di Ricerca in High Performance Computing, Big Data e Quantum Computing, Via Magnanelli 2, Bologna, Italy\label{aff26}
\and
INFN-Sezione di Genova, Via Dodecaneso 33, 16146, Genova, Italy\label{aff27}
\and
Dipartimento di Fisica, Universit\`a di Genova, Via Dodecaneso 33, 16146, Genova, Italy\label{aff28}
\and
ESAC/ESA, Camino Bajo del Castillo, s/n., Urb. Villafranca del Castillo, 28692 Villanueva de la Ca\~nada, Madrid, Spain\label{aff29}
\and
INFN-Sezione di Bologna, Viale Berti Pichat 6/2, 40127 Bologna, Italy\label{aff30}
\and
Max Planck Institute for Extraterrestrial Physics, Giessenbachstr. 1, 85748 Garching, Germany\label{aff31}
\and
Universit\"ats-Sternwarte M\"unchen, Fakult\"at f\"ur Physik, Ludwig-Maximilians-Universit\"at M\"unchen, Scheinerstr.~1, 81679 M\"unchen, Germany\label{aff32}
\and
Department of Physics "E. Pancini", University Federico II, Via Cinthia 6, 80126, Napoli, Italy\label{aff33}
\and
INAF-Osservatorio Astronomico di Capodimonte, Via Moiariello 16, 80131 Napoli, Italy\label{aff34}
\and
Dipartimento di Fisica, Universit\`a degli Studi di Torino, Via P. Giuria 1, 10125 Torino, Italy\label{aff35}
\and
INFN-Sezione di Torino, Via P. Giuria 1, 10125 Torino, Italy\label{aff36}
\and
INAF-Osservatorio Astrofisico di Torino, Via Osservatorio 20, 10025 Pino Torinese (TO), Italy\label{aff37}
\and
INAF-IASF Milano, Via Alfonso Corti 12, 20133 Milano, Italy\label{aff38}
\and
INAF-Osservatorio Astronomico di Roma, Via Frascati 33, 00078 Monteporzio Catone, Italy\label{aff39}
\and
INFN-Sezione di Roma, Piazzale Aldo Moro, 2 - c/o Dipartimento di Fisica, Edificio G. Marconi, 00185 Roma, Italy\label{aff40}
\and
Centro de Investigaciones Energ\'eticas, Medioambientales y Tecnol\'ogicas (CIEMAT), Avenida Complutense 40, 28040 Madrid, Spain\label{aff41}
\and
Port d'Informaci\'{o} Cient\'{i}fica, Campus UAB, C. Albareda s/n, 08193 Bellaterra (Barcelona), Spain\label{aff42}
\and
Institute for Theoretical Particle Physics and Cosmology (TTK), RWTH Aachen University, 52056 Aachen, Germany\label{aff43}
\and
Deutsches Zentrum f\"ur Luft- und Raumfahrt e. V. (DLR), Linder H\"ohe, 51147 K\"oln, Germany\label{aff44}
\and
INFN section of Naples, Via Cinthia 6, 80126, Napoli, Italy\label{aff45}
\and
Institute for Astronomy, University of Hawaii, 2680 Woodlawn Drive, Honolulu, HI 96822, USA\label{aff46}
\and
Dipartimento di Fisica e Astronomia "Augusto Righi" - Alma Mater Studiorum Universit\`a di Bologna, Viale Berti Pichat 6/2, 40127 Bologna, Italy\label{aff47}
\and
Instituto de Astrof\'{\i}sica de Canarias, E-38205 La Laguna, Tenerife, Spain\label{aff48}
\and
European Space Agency/ESRIN, Largo Galileo Galilei 1, 00044 Frascati, Roma, Italy\label{aff49}
\and
Universit\'e Claude Bernard Lyon 1, CNRS/IN2P3, IP2I Lyon, UMR 5822, Villeurbanne, F-69100, France\label{aff50}
\and
Institut de Ci\`{e}ncies del Cosmos (ICCUB), Universitat de Barcelona (IEEC-UB), Mart\'{i} i Franqu\`{e}s 1, 08028 Barcelona, Spain\label{aff51}
\and
Instituci\'o Catalana de Recerca i Estudis Avan\c{c}ats (ICREA), Passeig de Llu\'{\i}s Companys 23, 08010 Barcelona, Spain\label{aff52}
\and
UCB Lyon 1, CNRS/IN2P3, IUF, IP2I Lyon, 4 rue Enrico Fermi, 69622 Villeurbanne, France\label{aff53}
\and
Department of Astronomy, University of Geneva, ch. d'Ecogia 16, 1290 Versoix, Switzerland\label{aff54}
\and
Universit\'e Paris-Saclay, CNRS, Institut d'astrophysique spatiale, 91405, Orsay, France\label{aff55}
\and
INFN-Padova, Via Marzolo 8, 35131 Padova, Italy\label{aff56}
\and
Aix-Marseille Universit\'e, CNRS/IN2P3, CPPM, Marseille, France\label{aff57}
\and
INAF-Istituto di Astrofisica e Planetologia Spaziali, via del Fosso del Cavaliere, 100, 00100 Roma, Italy\label{aff58}
\and
INFN-Bologna, Via Irnerio 46, 40126 Bologna, Italy\label{aff59}
\and
School of Physics, HH Wills Physics Laboratory, University of Bristol, Tyndall Avenue, Bristol, BS8 1TL, UK\label{aff60}
\and
University Observatory, LMU Faculty of Physics, Scheinerstr.~1, 81679 Munich, Germany\label{aff61}
\and
INAF-Osservatorio Astronomico di Padova, Via dell'Osservatorio 5, 35122 Padova, Italy\label{aff62}
\and
INFN-Sezione di Milano, Via Celoria 16, 20133 Milano, Italy\label{aff63}
\and
Institute of Theoretical Astrophysics, University of Oslo, P.O. Box 1029 Blindern, 0315 Oslo, Norway\label{aff64}
\and
Jet Propulsion Laboratory, California Institute of Technology, 4800 Oak Grove Drive, Pasadena, CA, 91109, USA\label{aff65}
\and
Felix Hormuth Engineering, Goethestr. 17, 69181 Leimen, Germany\label{aff66}
\and
Technical University of Denmark, Elektrovej 327, 2800 Kgs. Lyngby, Denmark\label{aff67}
\and
Cosmic Dawn Center (DAWN), Denmark\label{aff68}
\and
Max-Planck-Institut f\"ur Astronomie, K\"onigstuhl 17, 69117 Heidelberg, Germany\label{aff69}
\and
Department of Physics and Astronomy, University College London, Gower Street, London WC1E 6BT, UK\label{aff70}
\and
Department of Physics, P.O. Box 64, University of Helsinki, 00014 Helsinki, Finland\label{aff71}
\and
Helsinki Institute of Physics, Gustaf H{\"a}llstr{\"o}min katu 2, University of Helsinki, 00014 Helsinki, Finland\label{aff72}
\and
Laboratoire d'etude de l'Univers et des phenomenes eXtremes, Observatoire de Paris, Universit\'e PSL, Sorbonne Universit\'e, CNRS, 92190 Meudon, France\label{aff73}
\and
SKAO, Jodrell Bank, Lower Withington, Macclesfield SK11 9FT, UK\label{aff74}
\and
Centre de Calcul de l'IN2P3/CNRS, 21 avenue Pierre de Coubertin 69627 Villeurbanne Cedex, France\label{aff75}
\and
University of Applied Sciences and Arts of Northwestern Switzerland, School of Computer Science, 5210 Windisch, Switzerland\label{aff76}
\and
Dipartimento di Fisica e Astronomia "Augusto Righi" - Alma Mater Studiorum Universit\`a di Bologna, via Piero Gobetti 93/2, 40129 Bologna, Italy\label{aff77}
\and
Department of Physics, Institute for Computational Cosmology, Durham University, South Road, Durham, DH1 3LE, UK\label{aff78}
\and
University of Applied Sciences and Arts of Northwestern Switzerland, School of Engineering, 5210 Windisch, Switzerland\label{aff79}
\and
Institute of Physics, Laboratory of Astrophysics, Ecole Polytechnique F\'ed\'erale de Lausanne (EPFL), Observatoire de Sauverny, 1290 Versoix, Switzerland\label{aff80}
\and
Telespazio UK S.L. for European Space Agency (ESA), Camino bajo del Castillo, s/n, Urbanizacion Villafranca del Castillo, Villanueva de la Ca\~nada, 28692 Madrid, Spain\label{aff81}
\and
Institut de F\'{i}sica d'Altes Energies (IFAE), The Barcelona Institute of Science and Technology, Campus UAB, 08193 Bellaterra (Barcelona), Spain\label{aff82}
\and
European Space Agency/ESTEC, Keplerlaan 1, 2201 AZ Noordwijk, The Netherlands\label{aff83}
\and
DARK, Niels Bohr Institute, University of Copenhagen, Jagtvej 155, 2200 Copenhagen, Denmark\label{aff84}
\and
Waterloo Centre for Astrophysics, University of Waterloo, Waterloo, Ontario N2L 3G1, Canada\label{aff85}
\and
Department of Physics and Astronomy, University of Waterloo, Waterloo, Ontario N2L 3G1, Canada\label{aff86}
\and
Perimeter Institute for Theoretical Physics, Waterloo, Ontario N2L 2Y5, Canada\label{aff87}
\and
Space Science Data Center, Italian Space Agency, via del Politecnico snc, 00133 Roma, Italy\label{aff88}
\and
Centre National d'Etudes Spatiales -- Centre spatial de Toulouse, 18 avenue Edouard Belin, 31401 Toulouse Cedex 9, France\label{aff89}
\and
Institute of Space Science, Str. Atomistilor, nr. 409 M\u{a}gurele, Ilfov, 077125, Romania\label{aff90}
\and
Instituto de F\'isica Te\'orica UAM-CSIC, Campus de Cantoblanco, 28049 Madrid, Spain\label{aff91}
\and
Institut de Recherche en Astrophysique et Plan\'etologie (IRAP), Universit\'e de Toulouse, CNRS, UPS, CNES, 14 Av. Edouard Belin, 31400 Toulouse, France\label{aff92}
\and
Universit\'e St Joseph; Faculty of Sciences, Beirut, Lebanon\label{aff93}
\and
Departamento de F\'isica, FCFM, Universidad de Chile, Blanco Encalada 2008, Santiago, Chile\label{aff94}
\and
Department of Physics and Helsinki Institute of Physics, Gustaf H\"allstr\"omin katu 2, University of Helsinki, 00014 Helsinki, Finland\label{aff95}
\and
Dipartimento di Fisica e Astronomia "G. Galilei", Universit\`a di Padova, Via Marzolo 8, 35131 Padova, Italy\label{aff96}
\and
Department of Physics, Royal Holloway, University of London, Surrey TW20 0EX, UK\label{aff97}
\and
Departamento de F\'isica, Faculdade de Ci\^encias, Universidade de Lisboa, Edif\'icio C8, Campo Grande, PT1749-016 Lisboa, Portugal\label{aff98}
\and
Instituto de Astrof\'isica e Ci\^encias do Espa\c{c}o, Faculdade de Ci\^encias, Universidade de Lisboa, Tapada da Ajuda, 1349-018 Lisboa, Portugal\label{aff99}
\and
Mullard Space Science Laboratory, University College London, Holmbury St Mary, Dorking, Surrey RH5 6NT, UK\label{aff100}
\and
Cosmic Dawn Center (DAWN)\label{aff101}
\and
Niels Bohr Institute, University of Copenhagen, Jagtvej 128, 2200 Copenhagen, Denmark\label{aff102}
\and
Universidad Polit\'ecnica de Cartagena, Departamento de Electr\'onica y Tecnolog\'ia de Computadoras,  Plaza del Hospital 1, 30202 Cartagena, Spain\label{aff103}
\and
Caltech/IPAC, 1200 E. California Blvd., Pasadena, CA 91125, USA\label{aff104}
\and
Dipartimento di Fisica e Scienze della Terra, Universit\`a degli Studi di Ferrara, Via Giuseppe Saragat 1, 44122 Ferrara, Italy\label{aff105}
\and
Istituto Nazionale di Fisica Nucleare, Sezione di Ferrara, Via Giuseppe Saragat 1, 44122 Ferrara, Italy\label{aff106}
\and
Universit\'e Paris Cit\'e, CNRS, Astroparticule et Cosmologie, 75013 Paris, France\label{aff107}
\and
INAF, Istituto di Radioastronomia, Via Piero Gobetti 101, 40129 Bologna, Italy\label{aff108}
\and
Astronomical Observatory of the Autonomous Region of the Aosta Valley (OAVdA), Loc. Lignan 39, I-11020, Nus (Aosta Valley), Italy\label{aff109}
\and
Universit\'e C\^{o}te d'Azur, Observatoire de la C\^{o}te d'Azur, CNRS, Laboratoire Lagrange, Bd de l'Observatoire, CS 34229, 06304 Nice cedex 4, France\label{aff110}
\and
Univ. Grenoble Alpes, CNRS, Grenoble INP, LPSC-IN2P3, 53, Avenue des Martyrs, 38000, Grenoble, France\label{aff111}
\and
Dipartimento di Fisica, Sapienza Universit\`a di Roma, Piazzale Aldo Moro 2, 00185 Roma, Italy\label{aff112}
\and
Aurora Technology for European Space Agency (ESA), Camino bajo del Castillo, s/n, Urbanizacion Villafranca del Castillo, Villanueva de la Ca\~nada, 28692 Madrid, Spain\label{aff113}
\and
Department of Mathematics and Physics E. De Giorgi, University of Salento, Via per Arnesano, CP-I93, 73100, Lecce, Italy\label{aff114}
\and
INFN, Sezione di Lecce, Via per Arnesano, CP-193, 73100, Lecce, Italy\label{aff115}
\and
INAF-Sezione di Lecce, c/o Dipartimento Matematica e Fisica, Via per Arnesano, 73100, Lecce, Italy\label{aff116}
\and
Institut d'Astrophysique de Paris, 98bis Boulevard Arago, 75014, Paris, France\label{aff117}
\and
ICL, Junia, Universit\'e Catholique de Lille, LITL, 59000 Lille, France\label{aff118}
\and
CERCA/ISO, Department of Physics, Case Western Reserve University, 10900 Euclid Avenue, Cleveland, OH 44106, USA\label{aff119}
\and
Departamento de F{\'\i}sica Fundamental. Universidad de Salamanca. Plaza de la Merced s/n. 37008 Salamanca, Spain\label{aff120}
\and
Aix-Marseille Universit\'e, Universit\'e de Toulon, CNRS, CPT, Marseille, France\label{aff121}
\and
Universit\'e de Strasbourg, CNRS, Observatoire astronomique de Strasbourg, UMR 7550, 67000 Strasbourg, France\label{aff122}
\and
Center for Data-Driven Discovery, Kavli IPMU (WPI), UTIAS, The University of Tokyo, Kashiwa, Chiba 277-8583, Japan\label{aff123}
\and
Max-Planck-Institut f\"ur Physik, Boltzmannstr. 8, 85748 Garching, Germany\label{aff124}
\and
Jodrell Bank Centre for Astrophysics, Department of Physics and Astronomy, University of Manchester, Oxford Road, Manchester M13 9PL, UK\label{aff125}
\and
California Institute of Technology, 1200 E California Blvd, Pasadena, CA 91125, USA\label{aff126}
\and
Department of Physics \& Astronomy, University of California Irvine, Irvine CA 92697, USA\label{aff127}
\and
Kapteyn Astronomical Institute, University of Groningen, PO Box 800, 9700 AV Groningen, The Netherlands\label{aff128}
\and
Departamento F\'isica Aplicada, Universidad Polit\'ecnica de Cartagena, Campus Muralla del Mar, 30202 Cartagena, Murcia, Spain\label{aff129}
\and
Instituto de F\'isica de Cantabria, Edificio Juan Jord\'a, Avenida de los Castros, 39005 Santander, Spain\label{aff130}
\and
Department of Computer Science, Aalto University, PO Box 15400, Espoo, FI-00 076, Finland\label{aff131}
\and
Universidad de La Laguna, Dpto. Astrof\'\i sica, E-38206 La Laguna, Tenerife, Spain\label{aff132}
\and
Ruhr University Bochum, Faculty of Physics and Astronomy, Astronomical Institute (AIRUB), German Centre for Cosmological Lensing (GCCL), 44780 Bochum, Germany\label{aff133}
\and
Department of Physics and Astronomy, Vesilinnantie 5, University of Turku, 20014 Turku, Finland\label{aff134}
\and
Serco for European Space Agency (ESA), Camino bajo del Castillo, s/n, Urbanizacion Villafranca del Castillo, Villanueva de la Ca\~nada, 28692 Madrid, Spain\label{aff135}
\and
ARC Centre of Excellence for Dark Matter Particle Physics, Melbourne, Australia\label{aff136}
\and
Centre for Astrophysics \& Supercomputing, Swinburne University of Technology,  Hawthorn, Victoria 3122, Australia\label{aff137}
\and
Department of Physics and Astronomy, University of the Western Cape, Bellville, Cape Town, 7535, South Africa\label{aff138}
\and
DAMTP, Centre for Mathematical Sciences, Wilberforce Road, Cambridge CB3 0WA, UK\label{aff139}
\and
Kavli Institute for Cosmology Cambridge, Madingley Road, Cambridge, CB3 0HA, UK\label{aff140}
\and
Institut d'Astrophysique de Paris, UMR 7095, CNRS, and Sorbonne Universit\'e, 98 bis boulevard Arago, 75014 Paris, France\label{aff141}
\and
Departement of Theoretical Physics, University of Geneva, Switzerland\label{aff142}
\and
Department of Physics, Centre for Extragalactic Astronomy, Durham University, South Road, Durham, DH1 3LE, UK\label{aff143}
\and
IRFU, CEA, Universit\'e Paris-Saclay 91191 Gif-sur-Yvette Cedex, France\label{aff144}
\and
INAF-Osservatorio Astrofisico di Arcetri, Largo E. Fermi 5, 50125, Firenze, Italy\label{aff145}
\and
Centro de Astrof\'{\i}sica da Universidade do Porto, Rua das Estrelas, 4150-762 Porto, Portugal\label{aff146}
\and
Instituto de Astrof\'isica e Ci\^encias do Espa\c{c}o, Universidade do Porto, CAUP, Rua das Estrelas, PT4150-762 Porto, Portugal\label{aff147}
\and
HE Space for European Space Agency (ESA), Camino bajo del Castillo, s/n, Urbanizacion Villafranca del Castillo, Villanueva de la Ca\~nada, 28692 Madrid, Spain\label{aff148}
\and
Department of Astrophysics, University of Zurich, Winterthurerstrasse 190, 8057 Zurich, Switzerland\label{aff149}
\and
INAF - Osservatorio Astronomico d'Abruzzo, Via Maggini, 64100, Teramo, Italy\label{aff150}
\and
Theoretical astrophysics, Department of Physics and Astronomy, Uppsala University, Box 516, 751 37 Uppsala, Sweden\label{aff151}
\and
Mathematical Institute, University of Leiden, Einsteinweg 55, 2333 CA Leiden, The Netherlands\label{aff152}
\and
Leiden Observatory, Leiden University, Einsteinweg 55, 2333 CC Leiden, The Netherlands\label{aff153}
\and
Institute of Astronomy, University of Cambridge, Madingley Road, Cambridge CB3 0HA, UK\label{aff154}
\and
Center for Astrophysics and Cosmology, University of Nova Gorica, Nova Gorica, Slovenia\label{aff155}
\and
Institute for Particle Physics and Astrophysics, Dept. of Physics, ETH Zurich, Wolfgang-Pauli-Strasse 27, 8093 Zurich, Switzerland\label{aff156}
\and
Department of Astrophysical Sciences, Peyton Hall, Princeton University, Princeton, NJ 08544, USA\label{aff157}
\and
Institut de Physique Th\'eorique, CEA, CNRS, Universit\'e Paris-Saclay 91191 Gif-sur-Yvette Cedex, France\label{aff158}
\and
International Centre for Theoretical Physics (ICTP), Strada Costiera 11, 34151 Trieste, Italy\label{aff159}
\and
Center for Computational Astrophysics, Flatiron Institute, 162 5th Avenue, 10010, New York, NY, USA\label{aff160}}

\abstract{Accurate modelling of redshift-space distortions (RSD) is essential for maximizing the cosmological information extracted from large galaxy redshift surveys. In preparation for the forthcoming analysis of the \Euclid spectroscopic data, we investigate three approaches to modelling RSD effects on the power spectrum multipoles of mock H$\alpha$ emission line galaxies. We focus on two one-loop perturbation theory models -- the effective field theory (EFT) and velocity difference generator (${\rm VDG_ \infty}$) -- which differ in their treatment of the real-to-redshift space mapping on small scales, and a third approach, the \bacco emulator, which adopts a hybrid strategy combining perturbation theory with high-resolution \nbody simulations. We assess the ability of these models to recover key cosmological parameters, including the expansion rate $h$, the cold dark matter density parameter $\omegac$, and the scalar amplitude $\As$, across four redshift bins spanning $0.9 \leq z \leq 1.8$. In each bin, we find that ${\rm VDG_ \infty}$ and \bacco outperform the EFT model across all scales up to $\kmax \lesssim 0.35 \kMpc$. While \bacco saturates in constraining power at intermediate scales and higher redshift, the ${\rm VDG_ \infty}$ model continues to improve parameter constraints beyond $\kmax \gtrsim 0.30 \kMpc$. The EFT model, although robust on large scales, exhibits significant parameter biases for $\kmax \gtrsim 0.25 \kMpc$, limiting its applicability to \Euclid-like H$\alpha$ samples. Among the full perturbation theory-based models, the enhanced treatment of small-scale RSD effects in ${\rm VDG_ \infty}$ improves cosmological parameter constraints by up to a factor of two.}

%
%

\titlerunning{Galaxy power spectrum modelling in redshift space}
\authorrunning{Euclid Collaboration: B. Camacho Quevedo et al.}
   
\maketitle
%
%
%
%

\section{Introduction}
\label{sec:intro}

    The study of the \gls{lss} through galaxy clustering is one of the primary sources of cosmological information at late times \citep{Colless2001,Beutler2012, LeFevre2013, York2000, Blake2011, Guzzo2014, Driver2011, Dawson2012, Dawson2016, DESI_DR2_cosmo, DESI_FS_cosmo, DESIBAO2024}. Together with weak gravitational lensing \citep{DESY3Cosmo, DESY3_cosmo_2025, KIDS_2023, KIDS_DESY3}, supernovae \citep{PantheonCosmo}, and complementary early-time observations via \gls{cmb} experiments \citep{Hinshaw2013, planck2020, ACT_DR6_cosmo}, galaxy clustering provides invaluable insights into both the geometry and the growth of structures of the Universe, with the potential to  disentangle the effects of different dark energy models from potential deviations from general relativity \citep{eBOSSFinal, DESI_FS_cosmo, DESI_FS, Semenaite2022, Carrilho-Moretti}.
    
    Ongoing Stage-IV spectroscopic surveys, such as the Dark Energy Spectroscopic Instrument \citep[DESI; ][]{DESI2016} and \Euclid \citep{Euclid2011,EuclidOverview}, will significantly enhance the information content that can be extracted from the \gls{lss}. These experiments reach a relatively uncharted epoch at $1\lesssim z\lesssim2$, when the Universe was only about half of its current age. Traditionally, information from spectroscopic galaxy surveys is extracted by modelling the \gls{2pcf} or its Fourier counterpart, the galaxy power spectrum \citep{Eisenstein2005, Eisenstein2007, Beutler2016, IvaSib1807, Sanchez2014}.
    The information encapsulated within these statistics is mostly carried by the \bao peak, determined by the sound horizon at the matter-radiation decoupling era \citep{Peacock2001, Eisenstein2005}, and \rsd, determined by the apparent anisotropic clustering induced by the peculiar velocities of galaxies \citep{Kaiser1987}. 
    Modelling each of these features presents its own unique challenges. For example, \bao analyses rely on reconstruction algorithms to sharpen the acoustic peak by predicting the galaxy velocity field \citep{Zeldovich1970, Eisenstein2007}.
    In turn, \rsd are particularly sensitive to nonlinear physics, requiring a complex modelling of the mildly nonlinear regime to accurately describe the observed clustering signal.
 
    Spectroscopic galaxy surveys do not provide direct access to the real-space position of galaxies, but rather to their angular position and redshift. The latter is sensitive not only to the recession velocity, due to the expansion of the Universe, but also to the galaxy peculiar velocity, which is sourced by its gravitational interaction with its surrounding. As a consequence, the observed redshift-space galaxy distribution corresponds to a distorted version of the underlying real-space distribution \citep[][]{Kaiser1987, hamilton:1992, fisher:1995, Scoccimarro1999, Scoccimarro2004b, Taruya2010, Senatore2014, Perko2016}, leading to a complex mapping involving both the density and velocity fields.
    Hence, accurately describing the real- to redshift-space mapping is critical to capture the effect of \rsd and reliably estimate the growth rate of cosmic structures. 
    
    In the context of galaxy clustering, models based on \glslink{pt}{perturbation theory} \citep[hereafter PT,][]{Bernardeau2002} are powerful tools for extracting cosmological information from full-shape measurements of the anisotropic galaxy power spectrum. Perturbative methods can be employed to model different, yet intertwined, effects that contribute to the shaping of the observed galaxy density field. First, it can be used to describe the nonlinear evolution of the dark matter density field over mildly nonlinear scales. Second, it can model the relationship between the galaxy and matter density fields \citep[usually refereed as galaxy bias expansion,][]{szalay:1988, Coles1993, FryGaz1993, scoccimarro/etal:2001, Smith2007, manera/etal:2010, Desjacques2009, frusciante/sheth:2012, schmidt/etal:2013}, which is essential as we can only observe biased tracers of the underlying dark matter field. Finally, \rsd and \uv limit effects become more important when approaching the strong nonlinear regime (around $k\sim 0.20\kMpc$), where \glsunset{pt}\gls{pt} begins to break down. These effects introduce significant distortions in the shape and amplitude of the power spectrum and their impact is modulated by a set of extra parameters, usually denoted as counterterms \citep{Senatore2014,DesJeoSch1812, IvanovEtal2020, PueblasScoccimarro2009}. Lastly, contributions from small scales (beyond $k=0.20\kMpc$), coupled with larger scales and acting as a stochastic field, must be modelled as noise parameters \citep{DekLah9907,TarSod9909,Mat9911}. Traditionally, this picture is complemented with a technique called \ir resummation, which serves as a way to model large-scale bulk flows smearing the \bao peak \citep{Eisenstein2007,BalMirSim1508,BlaGarIva1607}. 
    
    In recent years, different approaches to model the real- to redshift-space mapping have emerged. Among them, one of the most successful frameworks is the \eft, which treats both density and velocity fields in a fully perturbative way \citep{Carrasco2012, SenatoreZaldarriaga2015, Perko2016, IvanovEtal2020,DAmicoEtal2020}. This framework accounts for the small-scale \rsd signal by expanding the velocity difference moment generating function and introducing a set of counterterms to reproduce its effect at leading- and next-to-leading order in the power spectrum \citep{PueblasScoccimarro2009, Senatore2014,DesJeoSch1812, IvanovEtal2020}. Alternatively, in an equally successful class of models, the moment generating function can be described analytically using a physically-motivated functional form, which tries to reproduce the pairwise velocity distribution at some fixed physical separation \citep{Taruya2010, Sanchez2016}. Finally, beyond these purely analytical approaches, hybrid models have shown considerable success recently \citep{ZenAngPel2101,AriAngZen2104,KokDeRChe2107,PellejeroIbanez2022}. These types of models keep the traditional expansion of the redshift-space galaxy density field, but emulate each of the individual components of the expansion from high-resolution \nbody simulations.

In this work, we aim at comparing three different modelling approaches for the anisotropic galaxy power spectrum, all of them based on the one-loop galaxy bias expansion, but differing mainly in how they perform the real- to redshift-space mapping. Two of these models are fully analytical, falling in one of the two categories described above, while the third one is based on a hybrid perturbative framework. As a testing ground for the different models, we use a set of data vectors derived from mock galaxy catalogues designed to reproduce the clustering properties of the H$\alpha$ sample targeted by \Euclid \citep{Pozzetti2016}.

The range of validity of each model is quantified through three complementary performance metrics.
First, we calculate the \fob to quantify the accuracy of the model in recovering the input parameters. Second, we compute the \fom to measure the statistical power of the model in constraining cosmological parameters. Third, we evaluate the goodness of fit (\pvalue) to assess how well the best-fit model matches the input data vectors. These metrics are computed for each model as a function of the maximum wave mode, $\kmax$, included in the fit.
Our analysis focuses on the recovery of the dimensionless Hubble parameter $h$, defined in terms of the Hubble constant $H_0$ as $H_0=100\,h\,\Hunits$, 
    the cold dark matter density parameter $\omegac\equiv\Omega_{\rm c}h^2$, and the scalar amplitude $\As$.
    
This work is part of a series of validation tests to identify the optimal modelling framework for galaxy clustering measurements within the Euclid Consortium. In \cite{Euclid-Pezzotta} we perform an equivalent comparison for modelling the galaxy power spectrum in real space, while in Euclid Collaboration: Pardede et al. (in prep.) we extend it to the galaxy bispectrum analysis (both in real and redshift space). Likewise, equivalent comparisons in configuration space are presented in \cite{Euclid-Karcher} in terms of the \gls{2pcf}, with the extension to higher-order statistics presented in \cite{Euclid-Guidi} and Euclid Collaboration: Pugno et al. (in prep.), for the real- and redshift-space case, respectively. 

This paper is structured as follows. We begin by describing the data vectors and covariance matrices of data noise used for testing the models in Sect.~\ref{sec:data}. In Sect.~\ref{sec:theory} we describe the three modelling frameworks tested in this work. In Sect.~\ref{sec:perf_metrics} we describe the performance metrics and fitting strategies employed in the analysis. In Sect.~\ref{sec:EulerianValidation} we validate the two fully perturbative models to ensure a fair comparisons with the hybrid model. The complete comparison between the performance of the three models, focusing on their range of validity and constraining power, is presented in Sect.~\ref{sec:results}. Finally we conclude by highlighting the key findings of this work in Sect.~\ref{sec:conclusions}.


\section{Data}
\label{sec:data}

\subsection{Euclid simulations }
\begin{table}
  \caption{Cosmological parameters of the Flagship I simulations. The table shows the values of the Hubble parameter $h$, the physical densities of cold dark matter and baryons ($\omegac$ and $\omegab$), the total neutrino mass $M_\nu$, the scalar amplitude $\As$, and the index $\ns$ of the primordial power spectrum.
  }
  \centering
   \renewcommand{\arraystretch}{1.3}
  \begin{tabular}{|c|c|c|c|c|c|}
    \hline
    \rowcolor{blue!5}
    $h$ & $\omegac$ & $\omegab$ & $M_\nu\,[\rm{eV}]$ & $10^9 \As$ & $\ns$ \\ 
    \hline
    $0.67$ & $0.121203$ & $0.0219961$ & $0$ &  $2.09$ & $0.97$ \\
    \hline
  \end{tabular}
  \label{tab:flagship_cosmology}
\end{table}

To study the performance of the different models, we use dedicated mock catalogues of $\mathrm{H}\alpha$ galaxies with a flux limit $f_{\mathrm{H}\alpha} = 2 \times 10^{-16} \rm{erg\,cm^{-2}\,s^{-1}}$, to match the primary spectroscopic sample of the \Euclid mission. These catalogues are based on the Flagship I \nbody simulation \citep{Potter2017},\footnote{Similarly to \cite{Euclid-Pezzotta}, we use the Roman numeral `\,I\,' to differentiate it from the its more recent version, that is, Flagship II \citep{Euclid5}.
} which used the \texttt{PKDGRAV3} algorithm to evolve two trillion dark matter particles within a comoving box of size $\lbox=3780\Mpc$ and with a particle mass resolution of $m_{\rm p} \sim 2.398 \times 10^{\,9} \Ms$. Such combination of large volume and high mass resolution is unique, and it enables to accurately resolve haloes hosting the majority of the expected \Euclid $\mathrm{H}\alpha$ sample, whose typical mass is of the order of few $10^{10}\Ms$\,, within a volume exceeding the nominal final-mission volume of \Euclid by several times. The fiducial cosmology of the simulation is presented in Table \ref{tab:flagship_cosmology} \citep[see discussion in Appendix A of][]{Euclid-Pezzotta}. In agreement with the real-space analysis in \cite{Euclid-Pezzotta}, we used the same four snapshots of $\mathrm{H}\alpha$ galaxies in the redshift range $0.9<z<1.8$. 

In this paper, we neglect the impact of different contaminants and systematic errors that may affect \Euclid observations, such as target incompleteness, purity of the sample, and the impact of the angular footprint and radial selection function.\footnote{For a description of the \Euclid visibility mask and observational systematic effects in \Euclid, see  Euclid Collaboration: Granett et al. (in prep.) and \cite{Euclid-Monaco}.}
In addition, we focus our analysis only on the Model 3 \hod sample described in \cite{Pozzetti2016},\footnote{For the specific shape of the implemented \hod see Sect.~2.1 in \cite{Euclid-Pezzotta}.} since it displays a lower number density that more closely matches the one expected systematic-free sample of H$\alpha$ galaxies.

The total number of galaxies in each comoving snapshot, their number density, and the scale $k_{\rm sn}$ at which the Poissonian shot-noise contribution reaches the same amplitude of the power spectrum monopole ($\ell=0$), that is, $\bar{n}\,P_0\sim 1$, are listed in Table~\ref{tab:hod_samples}.

\begin{table}
\caption{Specifications of the Model 3 $\mathrm{H}\alpha$ mock galaxy samples used throughout this paper.
The columns indicate the redshift of the sample $z$, the total number of galaxies $N_{\rm g}$, the mean number density $\bar{n}$, and the scale $k_{sn}$ at which the Poissonian shot-noise contribution equals the signal of the $\ell=0$ galaxy power spectrum multipole. Lastly, $\eta$ indicates the normalization factor to scale the Flagship covariance to \Euclid-like errors, as explained in Sect.~\ref{sec:measAndCov}.}
  \centering
  \renewcommand{\arraystretch}{1.3}
  \begin{tabular}{|c|r|c|c|c|}
    \hline
    \rowcolor{blue!5}
    {$z$} & {$N_{\rm g}$\hspace{18pt}} & {$\bar{n}\,[\kcMpc]$} & {$k_{\rm sn}\,[\kMpc]$} & $\eta$\\
    \hline
    {0.9} & 110\,321\,755 & 0.0020 & 0.50 & 6.6 \\
    \hline
    {1.2} & 55\,563\,490 & 0.0010 & 0.42 & 5.9 \\
    \hline
    {1.5} & 31\,613\,213 & 0.0006 & 0.30 & 5.3 \\
    \hline
    {1.8} & 16\,926\,864 & 0.0003 & 0.26 & 3.3 \\
    \hline
  \end{tabular}
  \label{tab:hod_samples}
\end{table}

\subsection{Power spectrum multipole measurements}
\label{sec:MultiMeas}

From the H$\alpha$ mocks, we compute the Legendre multipoles of the galaxy power spectrum using the publicly available \texttt{PowerI4} code,\footnote{\url{https://github.com/sefusatti/PowerI4}.} which calculates power spectra in periodic cubic boxes employing the flat-sky approximation. This allows us to measure the \rsd signal along each of the three coordinate axes ($x$, $y$, $z$). 

The fundamental frequency of the power spectrum measurements is determined from the simulation box size as $\kF\equiv2\pi/\lbox \sim 0.0016 \kMpc$. In contrast, the maximum wave mode is defined by the Nyquist frequency, which for a density grid with $N_\mathrm{grid}$ linear bins is given by $\kN\equiv\pi \, N_{\rm grid}/\lbox$. For the three different directions, we adopt the same grid resolution $N_\mathrm{grid}=1024$, which leads to $\kN\sim0.85\kMpc$. These quantities define the wave mode range of our measurements, $k\in[\kF,\kN]$, which we sample using a linear binning of width $\Delta k=\kF$\,.

We obtain measurements of the power spectrum multipoles using each of the three directions of the cubic box as line of sight. In redshift space, these three measurements are highly correlated; albeit they have different line-of-sight velocity contributions, they share the same density field \citep[for a comprehensive discussion on how these effects are coupled together, see][]{Smith2020_averagedCovariance}. Instead of averaging the power spectra over the three axes -- this would unevenly weight the density field (common to all directions) and the velocity field -- we use only the measurement along the axis that shows the best goodness of fit when compared to the average over the three axes.
Precisely, for a given snapshot at redshift $z$\,, we define $\chi^2\equiv\paren{\,\chi^2_x,\,  \chi^2_y,\, \chi^2_z\,}$\,, where each element $\chi^2_i$ corresponds to the error-weighted mean square difference between the averaged multipoles and the ones with line of sight parallel to the $i$-th axis. We find that $\chi^2(0.9) = \paren{0.59, 0.55, 0.50}$\,, $\chi^2(1.2) = \paren{0.58, 0.57, 0.50}$\,, $\chi^2(1.5)=\paren{0.60, 0.53, 0.48}$, and $\chi^2(1.8)=\paren{0.50, 0.51, 0.51}$. We therefore adopt the $z$ axis as our baseline data vector, since it provides the lowest $\chi^2$ across most of our samples.

\begin{figure*}
    \centering
    \includegraphics[width=1.\linewidth]{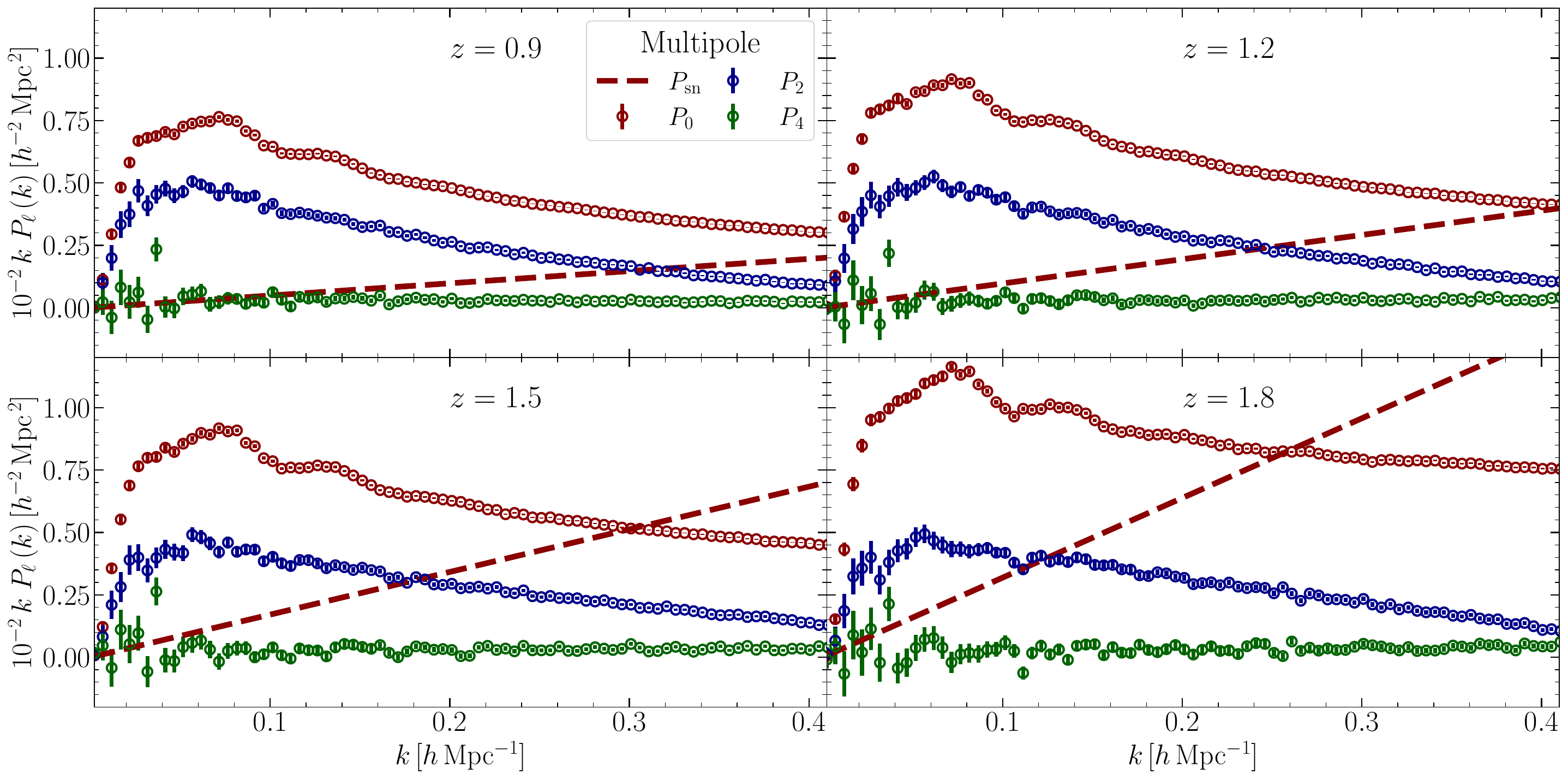}
    \caption{Power spectrum Legendre multipoles as measured from the Model 3 \hod samples built on the Flagship I comoving snapshots. Markers show each multipole along the $z$ axis, with different colours marking different multipoles as indicated in the legend. The error bars correspond to the Gaussian covariance for a comoving volume corresponding to the full box of the Flagship I simulation, as described in Sect.~\ref{sec:data}. Dashed red line refers to the Poissonian shot-noise contribution $P_{\rm sn}$. For visualization purposes, we show only one out of three data points.}
    \label{fig:Measured_Multipoles}
\end{figure*}

In Fig.~\ref{fig:Measured_Multipoles} we show the first three even Legendre multipoles along the $z$ axis for each of the considered redshifts. Additionally, we show the Poissonian shot-noise contribution $P_{\rm sn}$ to highlight the transition to the shot-noise dominated scales. At low redshift, shot-noise remains negligible up to $k=0.5\,\kMpc$, while at high redshift it matches the monopole amplitude already at $k=0.26\,\kMpc$, as indicated in Table~\ref{tab:hod_samples}.

\subsection{Power spectrum covariance }
\label{sec:measAndCov}

The covariance matrices of the power spectrum multipoles are generated using a theoretical model based on the Gaussian predictions, as described in \cite{Grieb2016}. 
In practice, we compute the diagonal and cross-diagonal covariance terms composed of all the possible combinations among the multipole orders $\ell = (0,2,4)$. 
For computing the theoretical covariance we use the nonlinear power spectra obtained from iteratively fitting the multipole measurements at fixed cosmology and high $\kmax$. The fit is repeated until converging to a stable set of values for the model parameters. This partially accounts for non-Gaussian contributions to the covariance. Other non-Gaussian terms are neglected for simplicity, since we are using a box geometry, and window effects are expected to be negligible \citep{Blot2019, WadekarIvanovScoccimarro2020}. 

Unless otherwise stated, the covariance matrix is computed assuming the full volume of the Flagship I simulation, 
$V_{\rm box}=(3.78\Gpc)^3$. The error bars displayed in Fig.~\ref{fig:Measured_Multipoles} correspond to the Flagship simulation volume and illustrate the extreme precision of the measurements implemented in this work. Notice that for scales $k>0.1\,\kMpc$ the error over the signal is smaller than $10^{-2}$ for all redshift bins.
Additionally, to assess the expected constraining power and model performance with more realistic errors, tiled to the one expected from the full \Euclid mission, we rescale the Flagship covariance matrices $C$, by the ratio between the volume of the Flagship I comoving outputs, $V_{\rm box}$, and the volume of the corresponding \Euclid redshift bin, $V_{\rm shell}$. This approach allows us to define a scaling factor
\be
        \eta=\frac{V_{\rm box}}{V_{\rm shell}}\,,
 \label{eq: eta_volume}
\ee
which is latter employed to compute a covariance $C_{\rm shell}$ for a given shell  as
\be
        C_{\rm shell} = \eta \, C\,.
\ee
In order to derive volume factors $\eta$ we follow the specifications defined in \citet[][see also \citealt{Euclid-Pezzotta}]{EuclidForecast2019}. We assume four non-overlapping redshift bins, centred at the snapshot redshift, with a width $\Delta z=(0.2,\, 0.2,\, 0.2,\, 0.3)$ and a total survey area of $15\,000$ deg$^2$. The $\eta$ values are shown in the last column of Table \ref{tab:hod_samples}. 


\section{Theoretical models}
\label{sec:theory}

In this study, we consider three different modelling frameworks for the redshift-space galaxy power spectrum multipoles. Two of these are based on \ept and differ in their treatment of small-scale \rsd. The first model uses a fully perturbative expansion of the real- to redshift-space mapping, whereas the second replaces this expansion with an analytic function derived from the resummation of various nonlinear terms. The third model is hybrid, uses a Lagrangian expansion of the redshift-space galaxy density field, but each term is measured and emulated using high-resolution \nbody simulations (see Sec. \ref{sec:BACCO}).

\subsection{Perturbation theory models}
\label{sec:pt_models}

\subsubsection{Redshift-space mapping}
\label{sec:rsd_basics}

The mapping from the real-space position $\xv$ of a given object to its observed redshift-space position $\sv$ is sourced by the peculiar velocity of the object. In the plane-parallel approximation this mapping is given as
\begin{equation}
\sv = \xv -f \, u_z(\xv)\, \hat{z}\,,
\label{eq:RSDmap}
\end{equation}
where $u_z(\xv)$ is the component of the normalised velocity field along the line of sight, defined through $u_z(\xv) \equiv v_z(\xv) / (- f\,{\cal H})$, where $f$ refers to the linear growth rate and ${\cal H}$ is the comoving Hubble parameter.

The redshift-space density field $\delta^s$ in Fourier space is then \citep{Scoccimarro1999}
\begin{equation}
\delta^s(\kv) = \int \diff^3 x \, \eul^{-\imag\kv \cdot \xv} \, \eul^{\imag f k_z u_z(\xv)} \, \left[ \delta(\xv) + f\, \nabla_z u_z(\xv) \right],
\label{eq:delta_rsd_fourier}
\end{equation}
where the term in squared brackets captures the large-scale Kaiser effect \citep{Kaiser1987}, while the exponential prefactor generates the small-scale Finger-of-God (FoG) suppression.

The corresponding redshift-space power spectrum is \citep{Scoccimarro2004a}
\begin{equation}
    \begin{split}
        P^s(k,\mu) = \int & \diff^3 r\,\eul^{-\imag\kv\cdot\,\rv} \, \bigg\langle \eul^{\imag fk \mu \,\Delta u_z} \, \\
        &\times \brackets{\delta(\xv) + f\, \nabla_z u_z(\xv)} \brackets{\delta(\xv') + f\, \nabla_z u_z(\xv')} \bigg\rangle \,,
    \end{split}
\label{eq:power_zspace}
\end{equation}
where $\Delta u_z\equiv u_z(\xv)-u_z(\xv')$, $\rv \equiv \xv-\xv'$ and $\mu$ is the cosine of the angle between the pair separation vector and the line of sight.

\subsubsection{EFT and VDG$_{\infty}$}
\label{sec:EFTvsVDG}

From Eq.~(\ref{eq:power_zspace}), we distinguish two approaches. The \eft model \citep{Senatore2014, Perko2016}, that expands the exponential prefactor $\eul^{\imag fk \mu \,\Delta u_z}$ and describes the nonlinear density and velocity fields perturbatively, with counterterms absorbing small-scale FoG damping and residual \uv deficiencies. The \eft anisotropic galaxy power spectrum is written as
\begin{equation}
  \label{eq:PEFT}
  P_{\rm gg,\, EFT}(\kv) = P_{\rm gg,\, SPT}^{\,\rm tree}(\kv) + P_{\rm gg,\, SPT}^{\,\rm 1\mbox{-}loop}(\kv) + P_{\rm gg}^{\,\rm stoch}(\kv) + P_{\rm gg}^{\,\rm ctr}(\kv)\,,
\end{equation}
where the terms represent the tree-level and one-loop \spt contributions, stochastic noise, and counterterms, respectively.

And the \vdg model, that describes the contribution from virialised velocities through an effective damping function, with the power spectrum given by
\be
\begin{split}
      P_{\rm gg,\, VDG_\infty}(\kv) =  W_{\infty}(\kv)\, \Big[&P_{\rm gg,\, SPT}^{\,\rm tree}(\kv) + P_{\rm gg,\, SPT}^{\,\rm 1\mbox{-}loop}(\kv) \\ 
      & - \Delta P(\kv) + P_{\rm gg}^{\,\rm ctr, LO}(\kv)\,\Big] \, + P_{\rm gg}^{\,\rm stoch}(\kv)\,,
\label{eq:PVDG}
\end{split}
\ee
where $W_{\infty}$ denotes the effective damping function for modelling \rsd, and is derived by taking the \vdg, leading to
\begin{equation}
  \label{eq:Winfty}
  W_{\infty}(\lambda) = \frac{1}{\sqrt{1- \lambda^2\,\avir^2}}\,\exp{\left(\frac{\lambda^2\,\sigma_v^2}{1 - \lambda^2\,\avir^2}\right)}\,,
\end{equation}
with $\lambda = - \imag\,f k \mu$, and $\avir$ being a free parameter to modulate the kurtosis of the velocity difference distribution \citep{Scoccimarro2004a, Eggemeier_RSDComparison}. The term $\Delta P(\kv)$ corrects for extra contributions generated by the perturbative expansion of the \vdgf in the \eft approach \citep[][]{Eggemeier-Comet, Eggemeier_RSDComparison}.

\subsubsection{Galaxy bias expansion and stochasticity}
\label{sec:gal_bias}

The galaxy overdensity $\deltag$ is related to the dark matter density field $\delta$ through a bias expansion. Up to cubic order in $\delta$, and adopting the parametrisation from \cite{Eggemeier2019}, we have (keeping only terms relevant to the power spectrum):
\begin{equation}
  \label{eq:deltag_expansion}
  \begin{split}
  \deltag(\xv) =\;&\bone\delta(\xv) + \frac{\btwo}{2}\delta^2(\xv) + \gtwo{\cal G}_2(\xv) + \gtwoone{\cal G}_{21}(\xv) \\ &+ \bdtwod\nabla^2\delta(\xv) + \varepsilon_g(\xv)\,,
  \end{split}
\end{equation}
where $\bone$ and $\btwo$ are the linear and quadratic biases, $\gtwo$ and $\gtwoone$ modulate the impact of the non-local-in-matter-density operators ${\cal G}_2$ and ${\cal G}_{21}$, and $\bdtwod$ is the higher-derivative bias. Explicit expressions for the $P_{gg}$ spectra in Eqs.~(\ref{eq:PEFT},\ref{eq:PVDG}) obtained from the expansion in Eq.~(\ref{eq:deltag_expansion}) are given in \cite{Eggemeier2019}. 

The stochastic field $\varepsilon_{\rm g}$ is uncorrelated with the large-scale density field, $\langle\varepsilon_{\rm g}\delta\rangle=0$, and its power spectrum at large scales is expanded as \citep{DesJeoSch1802,EggScoSmi2106}
\begin{equation}
  \label{eq:Pshot}
  P_{\varepsilon_{\rm g}\varepsilon_{\rm g}}(k) = \frac{1}{\bar{n}}\left(\npzero + \nptwozero\,k^2 + \ldots\right)\,,
\end{equation}
where $\bar{n}$ represents the mean number density of the tracer population, whereas $\npzero$ and $\nptwozero$ are two free parameters modulating the constant and leading-order scale-dependent stochastic correction to take into account deviations from purely Poisson shot noise and exclusion effects. An additional stochastic component from the redshift-space mapping, $\varepsilon_{\nabla_z u_z}$, contributes a term \citep{Eggemeier_RSDComparison}
\begin{equation}
  \label{eq:Pshot_vel}
  P_{\varepsilon_{\rm g}\varepsilon_{\nabla_z u_z}}(k,\mu) = \frac{\nptwotwo}{\bar{n}} {\cal L}_2(\mu)\,k^2 + \ldots\,,
\end{equation}
where ${\cal L}_2(\mu)$ denotes the second-order Legendre polynomial. The total stochastic power spectrum is then
\begin{equation}
  \label{eq:Pstoch}
  P_{\rm gg}^{\,\rm stoch}(k,\mu) = P_{\varepsilon_{\rm g}\varepsilon_{\rm g}}(k) + P_{\varepsilon_{\rm g}\varepsilon_{\nabla_z u_z}}(k,\mu)\,.
\end{equation}

\subsubsection{Counterterms}
\label{sec:counterterms}

To mitigate the breakdown of \pt on small scales, both the \eft and \vdg models introduce a set of counterterms. At leading order we parametrise them as
\begin{equation}
  \label{eq:PctrLO}
  P_{\rm gg}^{\,\rm ctr, LO}(k,\mu) = -2 \sum_{n=0}^{2} c_{2n}\,{\cal L}_{2n}(\mu)\,k^2\,P_{\rm L}(k)\,,
\end{equation}
where $P_{\rm L}$ corresponds to the linear matter power spectrum. The angular dependence of the counterterms is parametrised in terms of the Legendre polynomials of order $2n$ -- in practice it introduces the parameters $\czero$, $\ctwo$, and $\cfour$, with adjustable amplitude. Since the contribution from higher-derivative bias scales identically to the $c_0$ counterterm, we let $c_0$ absorb the dependence on $b_{\nabla^2\delta}$.
 
The \vdg models has hence 3 counter-terms. For the \eft model, we include an extra next-to-leading order term with free amplitude $\cnlo$
\begin{equation}
  \label{eq:PctrNLO}
  P_{\rm gg}^{\,\rm ctr,NLO}(k,\mu) = \cnlo\,(\mu\,k\,f)^4\,P_{\rm gg, SPT}^{\,\rm tree}(\kv)\,.
\end{equation}
The total counterterm correction for the \eft model is therefore
\begin{equation}
  \label{eq:Pctr}
  P_{\rm gg}^{\,\rm ctr}(k,\mu) = P_{\rm gg}^{\,\rm ctr,LO}(k,\mu) + P_{\rm gg}^{\,\rm ctr,NLO}(k,\mu)\,.
\end{equation}

\subsubsection{Infrared resummation and bias relations}
\label{sec:IR-resummation}

To account for the smearing of \bao features by large-scale bulk flows \citep{Crocce2006b, Crocce2008}, we adopt an \ir resummation formalism, where the linear power spectrum is decomposed into a smooth component $\Pnw$ and a wiggly component $\Pw$: $\Plin(k)=\Pnw(k)+\Pw(k)$. Then the \ir-resummed power spectrum is computed as the sum of the smooth and the anisotropically damped wiggly component, following the treatment in \cite{IvaSib1807}.

Finally, for the \ept models, we assess the validity of physically-motivated relations to reduce the number of free parameters. We use the co-evolution relations for the non-local biases \citep{Fry1996, Chan2012}
\be
\gtwoone^{\rm coev} = \frac{2}{21}(\bone - 1 ) + \frac{6}{7}\gtwo\,,
\label{eq:g21Coevol}
\ee
and for $\gtwo$ we employ the excursion set relation \citep{EggScoCro2011}:
\be
\gtwo^\mathrm{ex-set} = 0.524 -0.547\,\bone + 0.046\,\bone^2\,.
\label{eq:g2ExSet}
\ee

\subsection{Hybrid model: the \bacco\ emulator}
\label{sec:BACCO}

The hybrid approach \citep{Modi_2020, PellejeroIbanez2022} comprises two components: a transformation from Lagrangian to Eulerian coordinates using the displacement field ${\bm \psi}\,(\qv)\,$ from \nbody simulations, and a Lagrangian bias model $F\brackets{\deltainit(\qv)}$. We adopt a second-order perturbative expansion for the bias model as \citep{Matsubara2008, DesJeoSch1802}
\begin{equation}
\begin{split}
     F\,\brackets{\deltainit(\qv)} = & \,1 + \bone^{\,{\cal L}}\,\deltainit(\qv) + \btwo^{\,{\cal L}}\, \brackets{ \deltainit^{\,2}(\qv)-\ave{\deltainit^{\,2}(\qv)}} \\
     & + b_s^{\,{\cal L}} \brackets{s^{\,2}(\qv) - \ave{ s^{\,2}(\qv)} } + b_{\nabla}^{\,{\cal L}}\, \nabla^{\,2} \deltainit(\qv) \, ,
	\label{eq:model}
\end{split}
\end{equation}
where $s^{\,2} \equiv s_{ij}s^{\,ij}$ is the scalar traceless tidal tensor. The Eulerian galaxy density field is then
\begin{equation}
1+\deltag(\xv) = \int \diff^3 q \, F[\deltainit(\qv)] \, \dirac\brackets{\xv-\qv-{\bm \psi}(\qv)}\,,
\label{eq:galmapp}
\end{equation}
where $\dirac$ represents the Dirac delta function.

The real- to redshift-space mapping is implemented by modifying the displacement field \citep{PellejeroIbanez2022}
\begin{equation}
{\bm \psi}(\qv) \rightarrow {\bm \psi}(\qv) + \, \frac{v_z(\xv)}{\cal H}\hat{z}\,,
\label{eq:galDispRSD}
\end{equation}
where the velocity field $\vv(\xv)$ is constructed from velocities measured in \nbody simulations.

To account for intra-halo satellite velocities, we introduce a damping function \citep[following][]{Orsi_2018} -- as in Eq.~(\ref{eq:PVDG}), it plays the role of damping the final galaxy power spectrum -- with shape
\begin{equation}
    W(k,\mu) \equiv \brackets{(1 - f_\text{sat}) + f_\text{sat} \frac{\lambda_\text{sat}^2}{\lambda_\text{sat}^2 + k^2 \mu^2}}^2\,.
\label{eq:w_bacco}
\end{equation}
Here $f_{\rm sat}$ is the satellite fraction and $\lamsat$ their velocity dispersion. Lastly, the stochastic component is modelled as $P_{\rm gg}^{\,\rm stoch}(k,\mu) = P_{\varepsilon_{\rm g}\varepsilon_{\rm g}}(k)$, with the same shape as in Eq.~(\ref{eq:Pshot}).

\subsection{Alcock--Paczynski distortions}
\label{sec:AP}
 
To account for geometric distortions due to discrepancies between the true and the assumed fiducial cosmology, we apply \gls{ap} corrections. The line-of-sight and transverse distortion parameters are defined as the ratios of the Hubble parameter ($H$) and the comoving transverse distance ($D_{\rm M}$) in the true and fiducial cosmologies
\begin{equation}
  q_{\parallel} \equiv \frac{H'(z)}{H(z)}\, ,\quad q_{\perp} \equiv \frac{D_{\mathrm{M}}(z)}{D'_{\mathrm{M}}(z)}\,,
\end{equation}
where fiducial values are indicated by primed quantities.

The power spectrum multipoles in the fiducial cosmology are computed as \citep{Bal9605}
\begin{equation}
  \label{eq:pipe.PwAP}
  P'_{\ell}(k') = \frac{2\ell + 1}{2\,q_{\perp}^2\,q_{\parallel}} \int_{-1}^1 \diff\mu'{\cal L}_{\ell}(\mu') \, P\brackets{k(k',\mu'),\, \mu(\mu')}\,,
\end{equation}
with the wavenumber and angle mapping given by
\begin{align}
  k &= \frac{k'}{q_{\perp}} \left[1 + \mu'\,^2 \left(F^{-2}-1\right)\right]^{\,1/2}, \\
  \mu &= \frac{\mu'}{F} \left[1 + \mu'\,^2 \left(F^{-2}-1\right)\right]^{\,-1/2},
\end{align}
where $F \equiv q_{\parallel}\,/\,q_{\perp}$.

\section{Analysis procedure}
\label{sec:perf_metrics}

In this section we outline the Bayesian likelihood analysis. We describe the parameter space and priors, and the performance metrics used to evaluate each model's accuracy, precision, and goodness of fit as a function of scale cut.

\subsection{Likelihood analysis}

We sample the posterior distribution using the nested sampling algorithm \texttt{MultiNest} via its \texttt{Python} wrapper \texttt{PyMultiNest} \citep{Skilling2006},\footnote{\url{http://johannesbuchner.github.io/PyMultiNest}} configured with 600 live points, a sampling efficiency of 0.5, and an evidence tolerance of 0.3. Posteriors are analyzed with \texttt{getdist} \citep{GetDist}.\footnote{\url{https://getdist.readthedocs.io}}

We assume a Gaussian likelihood, which up to a normalisation constant can be expresses as:
\be
-2\ln\mathcal{L}\,(\thetav) = \sum_{i,j=1}^{N_{\rm bins}}\brackets{P^{\,\rm th}_i(\thetav)-P^{\,\rm data}_i\,}\,C^{\,-1}_{ij}\,\brackets{P^{\,\rm th}_j(\thetav)-P^{\,\rm data}_j\,}\,,
 \label{eq:logL}
\ee
where $P^{\,\rm th}$ and $P^{\,\rm data}$ are the theoretical and measured power spectrum multipoles, $C_{ij}$ is the data covariance (Sect.~\ref{sec:measAndCov}), and $N_{\rm bins}$ is the total number of $k$ bins. We test each model up to a maximum wave mode $\kmax$, applying a scale cut of ($\kmax$, $\kmax-\Delta k$, $\kmax-\Delta k$) for the $(P_0, P_2, P_4)$ multipoles, with $\Delta k=0.05\hMpc$ in our baseline analysis.

To achieve the required $\mathcal{O}(10^6)$ model evaluations, we use fast emulators: the \texttt{COMET} code \citep{Eggemeier-Comet, comet2}\footnote{\url{https://comet-emu.readthedocs.io}} for the \ept models and the \bacco\ emulator \citep{PellejeroIbanez2023}\footnote{\url{https://bacco.dipc.org/emulator.html}} for the hybrid approach, both providing evaluations in ~$10\,$ms.

\subsection{Parameter priors}
\label{sec:priors}

\begin{table}[h]
    \caption{List of parameters varied in our analysis. The first and second columns indicate the parameter type and name, while the third column reports the prior ranges adopted. The symbol $\mathcal{U}$ denotes a uniform distribution, with the minimum and maximum values specified in square brackets.
Since the emulation is performed in terms of $\sigma_8$, the \bacco prior on $A_{\mathrm{s}}$ is flat (as in the \ept models), but its boundaries are determined by the emulated values $\sigma_8^{\rm min}$ and $\sigma_8^{\rm max}$. In the table, we explicitly display these $\sigma_8$ limits in the prior column.
    }
  \renewcommand{\arraystretch}{1.4}
  \centering
  \begin{tabular}{|c|c|c|}
    \hline
    \rowcolor{blue!5}
     & Parameter & Prior\\
     \hline
    \rowcolor{blue!5}
    \multicolumn{3}{|c|}{{\bf Eulerian bias} expansion} \\
    \hline
    \multirow{3}{*}{Cosmology} & $h$ & \small{$\mathcal{U}\,[0.6,0.8]$}\\
    \cline{2-3}
    & $10^9\As$ & \small{$\mathcal{U}\,[0.85,2.95]$} \\
    \cline{2-3}
    & $\omega_\mathrm{c}$ & \small{$\mathcal{U}\,[0.085,0.15]$} \\
    \hline
    \multirow{4}{*}{Bias} & $\bone$ & \small{$\mathcal{U}\,[0.25,4]$}\\
    \cline{2-3}
    & $\btwo$ & \small{$\mathcal{U}\,[-5,10]$}\\
    \cline{2-3}
    & $\gamma_{\, 2}$ & \small{$\mathcal{U}\,[-4,4]$ or fix to Eq.~\eqref{eq:g2ExSet}}\\
    \cline{2-3}
    & $\gamma_{\, 21}$ & \small{$\mathcal{U}\,[-8,8]$ or fix to Eq.~\eqref{eq:g21Coevol}}\\
    \hline
    \multirow{3}{*}{Counterterms} & $\czero\,\big[({\rm Mpc}/h)^2\big]$ & \small{$\mathcal{U}\,[-500,500]$}\\
    \cline{2-3}
    & $\ctwo\,\big[({\rm Mpc}/h)^2\big]$ & \small{$\mathcal{U}\,[-500,500]$}\\
    \cline{2-3}
    & $\cfour\,\big[({\rm Mpc}/h)^2\big]$ & \small{$\mathcal{U}\,[-500,500]$}\\
    \hline
    EFT only & $\cnlo\,\big[({\rm Mpc}/h)^4\big]$ & \small{$\mathcal{U}\,[-500,500]$}\\
    \hline
    \vdg only & $\avir\,\big[{\rm Mpc}/h\big]$ & \small{$\mathcal{U}\,[0,20]$}\\
    \hline
    \multirow{3}{*}{Shot-noise} & $\npzero$ & \small{$\mathcal{U}\,[-1,2]$} \\
    \cline{2-3}
    & $\nptwozero\,\big[({\rm Mpc}/h)^2\big]$ & \small{$\mathcal{U}\,[-500,500]$}\\
    \cline{2-3}
    & $\nptwotwo\big[({\rm Mpc}/h)^2\big]$ & \small{$\mathcal{U}\,[-500,500]$}\\
    \hline
    \rowcolor{blue!5}
    \multicolumn{3}{|c|}{{\bf Hybrid Lagrangian bias} expansion}\\
    \hline
    \multirow{3}{*}{Cosmology} & $h$ & \small{$\mathcal{U}\,[0.6,0.8]$}\\
    \cline{2-3}
    & $A_\mathrm{s}$ & \small{$\mathcal{U}\,[\sigma_8^{\rm min} = 0.65, \sigma_8^{\rm max} = 0.9]$} \\
    \cline{2-3}
    & $\Omega_\mathrm{c}$ & \small{$\mathcal{U}\,[0.23,0.4]$} \\
    \hline
    \multirow{4}{*}{Bias} & $\bone^{\,{\cal L}}$ & \small{$\mathcal{U}\,[0.2,2]$} \\
    \cline{2-3}
    & $\btwo^{\,{\cal L}}$ & \small{$\mathcal{U}\,[-2,2]$} \\
    \cline{2-3}
    & $b_{s^2}^{\,{\cal L}}$ & \small{$\mathcal{U}\,[-2,2]$} \\
    \cline{2-3}
    & $b_{\nabla^2 \delta}^{\,{\cal L}}\,\big[({\rm Mpc}/h)^2\big]$ & \small{$\mathcal{U}\,[-4,4]$} \\
    \hline
    \multirow{2}{*}{FoG} & $\fsat$ & \small{$\mathcal{U}\,[0,1]$} \\
    \cline{2-3}
    & $\lamsat$ & \small{$\mathcal{U}\,[0,2]$} \\
    \hline
    \multirow{2}{*}{Shot-noise} & $\npzero$ & \small{$\mathcal{U}\,[-1,1]$} \\
    \cline{2-3}
    & $\nptwozero\,\big[({\rm Mpc}/h)^2\big]$ & \small{$\mathcal{U}\,[-3,3]$}\\
    \hline
  \end{tabular}
  \label{tab:priors}
\end{table}

We sample the cosmological parameters $\paren{h, \omegac, \As}$ 
with broad flat priors (Table~\ref{tab:priors}). The spectral index $\ns$ and baryon density $\omegab$ are fixed to their fiducial values.

The \ept model parameter space includes 11 nuisance parameters: ten shared parameters ($\bone,$ $\btwo,$ $\gtwo,$ $\gtwoone,$ $\czero,$ $\ctwo,$ $\cfour,$ $\npzero,$ $\nptwozero,$ $\nptwotwo$) and one model-specific parameter ($\cnlo$ for \eft; $\avir$ for \vdg).

The hybrid \bacco\ bias basis can be related to the \ept basis via the approximate relations \citep{Zennaro_2022}:
\begin{equation}
    \bone = \bone^{\,{\cal L}} + 1\,, \quad  \btwo = 2\btwo^{\,{\cal L}} + \frac{8}{21}\bone^{\,{\cal L}}\,, \quad \gtwo = b_{s^2}^{\,{\cal L}} - \frac{2}{7}\bone^{\,{\cal L}}.
    \label{eq:bacco2ept}
\end{equation}
While \bacco\ does not explicitly include $\gtwoone$ as a free parameter, this term is generated during the advection to Eulerian space. The hybrid model shares six nuisance parameters with the \ept models (assuming the $\gtwoone$ relation in Eq.~\ref{eq:g21Coevol}), plus $\npzero$ and $\nptwozero$. Its higher-derivative term $b_{\nabla^2 \delta}^{\,{\cal L}}$ partially reproduces the $\czero$ counterterm, while the damping parameters $f_{\rm sat}$ and $\lambda_{\rm sat}$ are specific to its treatment of FoG effects.

Nuisance parameters are varied under broad flat priors (Table~\ref{tab:priors}), providing a conservative setup to avoid prior-driven biases. The only exceptions are configurations validating bias relations (Sect.~\ref{sec:IR-resummation}), where $\gtwo$ and $\gtwoone$ are fixed to Eqs.~\eqref{eq:g2ExSet} and \eqref{eq:g21Coevol}, respectively. Prior ranges differ between the \ept models and \bacco, but are sufficiently wide that posteriors do not hit boundaries at intermediate scales. We explore potential boundary effects at large scales in Sect.~\ref{sec:RSD_comparison}.

\subsection{Performance metrics}
\label{sec:stat_metrics}
We assess model performance using three metrics.

\subsubsection{Figure of bias}
\label{sec:figure_of_bias}

We quantify parameter recovery bias using the figure of bias (\fob)
\begin{equation}
    {\rm FoB}(\thetav) \equiv \brackets{\paren{\ave{\thetav}-\thetav_{\,\rm fid}}^{\intercal}S^{-1}(\thetav)\,\paren{\ave{\thetav}-\thetav_{\,\rm fid}}}^\frac{1}{2},
    \label{eq:fob}
\end{equation}
where $\ave{\thetav}$ is the posterior mean, $\thetav_{\,\rm fid}$ are the fiducial values, and $S(\thetav)$ is the parameter covariance. We compute the \fob for $\thetav=\paren{h,\, \As,\, \omegac}$, marginalizing over nuisance parameters.

For a 3-parameter \fob, the $68\%$ and $95\%$ confidence limits correspond to values of 1.88 and 2.83, respectively. For analyses with \Euclid-like covariances, we rescale the $68\%$ and $95\%$ thresholds by $1/\sqrt{\eta}$, where $\eta$ is the volume ratio between the Flagship box and the \Euclid-like shell (Table~\ref{tab:hod_samples}).

\subsubsection{Figure of merit}
\label{sec:FoM}

We evaluate constraining power using the figure of merit (\fom) \citep{Wang2008}
\be
{\rm FoM}(\thetav) = \brackets{\det S\negthickspace\paren{\thetav}}^{\,-{1}/{2} },
\label{eq:fom}
\ee
where $\det S\negthickspace\paren{\thetav}$ is the determinant of the parameter covariance. A larger \fom indicates tighter constraints.

\subsubsection{Goodness of fit}
\label{sec:GoF}

We assess model goodness of fit fit using a \ppd test \citep{PPD-Doux}. This involves generating replicated data $P^{\,\rm rep}$ from the posterior and comparing the $\chi^2$ statistic between replicated and observed data. The $p$-value is
\be
    p = {\cal P}\brackets{\chi^2\Big(P^{\,\rm rep}, \thetav\Big) > \chi^2\Big(P^{\,\rm data}, \thetav\Big)\,\Big|\,P^{\,\rm data}},
    \label{eq:pvalue}
\ee
where the $\chi^2$ has the same form as Eq.~(\ref{eq:logL}). Small $p$-values ($\lesssim 0.05$) indicate poor model fit, while very large ($\gtrsim 0.95$) ones may suggest covariance overestimation.

\section{Baseline modelling choices for \eft and \vdg }

\label{sec:EulerianValidation}
\label{sec:fully_theory}
\begin{figure*}[h]
	 \centering
    \includegraphics[width=1.\linewidth]{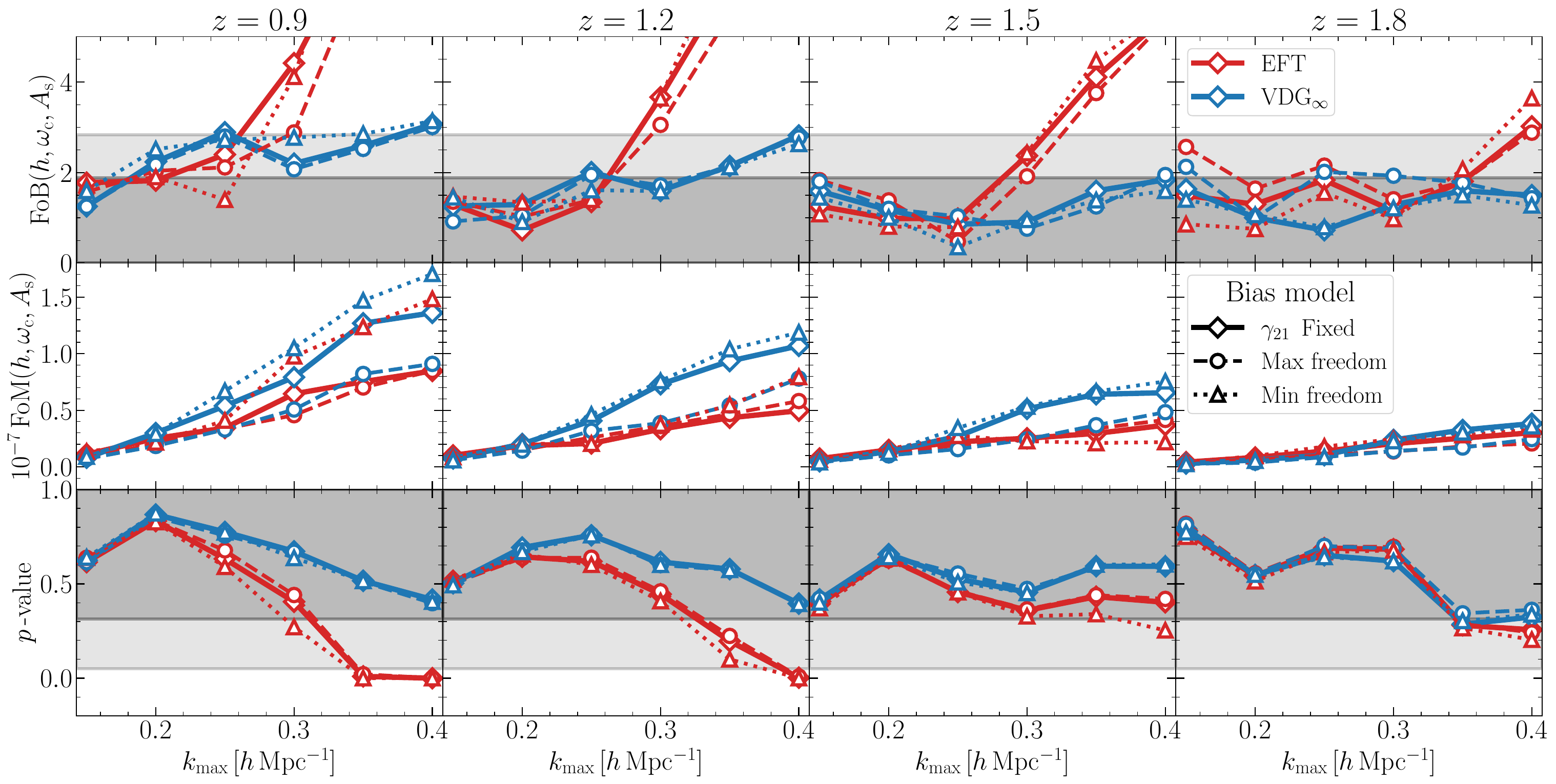}
    \caption{Performance metrics  -- \fob, \fom, and \pvalue are shown in the \emph{top}, \emph{middle}, and \emph{bottom row}, respectively -- for the two different \ept models, here marked with different colours, as a function of the maximum fitting scale $\kmax$. For each case, we show the maximal-freedom configuration (dashed lines with circle markers), the minimal-freedom one (dotted lines with triangle markers), and the one with only $\gtwoone$ fixed to the coevolution relation (solid lines with diamond markers). In all cases, we assume a covariance matrix matching the full volume of the Flagship comoving snapshots.
    \emph{Each column} shows results at different redshift bins. The grey bands in the \fob and \pvalue panels represent the $68\%$ and $95\%$ percentiles of the corresponding distribution. 
    }
    \label{fig:compare_EFTvsVDG_AllFreeAndg2g21Fix}
\end{figure*}

In this section we investigate different modelling choices for the \pt-based models, towards the definition of a baseline analysis configuration that will then be used in Sect.~\ref{sec:RSD_comparison}. This includes determining optimal nuisance parameters priors, scale cuts, and potentially fixing some of the tidal bias parameters to a given relation. For this exercise, we assume a covariance matrix corresponding to the full comoving volume of the mocks, in order to have the most stringent possible test.

\subsection{Bias model assumptions}
\label{sec:biasmodelassumption}

Figure~\ref{fig:compare_EFTvsVDG_AllFreeAndg2g21Fix} illustrates the relative performance of the \eft (red curves) and \vdg (blue curves) model as a function of the maximum wave mode $\kmax$ for three different bias configurations:
\begin{itemize}[leftmargin=8pt, topsep=0pt, itemsep=0.5pt]
    \item[-] {`maximal freedom'}, where all bias parameters are allowed to vary freely (dashed lines with circle markers);
    \item[-] {`minimal freedom'}, where both tidal biases $\gtwo$ and $\gtwoone$ are fixed to the excursion-set and co-evolution relations, respectively, as described in Sect.~\ref{sec:IR-resummation} (dotted lines with triangle markers);
    \item[-] {`intermediate'}, where only the third-order tidal bias $\gtwoone$ is fixed to the co-evolution relation (solid lines with diamond markers).
\end{itemize}

In the maximal-freedom case we observe that the inclusion of information from increasingly smaller scales impacts very differently the performance of the \eft and \vdg models. In fact, the \vdg model displays a significantly better  \pvalue and \fob compared to the \eft model at high $\kmax$, especially at lower redshifts, where the performance of the \eft model deteriorates substantially. Conversely, the \vdg model exhibits a \fob that stays nearly flat across the explored redshift range -- consistently staying within the 95\% confidence interval except at $z=0.9$ and $\kmax=0.4\,\kMpc$ -- while its constraining power displays a monotonic improvement. This trend is additionally reinforced by the \pvalue, which stabilises (around $0.5$) once the fit starts including scales with higher signal-to-noise. On the other hand, for the \eft case the \fob exceeds the $2\sigma$ threshold at a typical scale of $\kmax \sim 0.25$--$0.3\kMpc$, pointing to a premature breaking of the model. 
On larger scales, at $\kmax <  0.25 \, \kMpc$, the \fob and \fom of the two models are generally consistent to each other. 

Compared to the maximal-freedom case, the use of the intermediate and minimal-freedom case results in neither \eft nor \vdg showing any substantial improvement or degradation in terms of \fob and \pvalue across most redshift bins. However,  when considering the \vdg model, we notice a significant increase in the \fom, ranging from 50\% to 100\% improvement (as the $\kmax$ extends from $0.25 \kMpc$ to about $0.4\kMpc$), in a regime where both \pvalue and \fob are still within the $2\sigma$ threshold. On the other hand, the \eft model does not show a noticeable gain in the \fom when fixing only $\gtwoone$.

It is worth pointing out that recent full-shape analyses based on the EFT formalism use a maximum wave mode $\kmax\lesssim  0.25 \, \kMpc$ \citep{IvanovEtal2020, DESI_FS}, in agreement with the minimum scales where we find that no significant biases are introduced. Nonetheless, even in the maximal-freedom configuration, the maximally achievable \fom obtained with the \vdg model (at $\kmax \sim 0.35\,\kMpc$) is about twice as large as the corresponding one obtained with the \eft model (at $\kmax \sim 0.25\,\kMpc$). Such improvement is more evident when assuming bias relations. For completeness, in Appendix~\ref{app:2d_bias_exploration} we add a detailed exploration on the parameter degeneracies exhibited by these two \rsd models when assuming the bias relations previously discussed.

\begin{figure*}[h]
	 \centering
    \includegraphics[width=1.\linewidth]{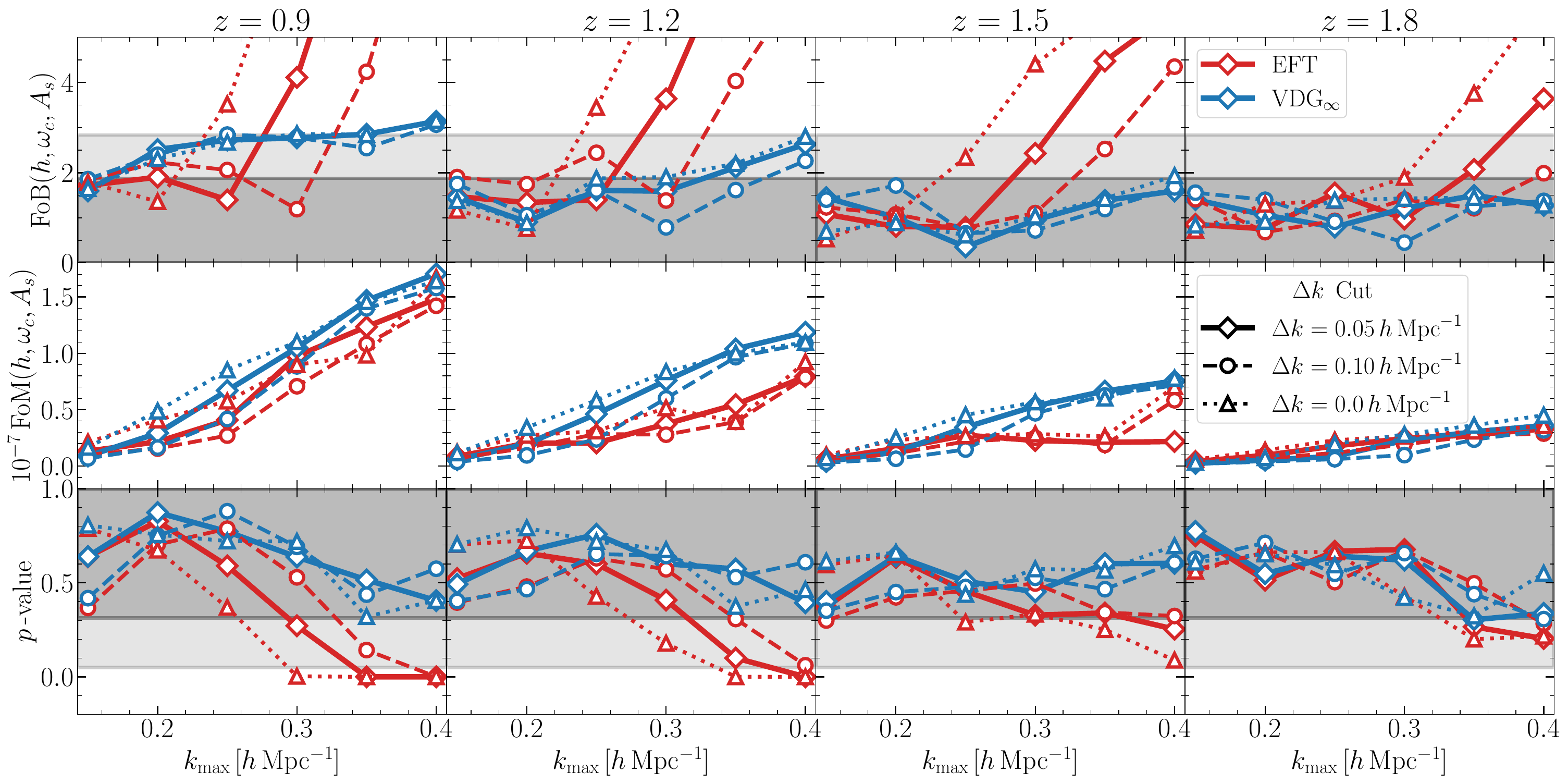}
    \caption{Same as in Fig.~\ref{fig:compare_EFTvsVDG_AllFreeAndg2g21Fix}, but only for the minimal-freedom bias configurations, and showing the impact of different values of $\Delta k$ in the determination of the $\kmax$ for the quadrupole and hexadecapole.}
    \label{fig:compare_DkCut}
\end{figure*}

\subsection{Test over different scale cuts per multipole}
\label{sec:p2p4_information}
In the previous section, we adopted scale cuts where the maximum wave mode considered for the quadrupole and hexadecapole differ by $\Delta k=0.05\kMpc$ from the one of the monopole. In this section we provide a justification for this choice.
Figure~\ref{fig:compare_DkCut} presents the cosmology fits for three configurations with $\Delta k \,=\, \left(0,\; 0.05,\; 0.1\right)\kMpc$. For this analysis, we adopt the minimal-freedom bias model, which, as demonstrated in Fig.~\ref{fig:compare_EFTvsVDG_AllFreeAndg2g21Fix}, gives a consistent performance to the maximal-freedom case in terms of \fob.

For the \eft model, the \fob shows a strong sensitivity to the amount of information contained in higher-order multipoles. Specifically, when using the same scale cut for all three multipoles ($\Delta k = 0$), the model becomes biased already at $\kmax \sim 0.25\kMpc$ for most redshifts, indicating a premature breaking of the \rsd modelling on those scales. This trend is reflected in the goodness-of-fit, where the \pvalue consistently worsens when considering smaller scales in the quadrupole and hexadecapole. Despite the dependency of the \fom on $\Delta k$ is weaker, it shows an improvement on scales where the model is still valid ($\kmax\leq0.25\kMpc$), particularly for the closest redshift. The combination of such trends indicates a slight preference for using $\Delta k = 0.10\kMpc$. 

In contrast, the \vdg model proves to be more robust against variations in $\Delta k$, showing no significant changes in terms of \fob or \pvalue. In this case, the \fom increases when incorporating additional information from higher-order multipoles, particularly at larger scales. Thus, the \vdg model slightly favours $\Delta k = 0\kMpc$.

For our baseline \rsd comparison, we set $\Delta k = 0.05\kMpc$, as it represents a compromise of the $\Delta k$ preferred by the two models. Finally, we verified that removing the hexadecapole from the fits only leads to marginal changes in the \fob and the \fom\,-- indicating, particularly for the \eft case, that their behaviour is driven mainly by the quadrupole. Nonetheless, the \pvalue consistently deteriorates, due to a weaker constraint of nuisance parameters.
Therefore, we choose to include the hexadecapole in our baseline comparison between the \eft and \vdg model. Further details on this comparison can be found in Appendix~\ref{sec:app_hexaInfo}.

\subsection{Priors on noise and stochastic terms}
\label{sec:noiseExploration}

In our main \rsd model comparison, presented in Sect.~\ref{sec:RSD_comparison}, we fix $\gtwoone$ to the co-evolution relation and neglect the anisotropic noise contribution by setting $\nptwotwo = 0$. This specific configuration enables a more direct comparison between the \ept-based models and the hybrid approach. 

\begin{figure*}[h]
	 \centering
    \includegraphics[width=1.\linewidth]{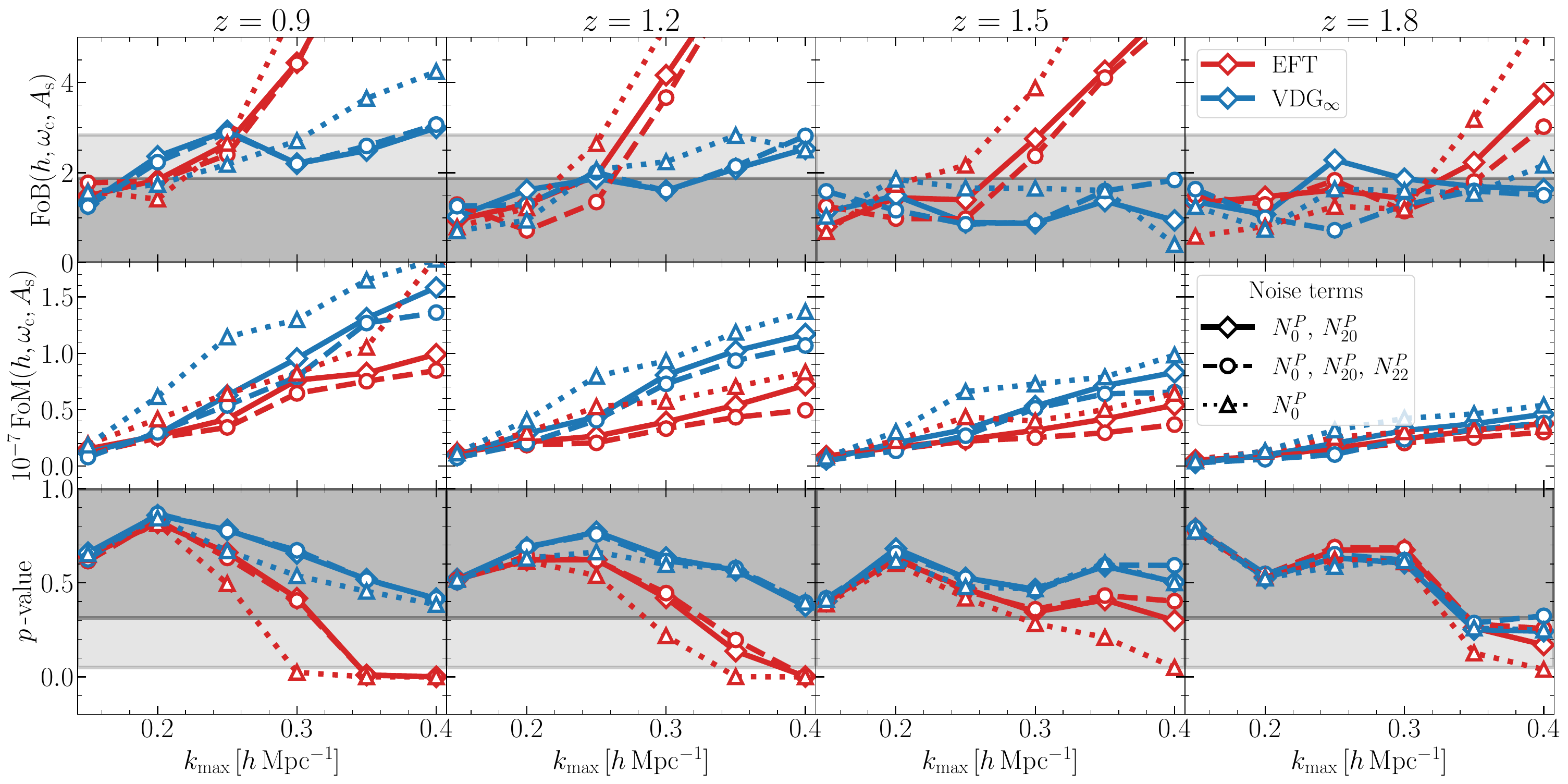}
    \caption{Same as in Fig.~\ref{fig:compare_EFTvsVDG_AllFreeAndg2g21Fix}, but only for the minimal-freedom bias configurations, and showing the impact of different configurations for the shot-noise parameters.}
    \label{fig:compare_NoiseImpact}
\end{figure*}

On the contrary, in Fig.~\ref{fig:compare_NoiseImpact} we present the performance of both the \eft and \vdg models under different configurations of the shot-noise parameters.
The circle markers and dashed lines correspond to the case where all noise terms are allowed to vary with uninformative priors. The solid lines show the baseline case that will be used in Sect.~\ref{sec:RSD_comparison}, for which $\nptwotwo=0$. Removing the additional degree of freedom associated with the anisotropic noise does not significantly alter the performance of the models. However, setting both the scale-dependent and anisotropic noise contributions to zero, as shown by the dashed lines and triangle symbols, introduces an extra bias in the cosmological parameters. This bias is particularly pronounced for smaller scales and lower redshifts. For the \eft model, this configuration not only increases the \fob but also degrades the \pvalue. In Sect.~\ref{sec:stocasticity_constraints} we discuss in more details the previous results.

\subsection{Baseline setup for BACCO}

In terms of the hybrid Lagrangian bias model, we use the \bacco code, specifically, the redshift-space hybrid bias emulator described in \cite{PellejeroIbanez2023}. This emulator has been tested on a variety of scenarios including \hod \citep{Beyond-2pt_2024}, SubHalo Abundance Matching \citep[SHAM,][]{PellejeroIbanez2023}, and observed data \citep{PellejeroIbanez2024b}.

For our specific analysis, we choose a $k$-binning for the data vectors much finer than the one used for the training of the original version of the \bacco emulator, which, convolved with the intrinsic noise in the training sample, leads to small but systematic biases in the model predictions. Consequently, we use a different version of \bacco that was trained on a finer grid. A detailed explanation of this issue is added in Appendix~\ref{app:bacco_performance}, where we also show how the current status of the \bacco code is not enough to deal with the Flagship I precision.\footnote{Alternatively, we could have chosen to re-bin the $P_\ell$ measurements, but since the errorbars shrinks accordingly with the increasing number of modes per bins, assuming the full Flagship volume still would have led to errorbars of the same order of the emulator precision.} Nonetheless, analysing a more realistic case such as the \Euclid-like volume, is suitable to test the hybrid model. As a consequence, for our main results, we opt for comparing the three different \rsd models only in the \Euclid-like scenario.

While the current version of \bacco already enables robust applications in realistic survey settings, there is still room for improvement, and ongoing developments aim to extend its precision and applicability.

\section{Results}
\label{sec:results}

In this section we summarise the main findings of our analysis for each of the \rsd models introduced in Sect.~\ref{sec:theory}. Our objective is to identify the smallest scale that can be adopted for each model configuration without introducing biases in the recovery of cosmological parameters, while still maximising the information extracted.

\subsection{RSD model comparison}
\label{sec:RSD_comparison}

\begin{figure*}[h]
	 \centering
    \includegraphics[width=1.\linewidth]{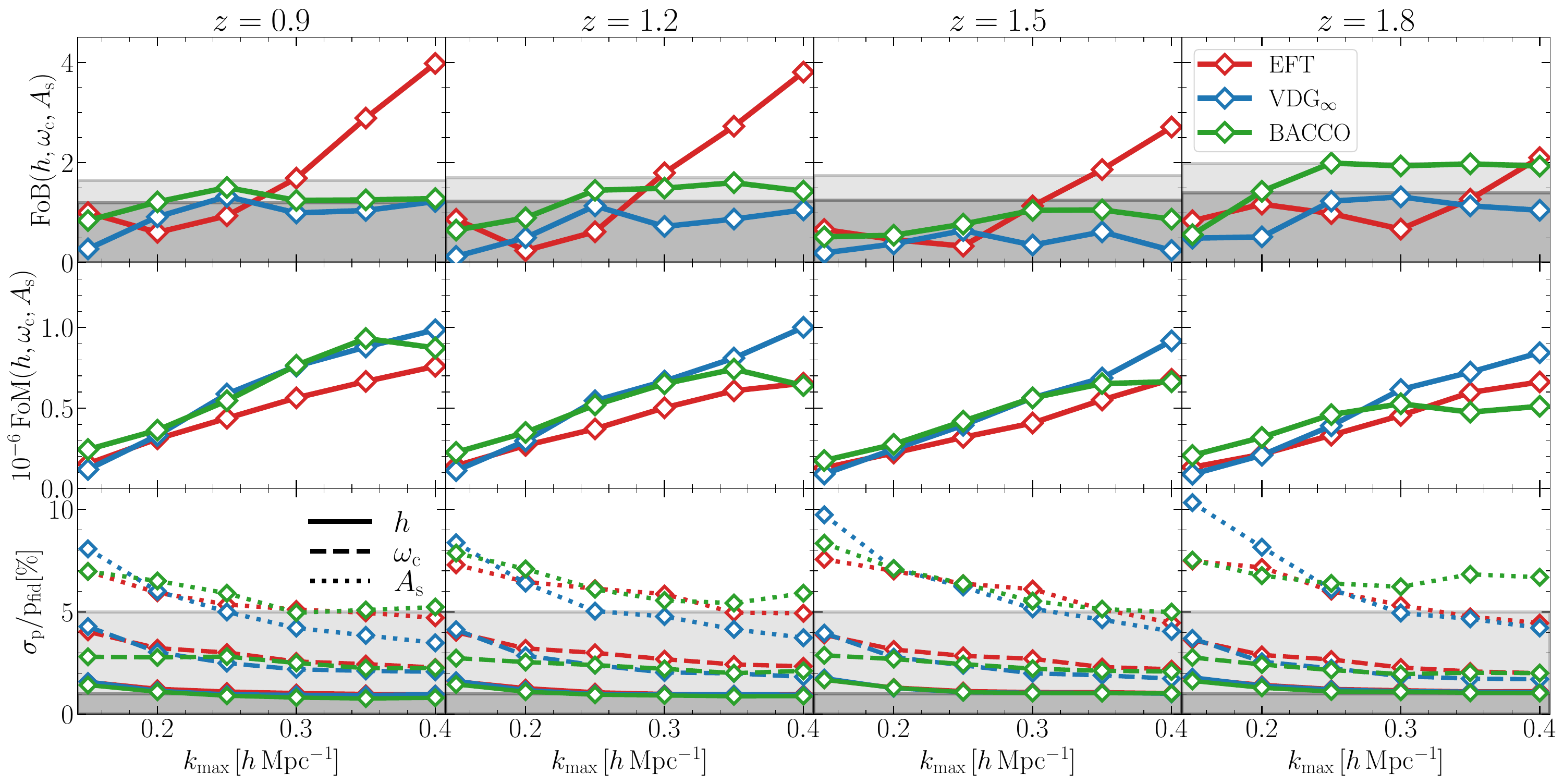}
    \caption{Performance metrics for the three \rsd models presented in Sect.~\ref{sec:theory} assuming a covariance matrix matching the expected volume by the end of the \Euclid mission. The \fob and \fom are shown in the \emph{first two columns}, while the \emph{bottom panel} shows the relative error of each parameter with respect to its fiducial value, as indicated in the legend. \emph{Each column} shows results for a different redshift bin. Different colours correspond to different models, as described in the legend. To have a consistent number of degrees of freedom, for the \ept-based models, we adopt the configuration with $\gtwoone$ fixed to the co-evolution relation and the anisotropic noise contribution $\nptwotwo$ set to 0. The $x$ axis shows the maximum wave mode $\kmax$ adopted in the fit. The grey bands in the \emph{\fob panel} represent the $68\%$ and $95\%$ percentiles of the corresponding \fob distribution, as explained in Sect.~\ref{sec:figure_of_bias}.}
    \label{fig:compare_EFTvsVDGvsBACCO_EucliVolume}
\end{figure*}

For this analysis we have rescaled the full volume of the Flagship I simulation to be closer to the \Euclid survey, as defined in Sect.~\ref{sec:measAndCov}. 
Since the covariance has been rescaled, we focus our analysis in the \fob and \fom, not considering the goodness-of-fit. Instead  of the \pvalue, we show the relative error on each cosmological parameter.

In Fig.~\ref{fig:compare_EFTvsVDGvsBACCO_EucliVolume} we compare the three different \rsd models described in Sect.~\ref{sec:theory} as a function of the maximum wave mode $\kmax$ included in the fit. 
For the two \ept models, we only consider the intermediate configuration, with $\gtwoone$ fixed to the co-evolution relation, and additionally set $\nptwotwo=0$, in order to partially reproduce the same default configuration of the hybrid model from the \bacco emulator (see Sect.~\ref{sec:BACCO}). 
The limits of the \fob are rescaled as described in Sec.~\ref{sec:figure_of_bias}, to make sure that they properly reflect the 68\% and 95\% \fob thresholds even after having rescaled the data covariance.

In terms of \fob, we observe that the three different \rsd models are well behaved up to a maximum wave mode $\kmax= 0.25\kMpc$, with deviations of at most $2\sigma_{\rm E}$ for all the considered redshifts. Pushing the models to smaller scales, at $\kmax>0.25\kMpc$, the \eft model starts showing a worsening trend in the \fob compared to \bacco and the \vdg models, particularly for the two closest redshifts, where the \fob crosses the $2\sigma_{\rm E}$ threshold immediately above $\kmax=0.25\kMpc$. For the snapshots at $z=1.5$ and $z=1.8$, the 
crossing of the $2\sigma_{\rm E}$ threshold occurs at slightly larger wave modes, namely at $\kmax=0.35\kMpc$ and $\kmax=0.4\kMpc$, respectively, due to the more linear behaviour at high $z$. Nonetheless, a growing trend in the \fob can still be observed starting from $\kmax\gtrsim 0.25\kMpc$. On the other hand, for most of the redshifts and scales explored in this work, both the \vdg and \bacco models return a \fob of at most $2\sigma_{\rm E}$, without any clear indication for a growing trend in the \fob.

In terms of \fom, \bacco consistently displays slightly better constraining power on scales $\kmax \lesssim 0.20\,\kMpc$, compared to the \eft and \vdg models. However we notice that over these scales 
some of the nuisance parameters of \bacco hit their prior boundaries, potentially affecting the marginalised posterior distribution of the cosmological parameters.\footnote{In Appendix \ref{app:BACCO_exploration} we describe some tests of the differences in the perturbative treatment and priors implemented in \bacco, compared to \vdg and \eft, finding that priors could lead to extra gains at larger scales.} It is also worth noticing that, across all redshifts, the gain in constraining power for both \bacco and the \vdg model roughly doubles when comparing the \fom at $\kmax \simeq 0.40\kMpc$ to the one at $\kmax = 0.25\kMpc$ -- beyond the range of validity of the \eft model. The flattening behaviour displayed by \bacco at $z=1.8$ (also present, though less prominent, at the other redshifts) and $\kmax>0.3\kMpc$, can be interpreted as a saturation of the information that can be extracted on these scales. In contrast, the \vdg model benefits from smaller-scale information to better constrain cosmological parameters, keeping a stable growing trend up to the minimum scale explored in this work.

The third row in Fig.~\ref{fig:compare_EFTvsVDGvsBACCO_EucliVolume} shows the constraining power on each of the cosmological parameters considered. This panel allows us to identify which parameters are better constrained in our fits. In particular, we notice that $h$ can be recovered with a precision of 1\% by all models, despite reaching a plateau in the information content when pushing the model to smaller scales. The matter density parameter $\omegac$ is recovered at the 3\% level.
Additionally, only with the \vdg model, it is possible to achieve a precision better that 5\% on $\As$ without introducing any significant bias. Finally, in the case of \bacco, we note that $\omegac$ is already well constrained from the first $k$-bin. However, its precision improvement is minimal when smaller scales are included in the analysis. Further, after $\kmax\gtrsim0.25\kMpc$ it follows the same trend as in \vdg and \eft. Such trend suggests that if affected by prior boundaries, the impact is only significant for the first $k$-bin. 

\begin{figure*}[h]
	 \centering
    \includegraphics[width=1.\linewidth]{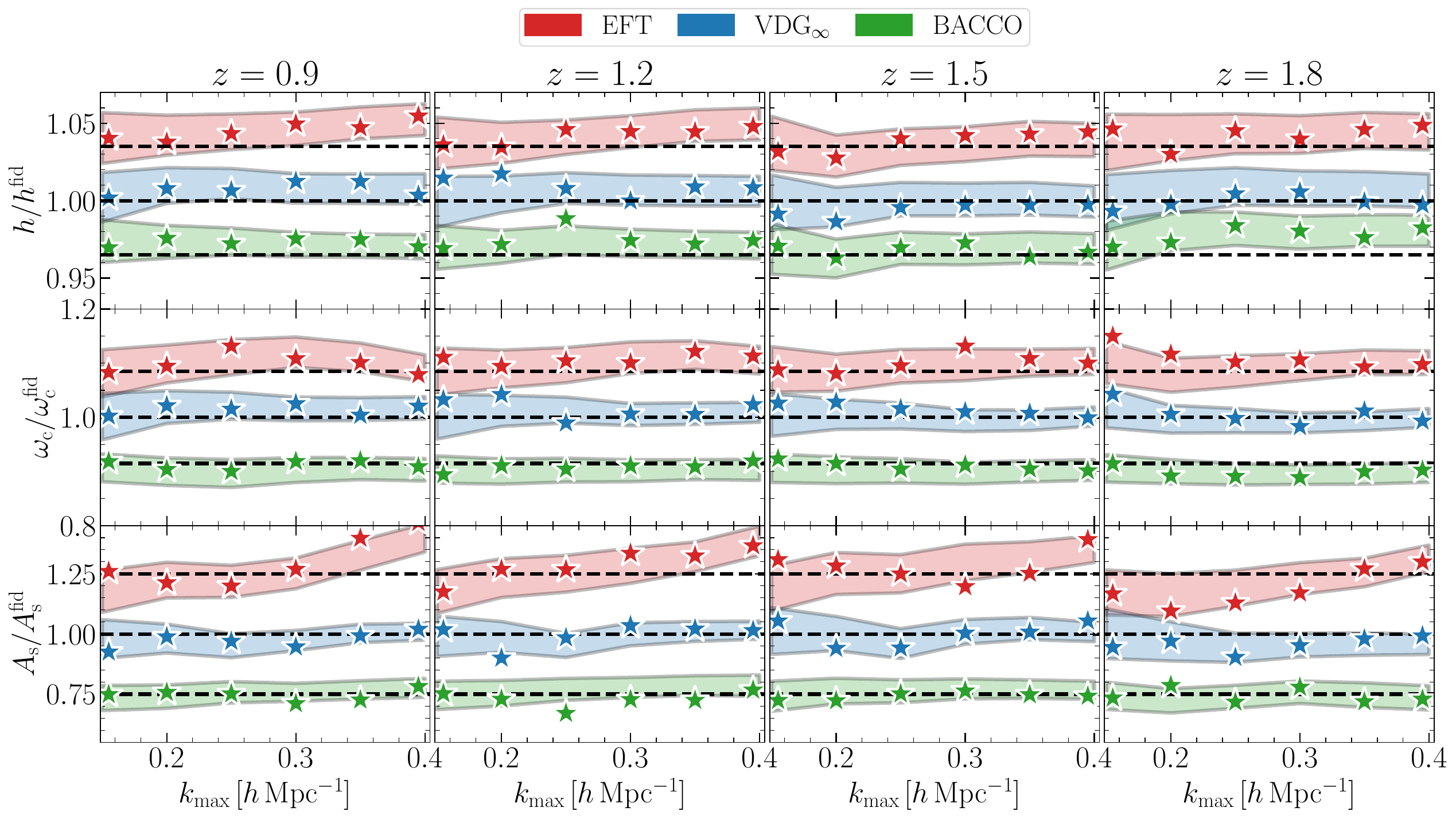}
    \caption{Evolution of the marginalised 1-dimensional posterior of each cosmological parameter, where shaded areas represent the 68\% confidence interval, centred around the mean values, of the posterior distribution normalised by the corresponding fiducial value. The normalised maximum-likelihood parameters, estimated by considering the position with the highest likelihood value obtained during the sampling process, are denoted with star symbols. The three cosmological parameters are shown in \emph{each one of the rows}, while the \emph{columns} display different redshifts. The colours correspond to the three \rsd models, as indicated in the legend. The dashed lines show the fiducial position for each of the cosmological parameters. For visualisation purposes, we have shifted the expected ratios by a constant value for each of the \rsd models, so that the different areas do not overlap.}
    \label{fig:compare_1D_EFTvsVDGvsBACCO_EucliVolume}
\end{figure*}

In Fig.~\ref{fig:compare_1D_EFTvsVDGvsBACCO_EucliVolume} we show the evolution as a function of $\kmax$ of the marginalised 1-dimensional posterior distribution for each one of the cosmological parameters. The shaded regions correspond to their 68\% confidence interval, centred at their mean marginalised posterior value, whereas star symbols correspond to the maximum-likelihood position as estimated from the collected samples. The different colours correspond to different \rsd models, as indicated in the legend. We find that the fiducial values of the three cosmological parameters are recovered within the $1\sigma$ region for most of the scale cuts both when using \bacco and the \vdg model, with the only exception at $z=1.8$, where \bacco displays a bias in $h$ beyond $\kmax=0.20\kMpc$. Otherwise, we observe that for the \eft model, particularly at the two closest redshifts, the parameters responsible for returning a larger \fob are $h$ and $\As$. 

Some recent works \citep{Holm-PriorsStudy, Carrilho2023} have shown that in state-of-art models (including several nuisance parameters per spectroscopic bin), the marginalised posterior parameter distributions is subject to strong prior-volume effects. Nonetheless, in a sample covering a large enough volume, as it is our case, this effect is negligible due to a more Gaussian likelihood \citep{Holm-PriorsStudy}.\footnote{In Appendix~\ref{sec:PriorsImpact} we test that more restrictive priors for the Flagship covariance case do not change our main conclusions.} For completeness, in Fig.~\ref{fig:compare_1D_EFTvsVDGvsBACCO_EucliVolume} we inspect the location of the maximum-likelihood values, finding that in the vast majority of cases explored in this work, it traces the mean of the posterior distribution, lying well within the $1\,\sigma$ marginalised contours. This result is in agreement with previous works, finding negligible prior-volume effects for the standard $\Lambda$CDM cosmology but becoming significant for extended cosmologies \citep{DESI_FS_cosmo, Carrilho2023, Moretti2023}.

\subsection{Posteriors of nuisance parameters for the \ept models}

In this section we focus on the two \ept models and compare them at the level of the nuisance parameters, precisely focusing on those that regulate the amplitude of \rsd and \uv effects plus those controlling the stochastic corrections.
For this test we adopt the most stringent error bars -- corresponding to the full Flagship I volume -- to better identify differences between the two models.

\subsubsection{Running of the counterterms}
\label{sec:counterter_tests}
\begin{figure}[h]
	 \centering
    \includegraphics[width=1.\columnwidth]{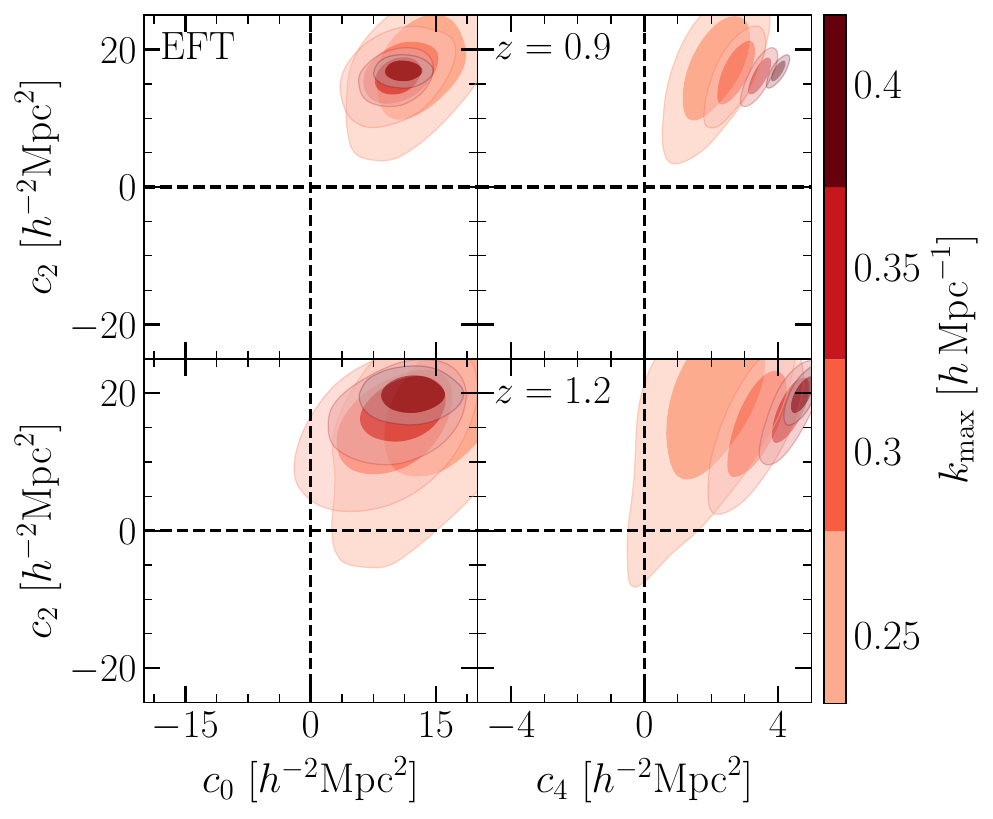}
    \includegraphics[width=1.\columnwidth]{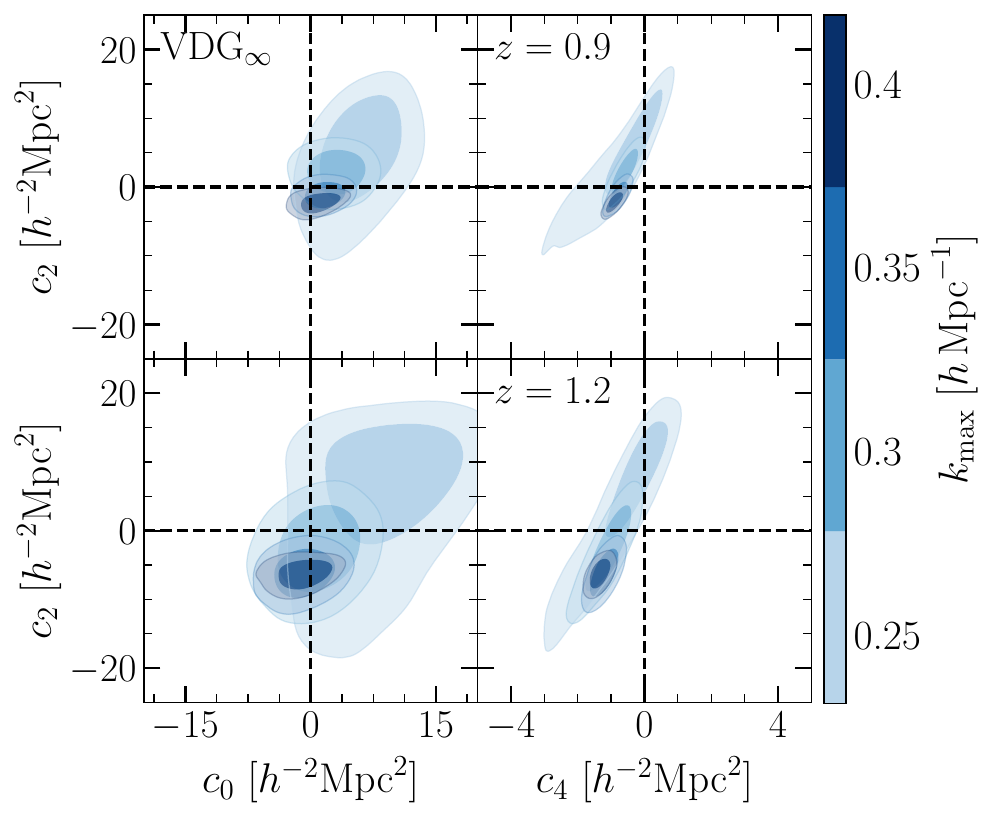}
\caption{2-dimensional posterior distribution of the leading-order counterterms $(\czero,\,\ctwo,\,\cfour)$ for both the \ept-based models. The \emph{top and bottom set of panels} corresponds to the \eft and \vdg model, respectively. In both sets, the \emph{first and second row} corresponds to the $z=0.9$ and $z=1.2$ snapshot, respectively. Dashed lines indicate the zero value.}
\label{fig:contour_counterterms}
\end{figure}
As explained in Sect.~\ref{sec:theory}, the counterterms play distinct roles in the two \ept-based models. In the \eft framework, these are mostly meant to capture the small-scale real- to redshift-space mapping, while additionally describing extra contributions coming from higher derivatives, halo exclusion, and velocity bias. On the other hand, in the \vdg model, the small-scale redshift-space mapping is captured by the analytical damping function $W_\infty$, leaving to the counterterms the role of describing effects other than \rsd. In Fig.~\ref{fig:contour_counterterms} we explore the behaviour of the counterterms at $\kmax>0.25\kMpc$ for the two closest redshift snapshots. Our goal is mostly to understand the behaviour of the counterterms in these configurations of the \eft model, where it becomes progressively more biased than the \vdg model, as indicated in Fig.~\ref{fig:compare_EFTvsVDGvsBACCO_EucliVolume}.

The first noticeable trend is that on scales $\kmax\geq 0.25\kMpc$ the 2-dimensional posteriors of the \eft model exhibit a drift in the $\ctwo$--$\cfour$ plane. This indicates that these parameters start absorbing residual effects at the expense of their main physical motivation. In contrast, the posteriors of the \vdg model converge to well localised positions, suggesting that these constraints benefit from the additional information carried by more nonlinear scales without introducing systematic biases. The second trend is related to the position at which the 2-dimensional $\ctwo$--$\czero$ constraints converge. For the \eft model, this position deviates significantly from zero, which is expected due to the role of the counterterms of modelling the small-scale \rsd effect. On the other hand, in the \vdg model, these parameters display a smooth convergence towards zero, particularly $\czero$, in a consistent way to results obtained in real-space.\footnote{See Fig.~F1 in \cite{Euclid-Pezzotta} for the corresponding real-space analysis.} This behaviour indicates that the damping function in the \vdg model is efficient in reproducing the impact from \rsd on small-scales, leaving mainly residual effects to be absorbed by counterterms.

\subsubsection{Constraints on stochasticity}
\label{sec:stocasticity_constraints}
\begin{figure}[h]
	 \centering
    \includegraphics[width=1.0\columnwidth]{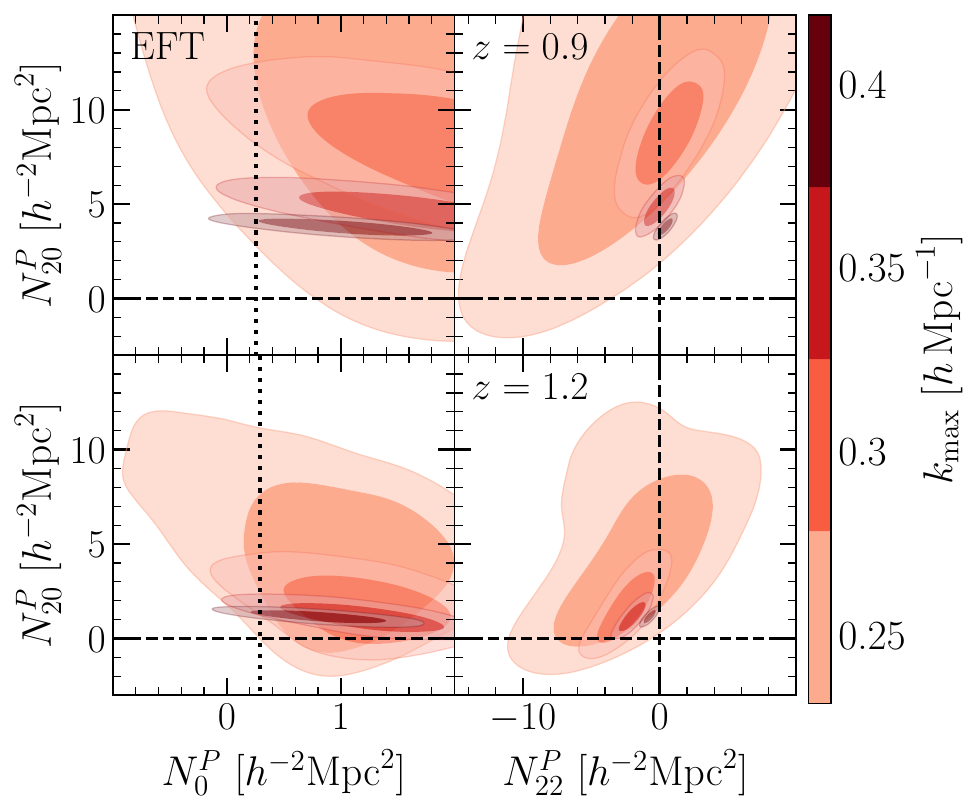}
    \includegraphics[width=1.0\columnwidth]{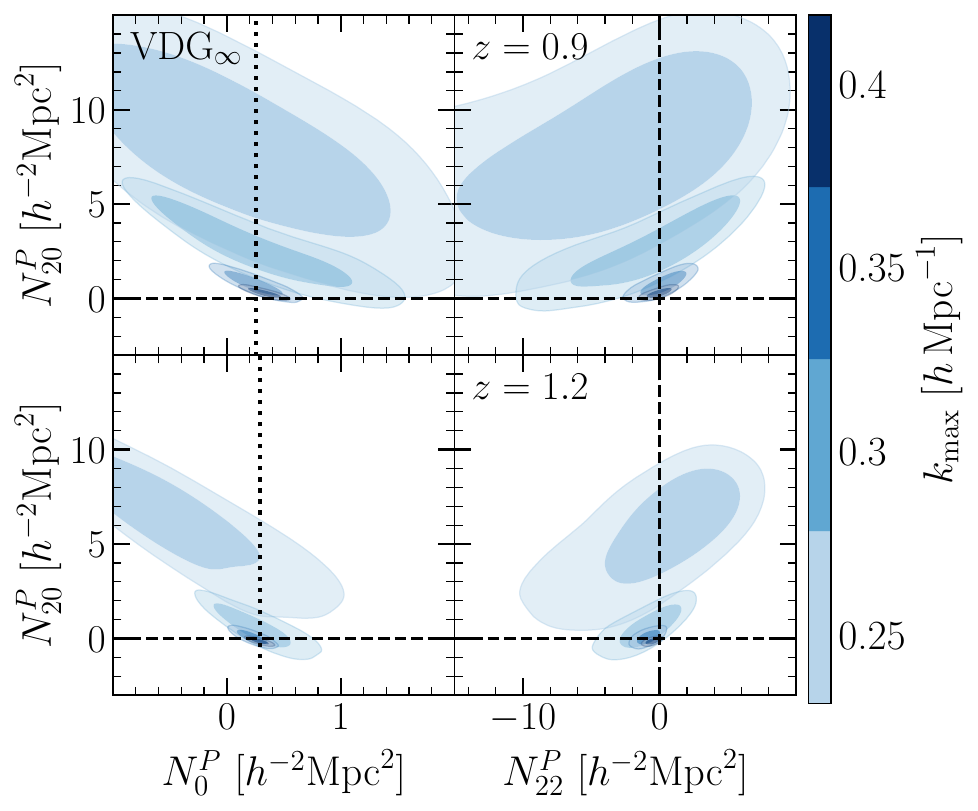}
\caption{2-dimensional posterior distribution of the shot-noise parameters $(\npzero,\,\nptwozero,\,\nptwotwo)$ for both \ept models. The \emph{top and bottom set of panels} corresponds to the \eft and \vdg models, respectively. In both sets, the \emph{first and second row} corresponds to the $z=0.9$ and $z=1.2$ snapshot, respectively.
Dotted lines in $\npzero$ show the large-scale value recovered from linear-theory fits of the $\Pgg/\Pmm$ ratio. Otherwise, for $\nptwozero$ and $\nptwotwo$, dashed lines indicate the zero value.
}
\label{fig:contour_Noiseterms}
\end{figure}
The results from the different shot-noise configurations observed in Sect.~\ref{sec:noiseExploration} can be better understood by inspection of the individual marginalised posterior of the shot-noise parameters. This is shown in Fig.~\ref{fig:contour_Noiseterms}, where we present marginalised 2-dimensional constraints of the parameters $(\npzero,\, \nptwozero,\, \nptwotwo\,)$. Although here we focus on the maximal-freedom case, the same conclusions hold when $\gtwoone$ is fixed to the co-evolution relation. Similarly to the tests from Sect.~\ref{sec:counterter_tests}, we explore scale cuts limited at $\kmax \geq 0.25 \, \kMpc$ and at the two closest redshifts, aiming to better understand whether such parameters compensate modelling failures.

We find that, while in the \eft model the constant deviation from purely Poissonian noise, $\npzero$, is recovered only at the $2\sigma$ level, in the \vdg model it is consistently constrained at the $1\sigma$ level. This may be one of the factors contributing to the enhanced constraining power of \vdg, given its degeneracy with $\bone$ and $\As$. As shown in Fig.~\ref{fig:compare_EFTvsVDGvsBACCO_EucliVolume}, $\As$ is precisely the parameter for which \vdg provides significantly tighter constraints compared to \eft. 

The second important observation is that, while the \eft model recovers a non-zero scale-dependent noise term ($\nptwozero \neq 0$), for \vdg this term is generally consistent with zero. This behaviour better aligns with findings from real-space analyses (see Fig.~8 in \citealt{Euclid-Pezzotta}).

Lastly, both models show that the anisotropic stochastic term $\nptwotwo$ is generally consistent with zero, though the \eft model at $z = 1.2$ displays a mild deviation from zero. Although the real values of these parameters depend on the specific galaxy sample under consideration, the constraints of $\npzero$ suggest that \vdg performs better than \eft in capturing the amplitude of the shot-noise corrections.

\subsubsection{Next-to-leading order effects}

\begin{figure*}
	 \centering
    \includegraphics[width=1.\columnwidth]{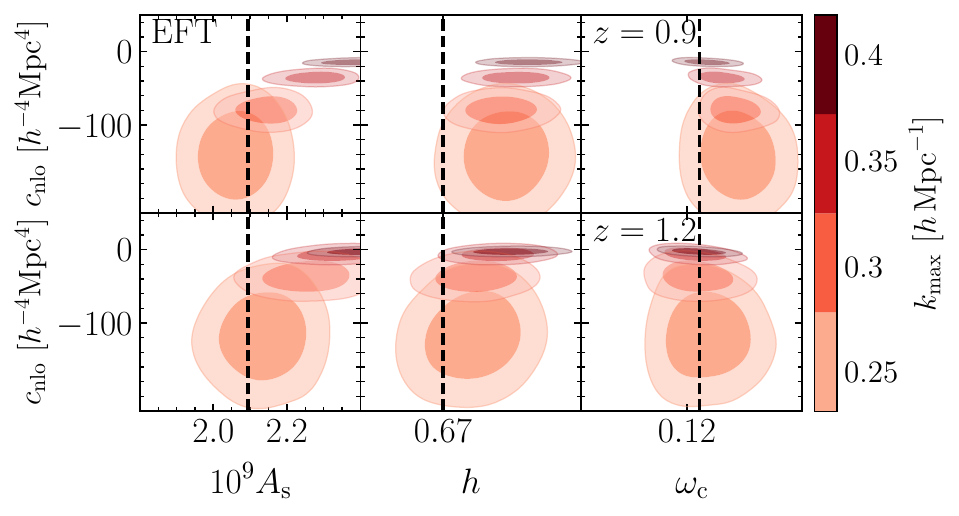}
    \includegraphics[width=1.\columnwidth]{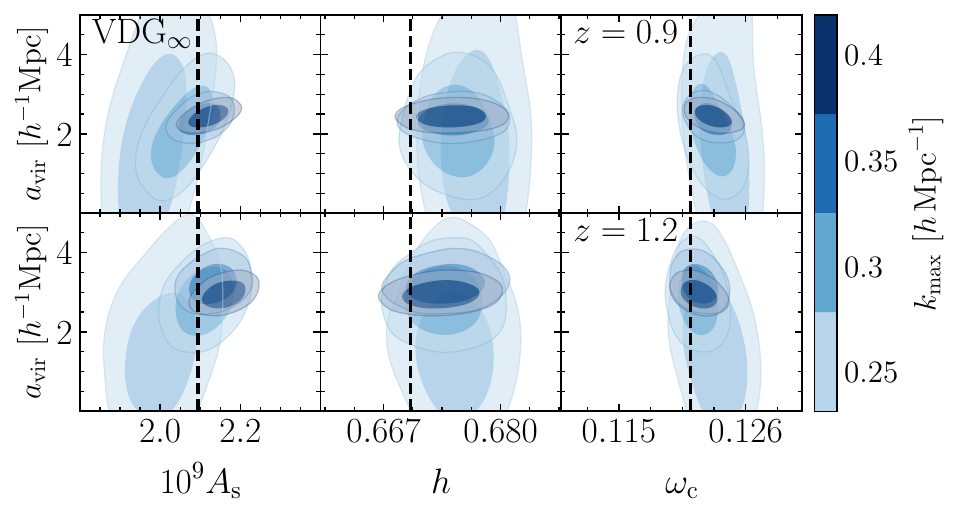}
\caption{2-dimensional posterior distribution of the cosmological parameters $(\As,\,h,\,\omegac)$, the next-to-leading order parameter $\cnlo$ of the \eft model, and the parameter $\avir$ of the \vdg model. The \emph{left- and right-hand side panels} correspond to the \eft and \vdg model posteriors, respectively. In each panel, the \emph{first and second row} corresponds to the $z=0.9$ and $z=1.2$ snapshots, respectively.}
\label{fig:contour_CnloAvir}
\end{figure*}

In the \eft model, higher-order corrections that are not fully reproduced by the leading-order expansion of the moment generating function are captured through the next-to-leading order parameter $\cnlo$. In the \vdg model this parameter is not necessary, since the exponential prefactor is modelled analytically as described in Eq. (\ref{eq:Winfty}). The large-scale limit of the VDG includes an extra degree of freedom represented by the parameter $\avir$, which is meant to reproduce the kurtosis of the pairwise velocity difference distribution.

In Fig.~\ref{fig:contour_CnloAvir} we show the 2-dimensional marginalised posterior distribution of $\cnlo$ and $\avir$ against the three cosmological parameters $(\As,\,h,\,\omegac)$. Consistently with the approach adopted in Sects.~\ref{sec:counterter_tests} and \ref{sec:noiseExploration}, we explore scale cuts with $\kmax\geq 0.25\kMpc$ for the two closest redshifts. In this case, we observe a notable drift in the posteriors of $\cnlo$, which is significantly correlated to shifts in the scalar amplitude $\As$. This behaviour suggests that, similarly to the lower-order counterterms, $\cnlo$ absorbs residual effects beyond its physical motivation. On the other hand, the $\avir$ parameter exhibits a stable trend as a function of $\kmax$, with no significant degeneracy with any cosmological parameter.


\section{Conclusions}
\label{sec:conclusions}

In this work, we compared different modelling prescriptions for the galaxy power spectrum in redshift space. Specifically, we tested the performance of three different redshift-space models: \bacco, \eft, and \vdg.  
To test their performance, we made use of a set of synthetic samples of H$\alpha$-emitting galaxies, populated in comoving snapshots of the Flagship I simulation and tailored to reproduce the clustering properties of the main spectroscopic sample that will be collected by \Euclid. The relative quality of the different models was assessed by means of three performance metrics: the \fom, the \fob, and the \pvalue, to quantify constraining power, accuracy, and goodness-of-fit, respectively. While the assumed \hod for building the catalogues may impact the performance of the different theoretical approaches \citep{DESI_HODSYS_FS}, the errors associated to the assumed \hod are currently under investigation in Euclid Collaboration: Gambardella et al. (in prep.).

The main conclusion from the comparison among the three different models is that, for all the metrics, \vdg and \bacco exhibit a consistent behaviour across most of the considered redshifts and scale cuts. However, at the highest redshifts, the \vdg model benefits the most from the information contained on smaller scales. In contrast, the \eft model shows an increasing trend in the \fob at $\kmax\geq 0.25\kMpc$ -- eventually leaving the $2\sigma$ region -- which is further associated with a smaller ($\sim 20\%$) \fom compared to \vdg and \bacco.

In detail, in terms of \fom, in the baseline analysis  (where both \ept models fix $\gtwoone$ and $\nptwotwo=0$)
\bacco consistently delivers a larger constraining power on cosmological parameters at $\kmax \lesssim 0.25 \kMpc$ compared to \eft and \vdg.
Moreover, the amount of information that can be extracted saturates at a typical scale of $\kmax \approx 0.35 \kMpc$. On the contrary, the \vdg model benefits more of the additional signal at $\kmax \geq 0.35 \kMpc$, displaying a monotonically increasing \fom over the whole range of scales considered.  
In terms of \fob, \bacco and \vdg hold inside the $1\sigma$ and $2\sigma$ region, respectively, when including more nonlinear information coming from smaller scales. Conversely, the \eft model shows a growing trend of the \fob at smaller scales, particularly for the two closest redshifts, suggesting a range of validity that stops at $\kmax \simeq 0.25 \kMpc$, in agreement with previous works. We further examined the constraints on individual cosmological parameters. The Hubble parameter $h$ is recovered with 1\% accuracy for all models. The parameter $\omegac$ is recovered within 3\% accuracy, with \bacco and \vdg showing similar precision. \vdg outperforms \bacco and \eft in constraining $\As$, achieving better than 5\% precision at all redshifts when including the smallest considered scales.

While validating our bias assumptions and assessing potential failures of the \ept models, we found that the inclusion of higher multipoles in the fit (our baseline setup) generally improves the constraining power (few percent) of the \vdg model without introducing significant biases. However, with the \eft model, the additional modelling of the quadrupole and hexadecapole leads to an increasing bias -- leaving the $2\sigma$ region at $\kmax>0.20\kMpc$ -- and a worse goodness-of-fit at smaller scales -- leaving the $2\sigma$ region at $\kmax\geq0.30\kMpc$ for the closest redshift. In terms of stochastic contributions, we observed a $k^2$-dependent factor, generally necessary in both models to avoid the presence of bias on the recovered cosmological parameters. Nevertheless, neglecting such term affects \vdg only at the smallest scales of the closest redshift bins, while enhancing its constraining power.

Counterterms are an essential ingredient in \eft to account for nonlinearities and \rsd at smaller scales. However, the impact of these terms seems to be limited. First, it can be seen that \eft already increases its \fob beyond the $2\sigma$ region at $\kmax \gtrsim 0.25 \kMpc$. Moreover, the amplitude of $\czero$, $\ctwo$, and $\cfour$ -- in general -- departs significantly from unity (on the order of tens, not expected for a perturbative approach). Otherwise, in the \vdg case, the corrections introduced by counterterms seem, in general, smaller, with $\czero$ generally compatible with zero. Additionally, higher-order corrections, encapsulated in the damping function in \vdg, are crucial for maintaining accuracy at smaller scales, ensuring robust performance with minimal biases (not going beyond the $2\sigma$ region). On the other hand, \eft's higher-order corrections are encapsulated in $\cnlo$, which is poorly constrained at intermediate scales.

In summary, our baseline analysis clarifies both the strengths and limitations of \bacco, \eft, and \vdg for modelling the anisotropic power spectrum. Our results show that the choice of model and the adopted scale range play a central role in optimising parameter constraints. In particular, improved mappings from real to redshift space, as implemented in \vdg and \bacco, substantially extend the range of validity of the theoretical predictions up to $\kmax\approx 0.40\kMpc$. While \bacco enables accurate analyses over a broad range of nonlinear scales, this comes at the cost of significantly higher computational demands compared to \eft and \vdg. Otherwise, although models such as \eft still require refinement to reach smaller scales, their fully theory-driven construction provides more flexibility when exploring modified gravity or alternative dark energy scenarios. Finally, we find that counterterms and higher-order corrections are essential for achieving accurate predictions; although stabilising them remains challenging, ongoing efforts toward physically motivated priors are a promising path forward. All these developments are essential for building increasingly precise and reliable theoretical frameworks for upcoming large-scale structure surveys.

\begin{acknowledgements}

\AckEC
BC and MC acknowledge support by the Spanish Ministry of Science and Innovation (MCIN/AEI/10.13039/501100011033) under grants PID2021-128989NB and PID2024-156844NB; and the programme Unidad de Excelencia Mar\'{\i}a de Maeztu, project CEX2020-001058-M. We acknowledge the use of resources from the Spanish Supercomputing Network (RES) provided by the Barcelona Supercomputing Center (BSC) in MareNostrum 5 under allocations AECT-2024-3-0020, 2025-1-0045, and 2025-2-0046. MC and MAB acknowledge support from project “Advanced Technologies for the exploration of the Universe”, part of Complementary Plan ASTROHEP, funded by the European Union - Next Generation (MCIU/PRTR-C17.I1).
This research made use of matplotlib, a Python library for publication quality graphics \citep{Hunter:2007}.

\end{acknowledgements}

%
%
\bibliography{refs.bib}

%

\begin{appendix}

\section{Hybrid model performance at higher precision level}
\label{app:bacco_performance}

The \bacco emulator employs neural networks to interpolate predictions from a hybrid bias model across a range of cosmological parameters, using outputs from cosmologically rescaled \nbody simulations (\citealt{AnguloWhite2010}). To ensure accuracy across cosmologies, we normalise the simulation outputs using a basis of \pt predictions, which reduces the dynamical range of the multipoles of the 15 bias terms.

To build the full hybrid model prediction, we multiply the simulation-based emulator outputs by their corresponding \pt predictions evaluated at the target cosmology. Two theoretical models are used to compute the baseline spectra: a linear smeared \bao model for matter-only statistics \citep{Angulo_2021}, and the Lagrangian Perturbation Theory (LPT) predictions implemented in \texttt{Velocileptors} \citep{Chen2020}. However, computing the analytical predictions is a computational bottleneck. To address this, we constructed separate emulators for each theoretical component.

Initially in \bacco, the \pt emulators were constructed using a coarser $k$-binning, consisting of 50 logarithmically spaced bins over the range $k \in \brackets{0.001, 1}\kMpc$. In contrast, the \nbody emulators used a finer binning of 100 logarithmically spaced bins over the same $k$-range. To reconcile this mismatch when building the hybrid prediction, we interpolated between the two sets of components. This approach proved sufficient for earlier applications involving survey data, where the power spectrum is typically measured on a linear binning with $\Delta k = 0.005\kMpc$. However, the Flagship simulation employs a much finer $k$-binning, with $\Delta k = 0.0016\kMpc$. This denser sampling requires the \pt emulators to be interpolated at most $k$-points, leading to small but systematic biases in the model predictions when using a coarser $k$-binning.

\begin{figure}[h]
	 \centering
    \includegraphics[width=1.\linewidth]{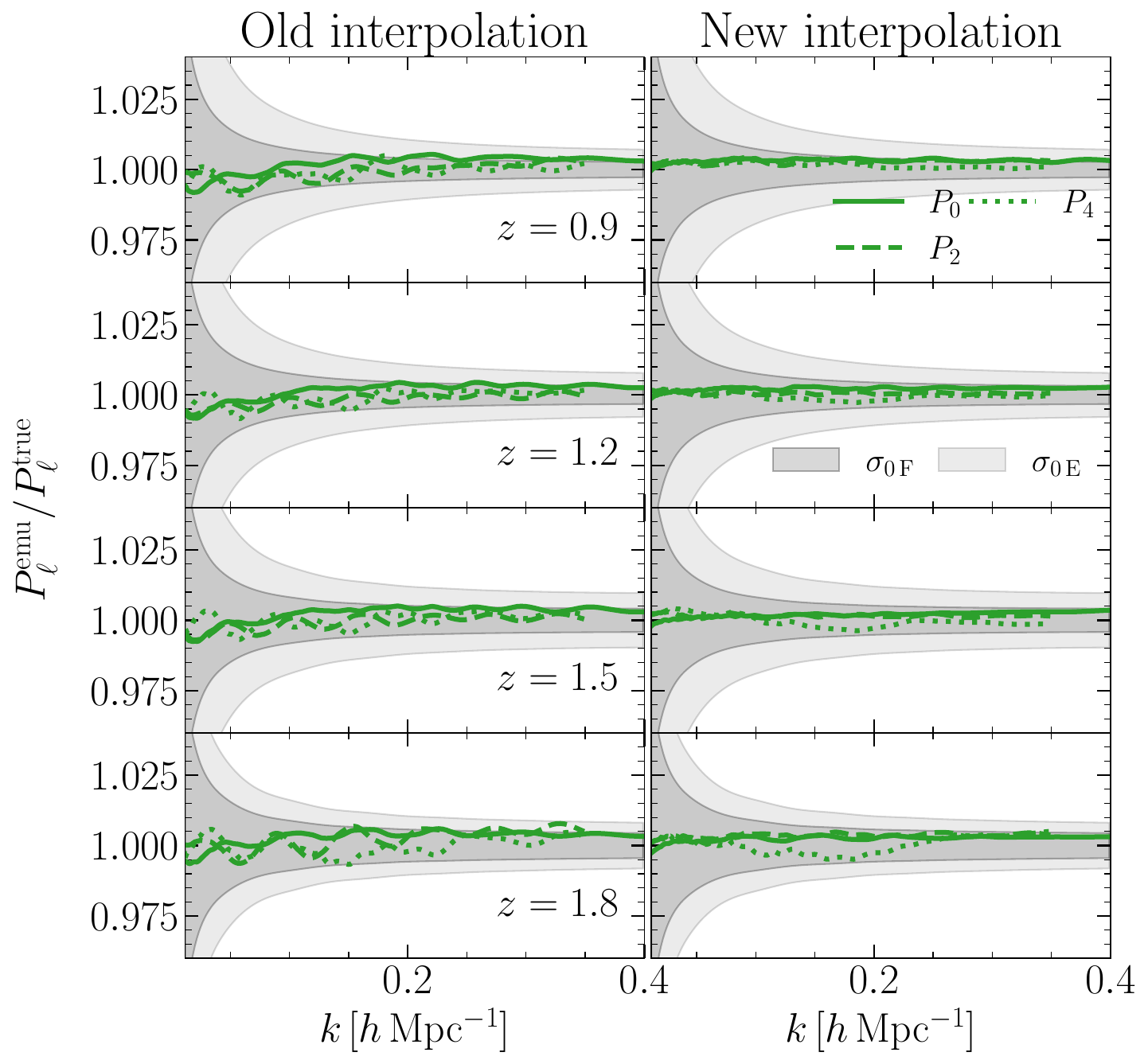}
    \caption{Accuracy of the previous and updated implementations of the \pt emulators in \bacco, compared to the direct (`true') evaluation of the \pt model. Grey shaded regions indicate the expected uncertainties for the monopole of the Flagship (F) and \Euclid (E) survey volumes. \emph{Rows} correspond to different redshifts, while \emph{columns} compare the old and new interpolation strategies. Oscillatory artifacts and scale-dependent tilts present in the older approach are significantly reduced with the updated method. Some artifacts are still present in the hexadecapole at intermediate scales and higher redshifts.}
    
    \label{fig:accuracyPTbacco}
\end{figure}

To mitigate these biases, we implemented two key improvements: (1) we regenerated the perturbation theory emulators using the same $k$-binning as the \nbody emulator, and (2) we adopted a modified neural-network architecture better suited for this task. In particular, we replaced the PCA-based input with the full power spectra, leveraging the smoothness of the perturbation theory predictions. Additionally, we increased the size of the hidden layer from 35 to 150 neurons to enhance the capacity of the model. These modifications led to a significant improvement in accuracy, as demonstrated in Fig.~\ref{fig:accuracyPTbacco}. In this figure, we illustrate the differences between the old and new interpolation strategies, shown relative to the expected monopole uncertainties for the Flagship and \Euclid volumes. In the left panels, the old strategy exhibits artificial oscillations -- visible as wiggles -- as well as scale-dependent tensions between large and small scales, particularly at the lowest and highest redshifts. Regardless of interpolation version, additional artifacts appear in the hexadecapole at redshifts $z \gtrsim 1.5$. Although these effects are smaller than the expected hexadecapole uncertainties, we adopt a conservative approach and limit our analysis to the monopole and quadrupole only.

While the updated interpolation method significantly reduces these issues, residual deviations of the same order as the Flagship volume uncertainties persist on scales around $k_{\rm max} = 0.4\kMpc$. For this reason, we restricted our \bacco analysis to the \Euclid volume. Achieving the precision required for a full Flagship-volume analysis will require further developments in our emulator methodology, which we leave for future work.

\section{2-dimensional posteriors exploration}
\label{app:2d_bias_exploration}
\begin{figure*}[h]
	 \centering
    \includegraphics[width=2.\columnwidth]{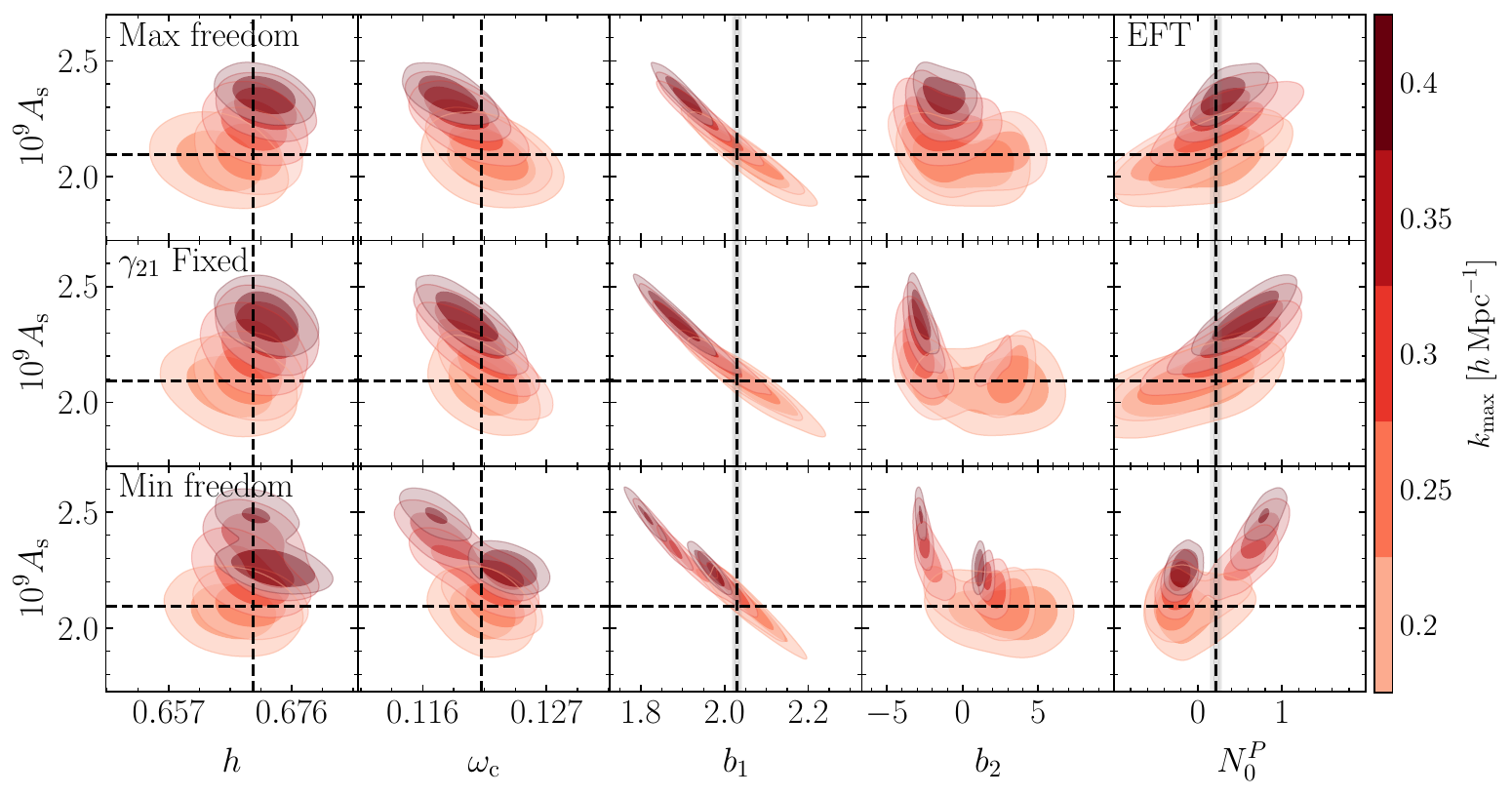}
    \includegraphics[width=2.\columnwidth]{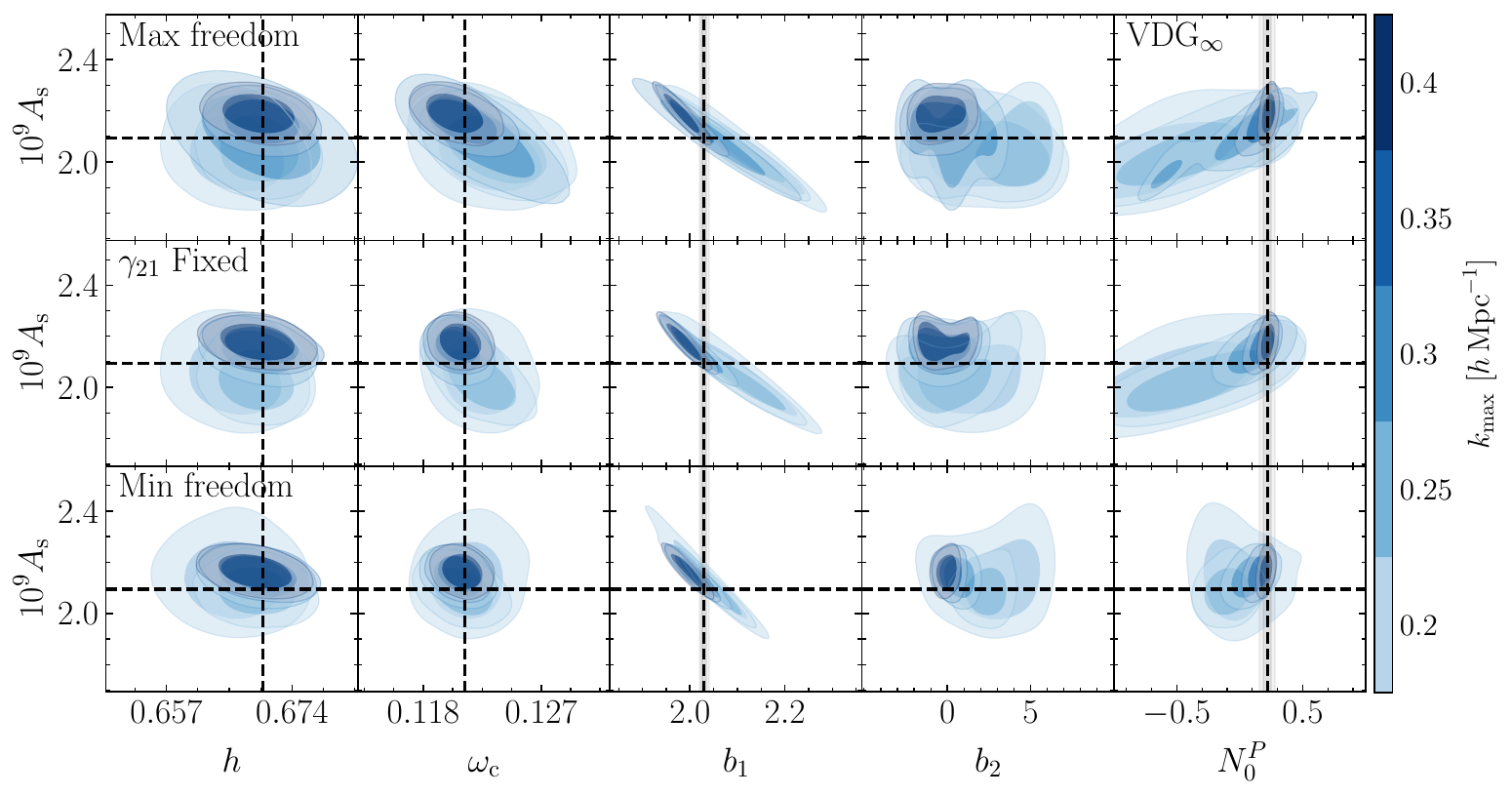}
\caption{2-dimensional marginalised posterior distributions for the $z=1.5$ snapshot. Cosmological parameters ($h,\, \omegac,\, \As$) are shown, along with the linear and second-order local biases and the scale-independent shot-noise parameter. The \emph{top and bottom panels} correspond to the \eft and \vdg models, respectively. Different colours indicate the maximum wave mode of the fit, as indicated by the colour bar. From top to bottom, \emph{different rows} show the impact of imposing relations on the tidal biases $\gtwo$ and $\gtwoone$. Fiducial values of the cosmological parameters are indicated with dashed vertical and horizontal lines. For $\bone$ and $\npzero$, dashed lines indicate the best-fit parameters obtained from the large-scale limit of the matter power spectrum (see text for details).}
 \label{fig:contour_AllBiasFree}
\end{figure*}

A more detailed examination of the influence of prior relations on non-local biases is presented in Fig.~\ref{fig:contour_AllBiasFree}, which directly shows the 2-dimensional marginalised posteriors for a subset of parameters. The top and bottom panels correspond to the \eft and \vdg models, respectively . Different colours indicate the maximum wave mode $\kmax$ of the fit, while each row refers to a different treatment of the tidal bias parameters. In addition to the cosmological parameters, we also show the posterior distribution of $\bone$ and $\npzero$, since these parameters are significantly degenerate with $\As$. Additionally, we include $\btwo$ to have a more comprehensive picture for the local bias parameters. The fiducial values of the cosmological parameters are shown with dashed lines, along with the large-scale limit values for $\bone$ and $\npzero$ recovered from real-space fits of the matter power spectrum \citep[see][for a detailed explanation on this derivation]{Euclid-Pezzotta}. For compactness, we only display the results for the $z=1.5$ snapshot, since it corresponds to the median redshift of \Euclid. 

In the \eft analysis (top panel), with all biases allowed to vary freely, we find that $\As$ is biased at the $2\sigma$ level. On the other hand, $h$ and $\omegac$ are biased at most at the $1\sigma$ level. As expected, the linear bias shows a strong degeneracy with $\As$ and exhibits significant bias when fitting smaller scales. Interestingly, the constant shot-noise parameter is generally recovered at the $1\sigma$ level, and $\btwo$ remains stable. When fixing $\gtwoone$, $\omegac$ is slightly better constrained at larger scales, but at the cost of introducing bi-modalities in $\btwo$ at intermediate scales and recovering negative values at smaller scales. Notice that such behaviour appears on scales where the \fob already starts to increase, indicating a breakdown of the model. Since the directions of these bi-modalities are orthogonal to $\As$, it is unlikely that they introduce additional biases in this parameter. Furthermore, the constraint on the constant noise parameter is also degraded in this regime. Finally, fixing both tidal biases introduces bi-modalities across all displayed parameters at smaller scales, with $\npzero$ also deviating from its fiducial value. As a final note, we recall that the bi-modalities in the cosmological parameters appear on scales where the model is already breaking down in terms of \fob, as shown in Fig.~\ref{fig:compare_EFTvsVDG_AllFreeAndg2g21Fix}. Moreover, fixing only $\gtwo$ to the excursion-set relation increases the constraining power on $\npzero$ and $\btwo$, but the impact on cosmological parameters is negligible. As a consequence -- and because previous works analysing the power spectrum plus bispectrum \citep[\eg][]{EggScoSmi2106} have shown that this is not, in general, a good approximation -- we do not include this case in the plot.

For the \vdg case (bottom panels of Fig.~\ref{fig:contour_AllBiasFree}), the main conclusion is that the posteriors exhibit an improved behaviour across all bias configurations, as there is no significant parameter drift from their fiducial values nor strong bi-modalities. The most notable feature is that fixing $\gtwoone$ reduces the degeneracies in the $\As$-$\omegac$ plane and significantly tightens the constraints on the $\npzero$ parameter compared to the maximal-freedom scenario (at the same $\kmax$). Furthermore, fixing both tidal biases simultaneously results in less biased and tighter posterior distributions, thereby lowering the \fob and substantially increasing the \fom in the minimal-freedom scenario. However, when compared to the intermediate configuration that only fixes $\gtwoone$, the additional gain appears marginal, as already observed in Fig.~\ref{fig:compare_EFTvsVDG_AllFreeAndg2g21Fix}.
For completeness, although this case is not displayed, we have tested that fixing $\gtwo$ does not by itself have any significant impact on the cosmological parameters, in the sense that the marginalised 2-dimensional posteriors do not change appreciably compared to the maximal-freedom case.

For the purpose of model validation and comparison, our focus is on recovering cosmological parameters without introducing additional biases arising from the breakdown of the adopted bias relations. As shown in Fig.~\ref{fig:contour_AllBiasFree}, the fiducial cosmological parameters are better recovered for both \pt models when $\gtwoone$ is fixed, compared to the minimal-freedom scenario, which introduces additional bi-modalities in the \eft case. Thus, we adopt fixing $\gtwoone$ as a robust baseline choice for the comparisons presented in Sect.~\ref{sec:results}.

Finally, we recall that similar conclusions hold for the four different redshifts explored in this work, for completeness we present violin plots with the 1D marginalised posterior distributions in Appendix~\ref{sec:app_1D_Exploration}.

\section{1-dimensional posteriors exploration}
\label{sec:app_1D_Exploration}

\begin{figure*}[h!]
	 \centering
    \includegraphics[width=1.\linewidth]{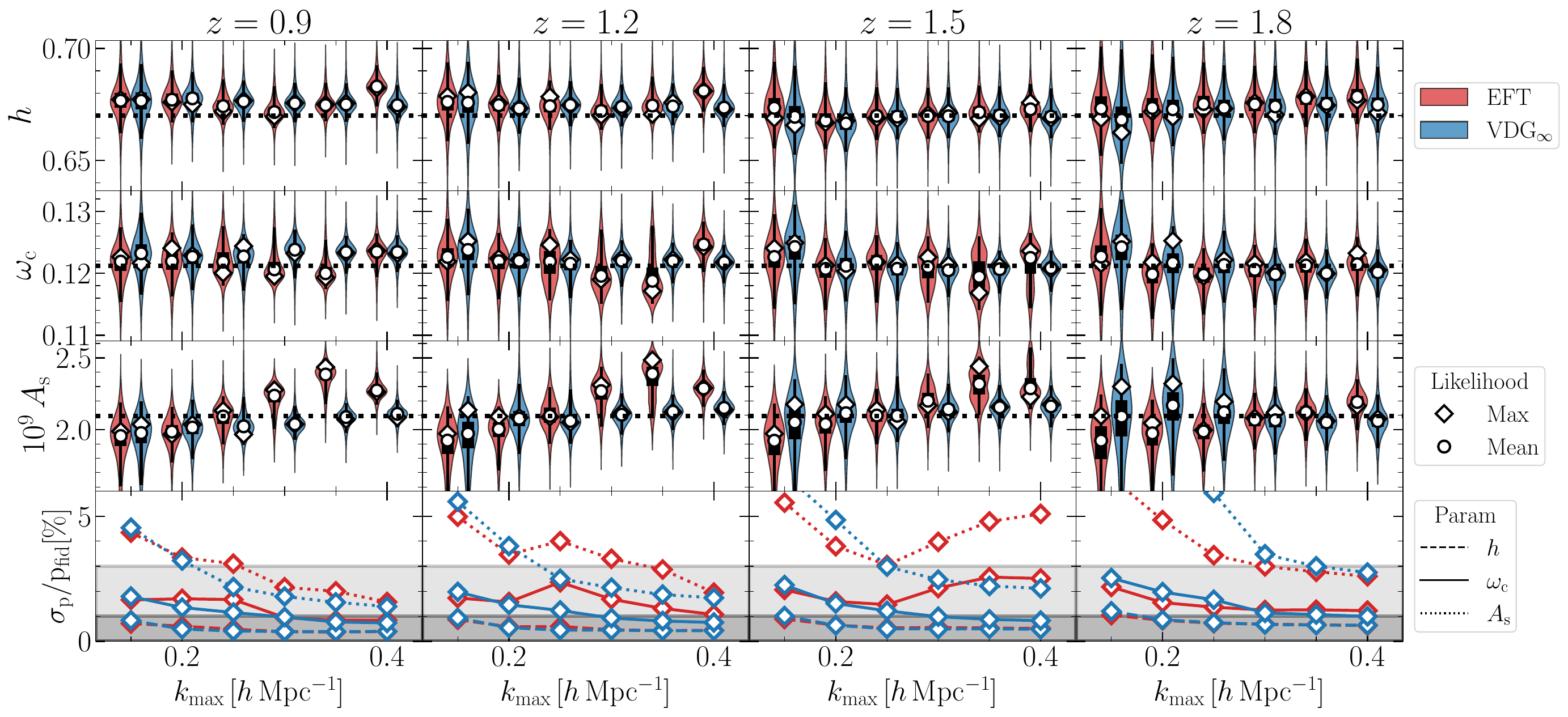}
\caption{Violin plot for the minimal-model fits as a function of the maximum wave mode, $\kmax$. In the \emph{first three rows}, the circles show the mean marginalised posterior of each parameter (\emph{the first row} corresponds to $h$, the \emph{second} to $\omegac$, and the \emph{third} to $\As$), while the diamonds correspond to the maximum-a-posteriori point where the full likelihood is maximised. The black boxes and whiskers indicate the $68\%$ and $95\%$ limits, respectively. The blue and red regions show the 1D marginalised distributions for each parameter, for either the \eft or \vdg models, as described in the legend.
The \emph{bottom panel} presents the ratio of the error on each parameter (as indicated on the labels) to its fiducial value. \emph{Each column} corresponds to a different redshift. The range displayed for each parameter is $h \pm 5\%$, $\omegac \pm 10\%$, and $A_s \pm 30\%$. For reference, the shaded regions in the \emph{bottom panel} show the $1\%$ and $3\%$ limits.}
\label{fig:violin_minimalModel}
\end{figure*}

In the first three rows of Fig.~\ref{fig:violin_minimalModel}, we show the violin plots for the three cosmological parameters corresponding to the minimal bias case. In the same figure, the fourth row shows the significance at which each parameter is recovered. This plot allows us to visualise, in a more compact way, the 1-dimensional marginalised posterior distributions for each parameter, each value of $\kmax$, each \rsd model, and each redshift.

In the first panel, focusing on the red symbols, which correspond to the \eft case, we see that in terms of bias and constraining power, $h$ is the most stably recovered parameter across all scales and redshifts. Moreover, in the bottom panel we see that it is the parameter whose constraining power saturates earliest. Looking at the second and third rows, we observe that the bi-modalities seen for \eft in Fig.~\ref{fig:contour_AllBiasFree} for the snapshot at $z=1.5$ also appear, to some degree, in all snapshots except the lowest redshift one, where only mild skewness remains. These bi-modalities are also responsible for reducing the gain in constraining power for both parameters, as illustrated in the bottom panel. Another interesting feature of this plot is that it allows us to assess the impact of projection effects (see Appendix~\ref{sec:PriorsImpact} for more details), since both the mean marginalised parameter and the maximum-a-posteriori estimate -- where the likelihood is maximised -- are shown. Notice that in most cases both the mean marginalised value and the maximum-a-posteriori value lie within the 68\% confidence interval.

Turning now to the blue symbols, which correspond to the \vdg case, we see that the posterior distributions are generally better behaved: they look more Gaussian, lie closer to the fiducial values, and tighten when adding information from smaller scales without biasing the parameters. One additional feature to note is that, at the highest redshift, the maximum-a-posteriori estimate lies slightly outside the 68\% interval relative to the mean marginalised value; however, this effect disappears once the model includes information from sufficiently small scales, leading to improved posterior distributions. Finally, the gain in constraining power is more stable for this model, and we recover most of the available information for each parameter by $\kmax = 0.4\kMpc$, since the errors tend to flatten at smaller scales -- with the only exception of $\As$ at the highest redshift.

\section{Hexadecapole information}
\label{sec:app_hexaInfo}

In our baseline comparison of the \rsd models presented in Sect.~\ref{sec:RSD_comparison}, we used information from the first three even multipoles ($P_0,\; P_2,\; {\rm and}\; P_4$). Here, we validate that this configuration is optimal for our study.

\begin{figure*}
	 \centering
    \includegraphics[width=2.\columnwidth]{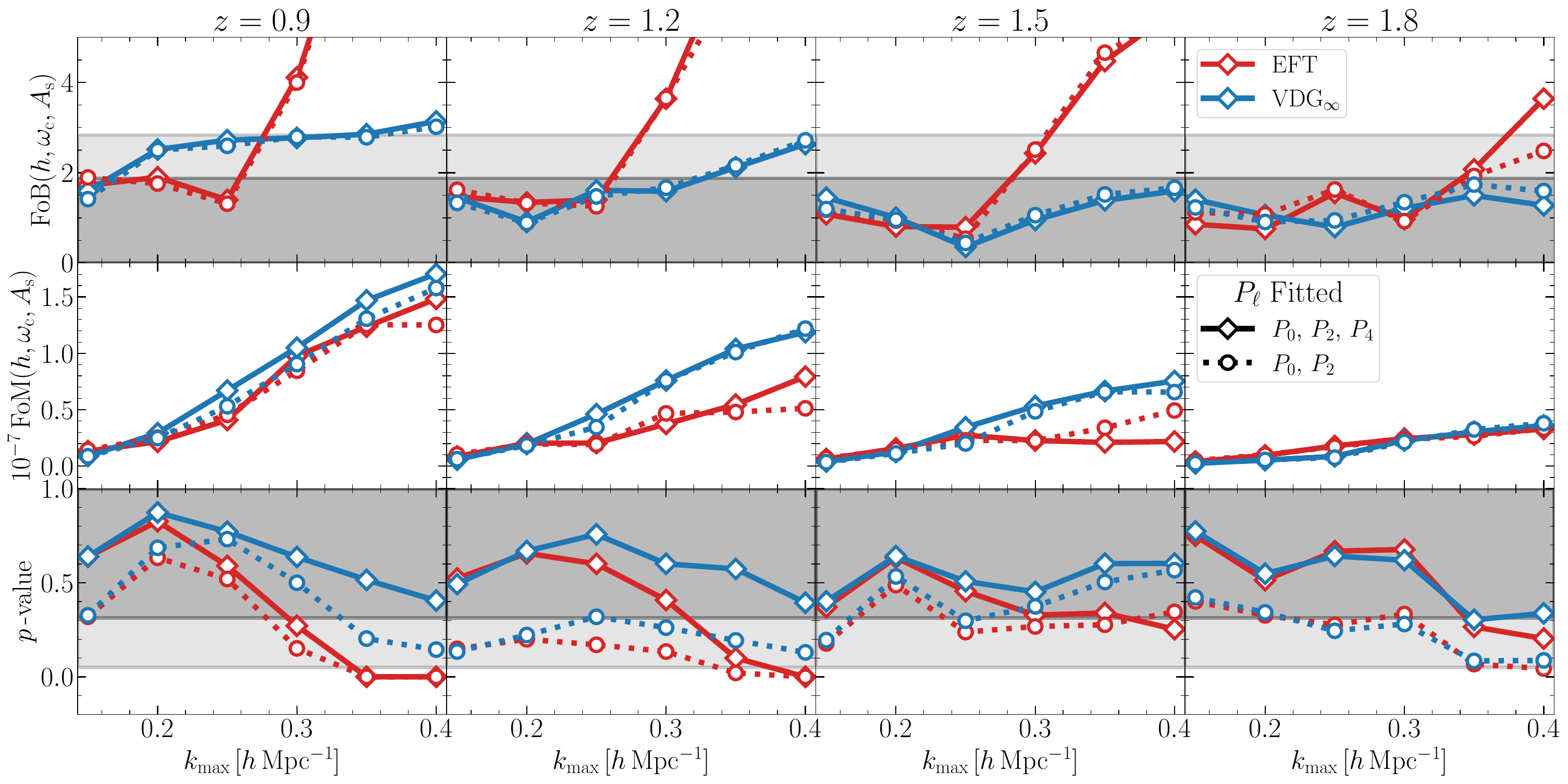}
    \caption{Same as Fig.~\ref{fig:compare_EFTvsVDG_AllFreeAndg2g21Fix}, but comparing the impact of neglecting the information from $P_4$, indicated by circles and dotted lines.}
    \label{fig:compare_EFTvsVDG_NoHexa}
\end{figure*}

In Fig.~\ref{fig:compare_EFTvsVDG_NoHexa}, we show that for the maximal bias case, neglecting the information from the hexadecapole leads to only marginal changes in the \fob and \fom. Nonetheless, the \pvalue is consistently degraded, indicating a poorer description of the dataset. This degradation in the goodness-of-fit arises because the hexadecapole helps to better constrain the nuisance parameters. Even though the cosmological constraints remain essentially unchanged, tighter constraints on the full parameter space allow the model to provide a more accurate description of the dataset as a whole.

\section{Comparison between \bacco and \vdg priors}
\label{app:BACCO_exploration}

The results discussed in Sect.~\ref{sec:RSD_comparison}, primarily those shown in Fig.~\ref{fig:compare_1D_EFTvsVDGvsBACCO_EucliVolume}, appear unexpected: the conclusions do not fully agree with our baseline analysis, and the information extracted by \bacco saturates at larger-than-expected scales. Here we describe several potential sources of spurious constraints in the analysis.

One of the main differences in the perturbation-theory scheme between \bacco and \vdg is the large-scale normalisation imposed on the $\delta^2\delta^2$ term, as described in Eq.~(34) of \cite{Pezzotta2021}. This correction is not applied in \bacco; instead, \bacco subtracts the mean of the $\delta^2$ field to build the $\delta^2\delta^2$ contribution \citep[see Eq.~1 in][]{PellejeroIbanez_2024}. We explicitly removed this normalisation in \vdg and repeated one of the analyses at large scales; we found that this difference is not the primary cause of the extra constraining power seen in \bacco at large scales. Nonetheless, and although the effect is minor, the normalisation in \vdg does impact mainly the recovered bias parameters.

Another significant difference between \vdg and \bacco is the absence of counterterms in the \bacco approach. We repeated one of the \vdg analyses while omitting the counterterms and found that the gain in constraining power on cosmological parameters was not significant. Therefore, we conclude that counterterms are unlikely to be responsible for the tighter cosmological constraints seen in \bacco.

An additional difference concerns the priors adopted in the \bacco runs, which were tighter for some nuisance and cosmological parameters. The cosmology priors were chosen to remain within the emulation range of \bacco; as discussed in \cite{Euclid-Pezzotta}, increasing the emulation range in \bacco is computationally expensive. In particular, the prior on the scale-dependent noise was noticeably tighter in \bacco. We repeated the \vdg analysis using the \bacco priors and found that, in this case, the priors do affect the results -- especially at larger scales and for the nearest redshifts -- by producing artificially tighter constraints. Nevertheless, for the remaining redshifts and scales the choice of priors did not significantly alter the conclusions.

\section{Priors impact in the \eft model}
\label{sec:PriorsImpact}
\begin{figure*}
	 \centering
    \includegraphics[width=2.\columnwidth]{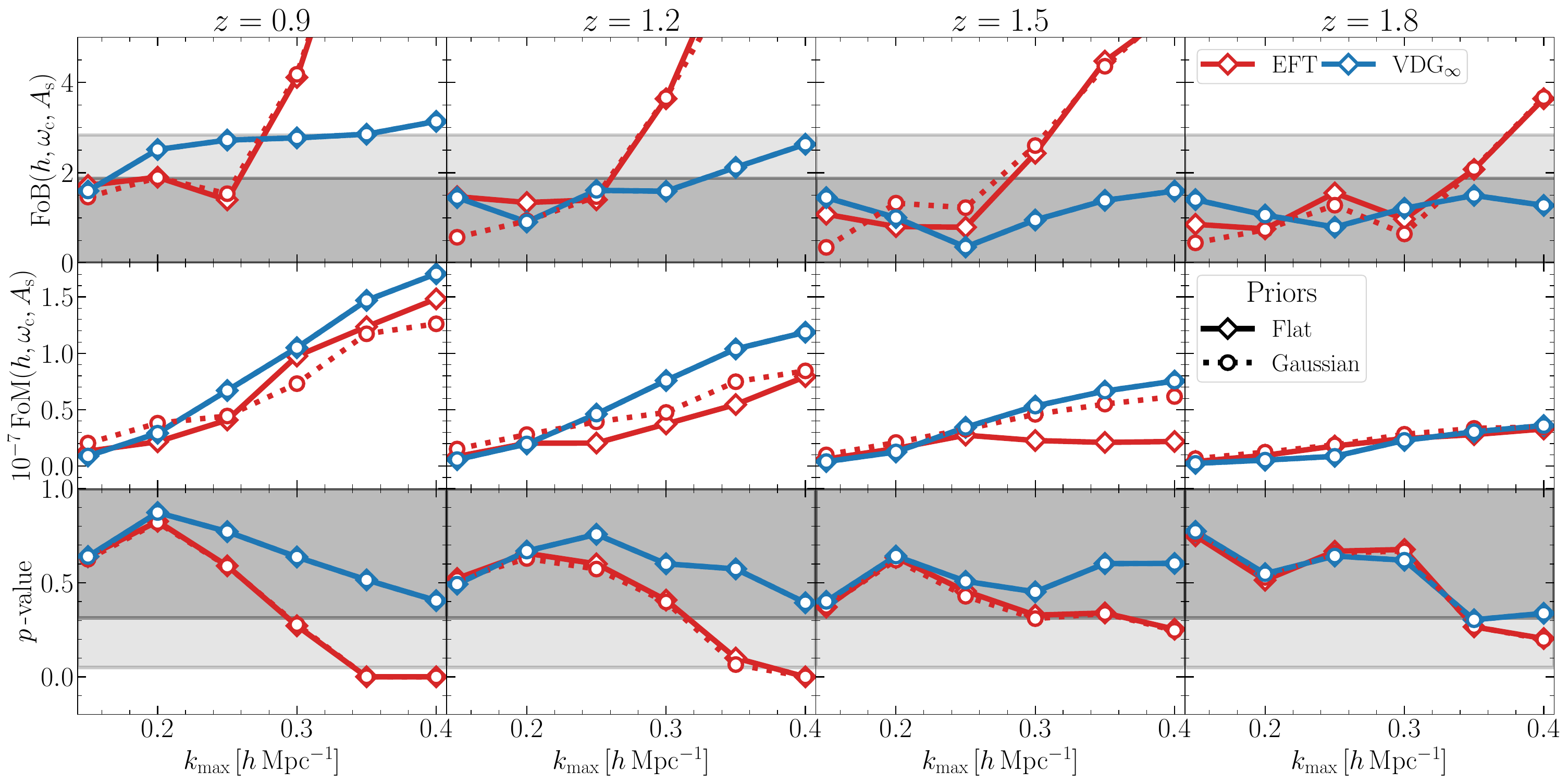}
    \caption{Same as Fig.~\ref{fig:compare_EFTvsVDG_AllFreeAndg2g21Fix}, but comparing the impact of imposing Gaussian priors, indicated by circles and dotted lines.}
    \label{fig:compare_EFTvsVDG_priorImpact}
\end{figure*}

As previously mentioned, several recent works \citep{Holm-PriorsStudy, Carrilho-Moretti} have shown that in the \eft modelling context, the marginalised posterior distributions of parameters can be strongly affected by prior volume effects. However, \cite{Holm-PriorsStudy} also demonstrated that when analysing a sufficiently large survey volume, these effects become negligible. To verify that our analysis is not impacted by such artefacts, we tested the effect of adopting Gaussian priors on all nuisance parameters -- as proposed in \cite{Philcox2022} -- and compared the results with those obtained using the broad flat priors adopted in our baseline analysis. This test was performed for the \eft model in the minimal-freedom configuration.

Figure~\ref{fig:compare_EFTvsVDG_priorImpact} shows that the impact at the level of summary statistics is small. Nevertheless, it is worth noting that imposing Gaussian priors on the nuisance parameters removes the bi-modalities present in $\npzero$, $\btwo$, and the counterterms. However, the disappearance of these bi-modalities does not translate into an improved recovery of the cosmological parameters.

In this work we have carried out only a simple assessment of the impact of prior assumptions, with the goal of confirming that our main conclusions remain robust when wide flat priors are applied. A more systematic study of prior effects is left for future work, particularly in the context of \Euclid survey specifications. In that configuration, the analysis may be more susceptible to prior volume effects -- especially when opening the parameter space to also include models of dynamical dark energy -- making it essential to understand their implications and how to mitigate them in order to avoid a prior-dominated inference.

\section{Full posteriors}
\label{app:full_rsd_posteriors}

Here we explicitly show the posterior distributions of the shared parameter space for the three \rsd models introduced in Sect.~\ref{sec:theory}.  

\begin{figure*}[h]
	 \centering
    \includegraphics[width=2. \columnwidth]{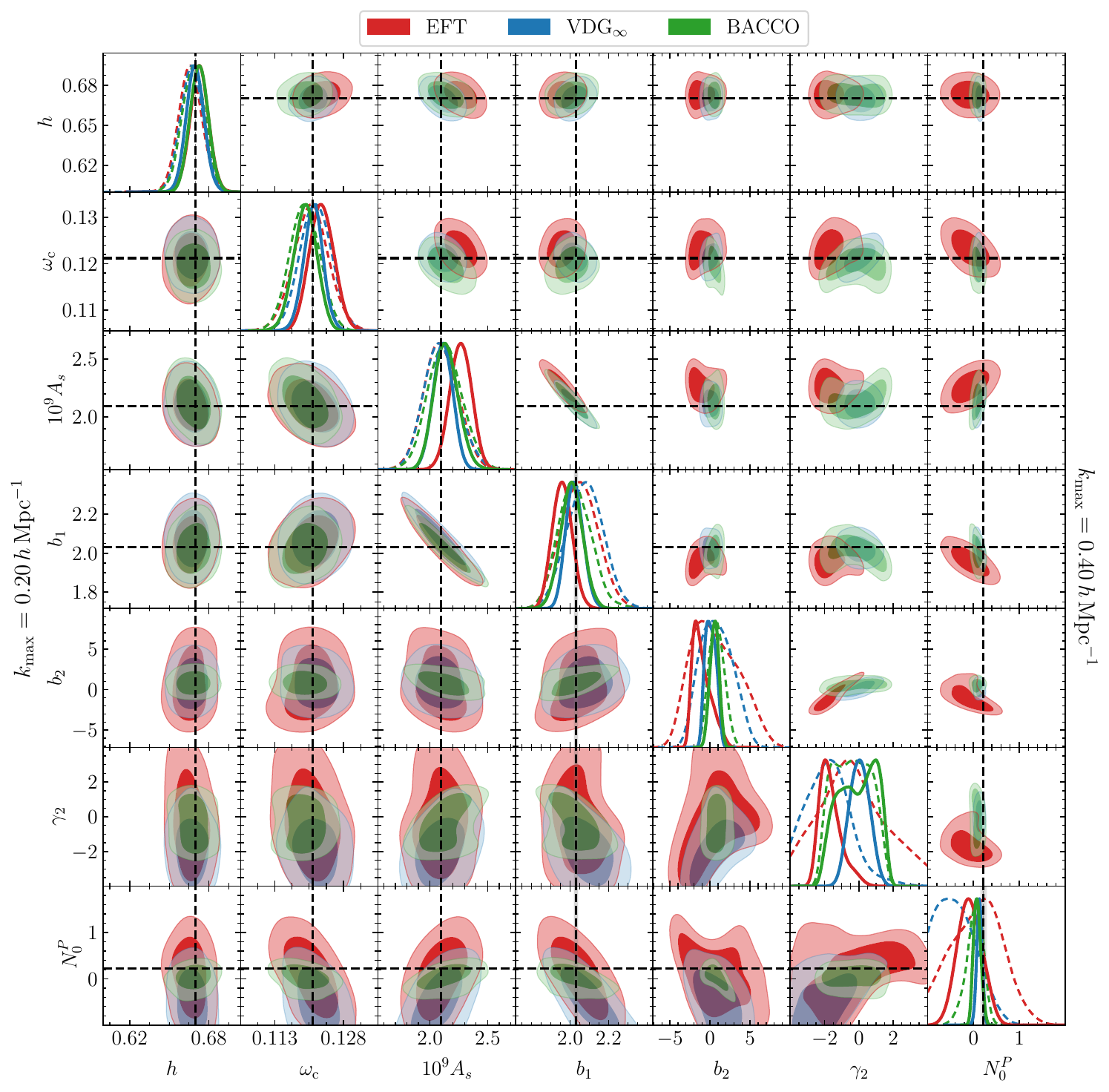}
\caption{Triangular plots showing the marginalised posterior distributions of the cosmological parameters $(h, \omegac, \As)$, the linear bias $\bone$, the second-order bias parameters $(\btwo, \gtwo)$, and the constant shot-noise parameter $(\npzero)$. Different colours correspond to different \rsd models, as indicated in the legend. The panels correspond to the $z=1.5$ snapshot. The \emph{lower and upper triangles} show the configurations with $\kmax = 0.20\kMpc$ and $\kmax = 0.40\kMpc$, respectively. In the 1-dimensional contours, dashed and solid lines correspond to the two scale cuts. Dashed black lines for the cosmological parameters indicate their fiducial values. For $\bone$ and $\npzero$, the dashed lines denote the best-fit parameters obtained from fitting only the large-scale limit of the $\Pgg/\Pmm$ ratio, as performed in previous works.}
\label{fig:triangle_EFTvsVDGvsBACCO_EuclidVolume}
\end{figure*}

A key aspect of the cross-validation between the different perturbative models is the level of consistency observed within their common parameter space. Below we summarise the main findings of this comparison, while Fig.~\ref{fig:triangle_EFTvsVDGvsBACCO_EuclidVolume} provides an illustrative example of the full posterior distributions.

At intermediate scales ($\kmax=0.20\kMpc$), the three models display very similar behaviour for the cosmological parameters. In contrast, \eft and \vdg yield significantly weaker constraints on the bias and noise parameters. This is likely related to \bacco hitting the prior boundaries for $\btwo$ and $\gtwo$, although this effect does not appear to significantly impact the cosmological parameters.

When nonlinear scales are included ($\kmax=0.40\kMpc$), both the \bacco and \vdg models remain stable and successfully recover the fiducial cosmological parameters. Notably, their posterior distributions for the bias and noise parameters also converge to the same region of parameter space, reinforcing the consistency between these two models at smaller scales. In contrast, the \eft model constrains the higher-order bias parameters in a region that differs from the one preferred by \vdg and \bacco, consistent with the shifts observed in the cosmological posteriors in the main text (see Figs.~\ref{fig:compare_EFTvsVDGvsBACCO_EucliVolume} and \ref{fig:compare_1D_EFTvsVDGvsBACCO_EucliVolume}). Although the constant-noise parameter inferred by \eft is relatively close to the values recovered by \vdg and \bacco, the \eft constraints remain weaker overall.

We notice that  the posteriors of \vdg and \eft behave similarly at larger scales in the whole parameter space, but \vdg converges to the posteriors of \bacco at smaller scales; this suggests that the strength of the \vdg model becomes particularly evident in the small-scale regime, where the improved modelling of \rsd offers significant advantages. Finally, the level of agreement between \vdg and \bacco at these scales works as a cross-validation of both models in the mildly nonlinear regime.

\end{appendix}

\end{document}